\documentclass{JINST}

\usepackage{amsfonts}
\usepackage{amsmath}
\usepackage{epsfig}
\title{Alignment of the CMS Silicon Strip Tracker during
  stand-alone Commissioning}

{
\author{W.~Adam, T.~Bergauer, M.~Dragicevic, M.~Friedl, R.~Fr\"{u}hwirth, S.~H\"{a}nsel, J.~Hrubec, M.~Krammer, M.Oberegger, M.~Pernicka
% ...
\\{Brown University, Providence, Rhode Island, USA}}

    \author{\\{$^a$}{Also at CERN, European Organization for Nuclear Research, Geneva, Switzerland}\\
    {$^b$}{Now at University of Hamburg, Institute for Experimental Physics, Hamburg, Germany}\\
   {{$^c$}{On leave from IFIN-HH, Bucharest, Romania}}\\
   {{$^d$}{Now at LLR-Ecole Polytechnique, France}}\\
   {{$^e$}{Now at Northeastern University, Boston,  USA}}\\
   {{$^f$}{Now at IFIC, Centro mixto U. Valencia/CSIC, Valencia, Spain}}\\
   {{$^g$}{Now at Universiteit Antwerpen, Antwerpen, Belgium}}\\
   {{$^h$}{Now at ETH Zurich, Zurich, Switzerland}}\\
   {{$^i$}{Also Peking University, China}}\\
   {{$^j$}{Now at Imperial College, London, UK}}\\
   {{$^k$}{Now at Universit\'{e} catholique de Louvain, UCL, Louvain-la-Neuve, Belgium}}\\
   {{$^l$}{Now at Eidgen\"{o}ssische Technische Hochschule, Z\"{u}rich, Switzerland}}\\
   {{$^m$}{Now at University of California, Davis, California, USA}}\\
   {{$^n$}{Now at Kansas University, USA}}\\
   {{$^o$}{Also at Research Institute for Particle and Nuclear Physics, Budapest, Hungary}}\\
   {{$^p$}{Now at University of Liverpool, UK}}\\
   {{$^q$}{Now at Massachusetts Institute of Technology, Cambridge, Massachusetts, USA}}\\
   {{$^r$}{Now at Stanford University, Stanford, California, USA}}\\
   {{$^s$}{Now at Harvard University, Cambridge, Massachusetts, USA} }\\
   {{$^t$}{Now at University of California, Los Angeles, California, USA}}\\
   {$^*$}{Corresponding author, E-mail: \email{gero.flucke@cern.ch}}\\
      {$^\dagger$}{Corresponding author, E-mail: \email{andrei.gritsan@cern.ch}}
 }
}

\abstract{
The results of the CMS tracker alignment analysis are presented
using the data from cosmic tracks, optical survey information, and the
laser alignment system at the Tracker Integration Facility at CERN.
During several months of operation in the spring and summer of 2007,
about five million cosmic track events were collected with a partially
active CMS Tracker. This allowed us to perform first alignment of the active
silicon modules with the cosmic tracks using three different statistical
approaches; validate the survey and laser alignment system performance;
and test the stability of Tracker structures under various stresses
and temperatures ranging from $+15^\circ$C to $-15^\circ$C.
Comparison with simulation shows that the achieved alignment precision
in the barrel part of the tracker leads to residual distributions similar
to those obtained with a random misalignment of $50$ ($80$)~$\mu$m
in the outer (inner) part of the barrel.
}


\begin{document}


\section{Introduction}
\label{sec:intro}

The all-silicon design of the CMS tracker poses new challenges in
aligning a system with more than 15,000 independent modules.
It is necessary to understand the alignment of the silicon modules 
to close to a few micron precision. 
Given the inaccessibility of the interaction region,
the most accurate way to determine the silicon detector positions
is to use the data generated by the silicon detectors themselves
when they are traversed in-situ by charged particles.
Additional information about the module positions is provided
by the optical survey during construction and by the Laser
Alignment System during the detector operation.

\subsection{CMS Tracker Alignment during Commissioning}
\label{sec:intro-tif}

A unique opportunity to gain experience in alignment of the CMS
silicon strip tracker~\cite{trackertdr,detector-paper} 
ahead of the installation in the underground cavern comes from 
tests performed at the Tracker Integration Facility (TIF).
During several months of operation in the spring and summer of
2007, about five million cosmic track events were collected.
The tracker was operated with different coolant temperatures 
ranging from $+15^\circ$C to $-15^\circ$C. About 15\% of the 
silicon strip tracker was powered and read-out simultaneously.
An external trigger system was used to trigger on cosmic track events.
The silicon pixel detector was only trial-inserted at TIF and was 
not involved in data taking.

This note primarily shows alignment results with the
track-based approach, where three statistical algorithms have
been employed showing consistent results. Assembly precision
and structure stability with time are also studied.
The experience gained in analysis of the TIF data will help 
evolving alignment strategies with tracks, give input into the stability of the
detector components with temperature and assembly progress,
and test the reliability of the optical survey information and the laser 
alignment system in anticipation of the first LHC beam collisions.

\subsection{CMS Tracker Geometry}
\label{sec:intro-geometry}

The CMS tracker 
is the largest silicon detector 
ever constructed.
Even with about 15\% of the silicon strip tracker activated during
the TIF test, more than 2,000 individual modules were read out.

\begin{figure}[bh] 
\begin{center}
\includegraphics*[angle=270,width=0.75\textwidth]{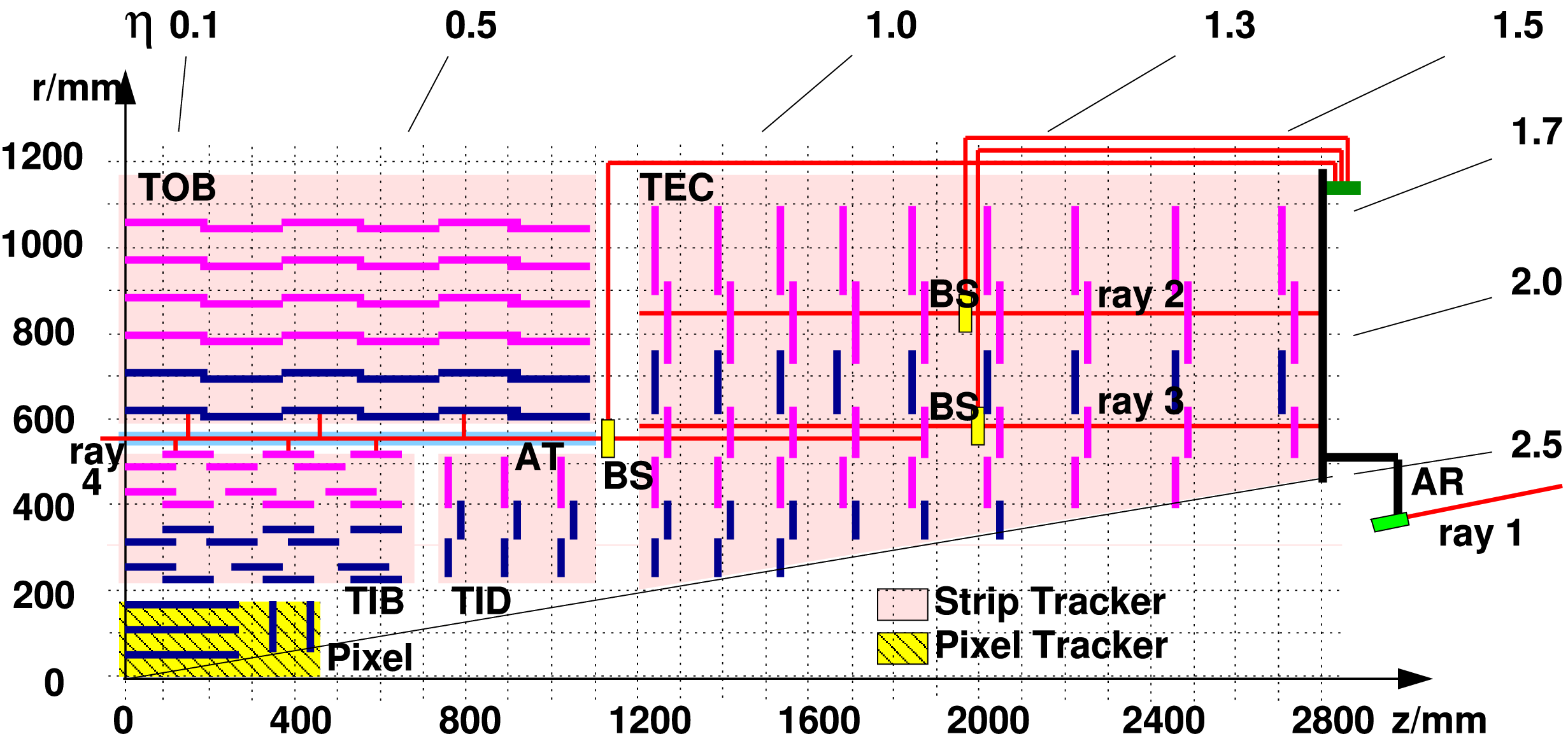}
\caption{\sl 
A quarter of the CMS silicon tracker in an $rz$ view. Single module positions
are indicated as purple lines and dark blue lines
indicate pairs of $r\phi$ and stereo modules.
The path of the laser rays, the beam splitters (BS) and the alignment tubes (AT)
of the Laser Alignment System are shown.
\label{fig:stripLayout}
}
\end{center}
\end{figure}

The strip detector of CMS is composed of four sub-detectors,
as illustrated in Fig.~\ref{fig:stripLayout}:
the Tracker Inner and Outer Barrels (TIB and TOB),
the Tracker Inner Disks (TID),
and the Tracker Endcaps (TEC).
They are all concentrically arranged around the nominal LHC beam axis
that coincides with the $z$-axis. The right handed, orthogonal CMS
coordinate system is completed by the $x$- and $y$-axes where the latter
is pointing upwards. The polar and azimuthal angles $\phi$ and $\theta$ are
measured from the positive $x$- and $z$-axis, respectively, whereas the
radius $r$ denotes the distance from the $z$-axis.

\begin{figure}[tb]
\begin{center}
\epsfig{figure=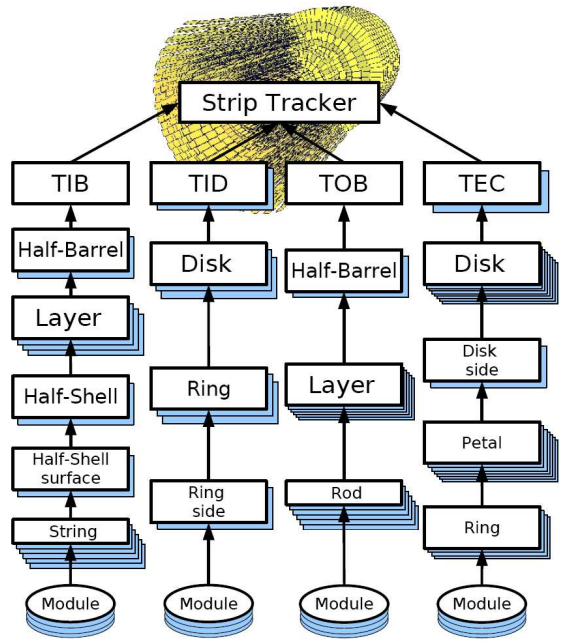,width=0.49\linewidth}
\caption{\sl 
  Hierarchy of the CMS silicon strip detector
  structures.
\label{fig:stripHierarchy}
}
\end{center}
\end{figure}

The TIB and TOB are composed of four and six layers, respectively. Modules are
arranged in linear structures parallel to the $z$-axis,
which
are named ``strings'' for TIB (each containing three modules) and ``rods'' 
for TOB  (each containing six modules). 
The TID has six
identical disk structures. 
The modules are arranged on both sides of ring-shaped concentric
structures, numbering three per disk.
Both TECs are built from nine disks, with eight ``front'' and ``back''
``petals'' alternatingly mounted on either side, with
a petal being a wedge-shaped structure covering a narrow $\phi$ region
and consisting of up to 28 modules, ordered in a ring structure as well.
We outline the hierarchy of the Strip detector structures in 
Fig.~\ref{fig:stripHierarchy}.

Strips in the $r\phi$ modules have their direction parallel to the beam axis
in the barrel and radially in the endcaps. There are also stereo modules
in the first two layers or rings of all four sub-detectors (TIB, TOB,
TID, TEC) and also in ring five of the TEC. The stereo modules are mounted
back-to-back to the $r\phi$ modules with a stereo angle of 100~mrad and
provide, when combining measurements with the $r\phi$ modules, 
a measurement of $z$ in the barrel or $r$ in the endcap.
A pair of an $r\phi$ and a stereo module is also called a double-sided module.
The strip pitch varies from 80 to 205~$\mu$m depending on the module,
leading to single point resolutions of up to
$23-53$~$\mu$m in the barrel~\cite{detector-paper}.

\section{Input to Alignment}
\label{sec:input}

In this section we discuss the input data for the alignment procedure of the CMS Tracker:
charged particle tracks,
optical survey prior to and during installation,
and laser alignment system measurements.


\subsection{Charged Particle Tracks}
\label{sec:input-tracking}

Track reconstruction and performance specific to the Tracker Integration
Facility configuration are discussed in detail in Refs.~\cite{tiftracking,
tifperformance}.

\begin{figure}[bt]
\begin{center}
\centerline{
\epsfig{figure=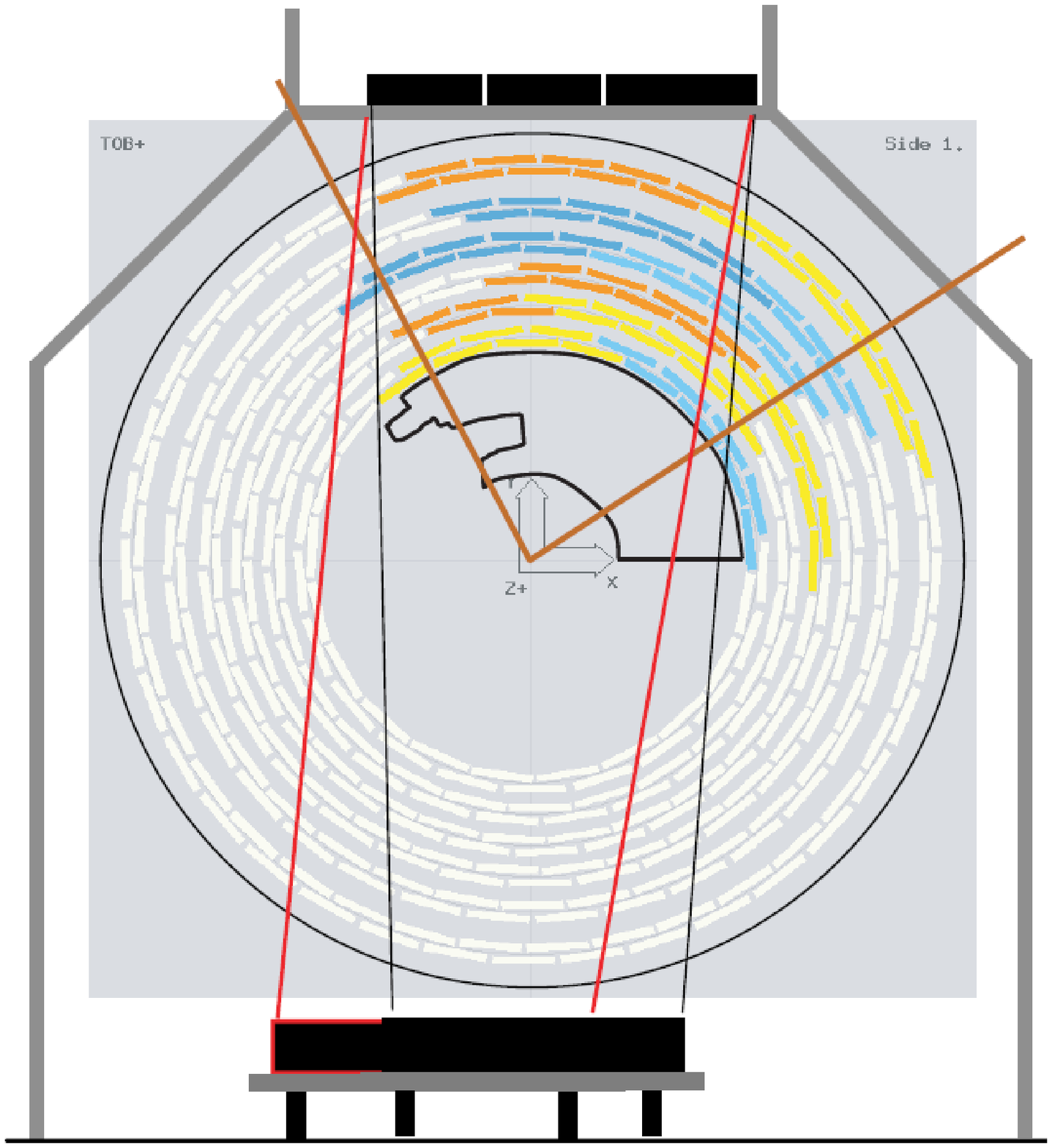,height=5.50cm}
\epsfig{figure=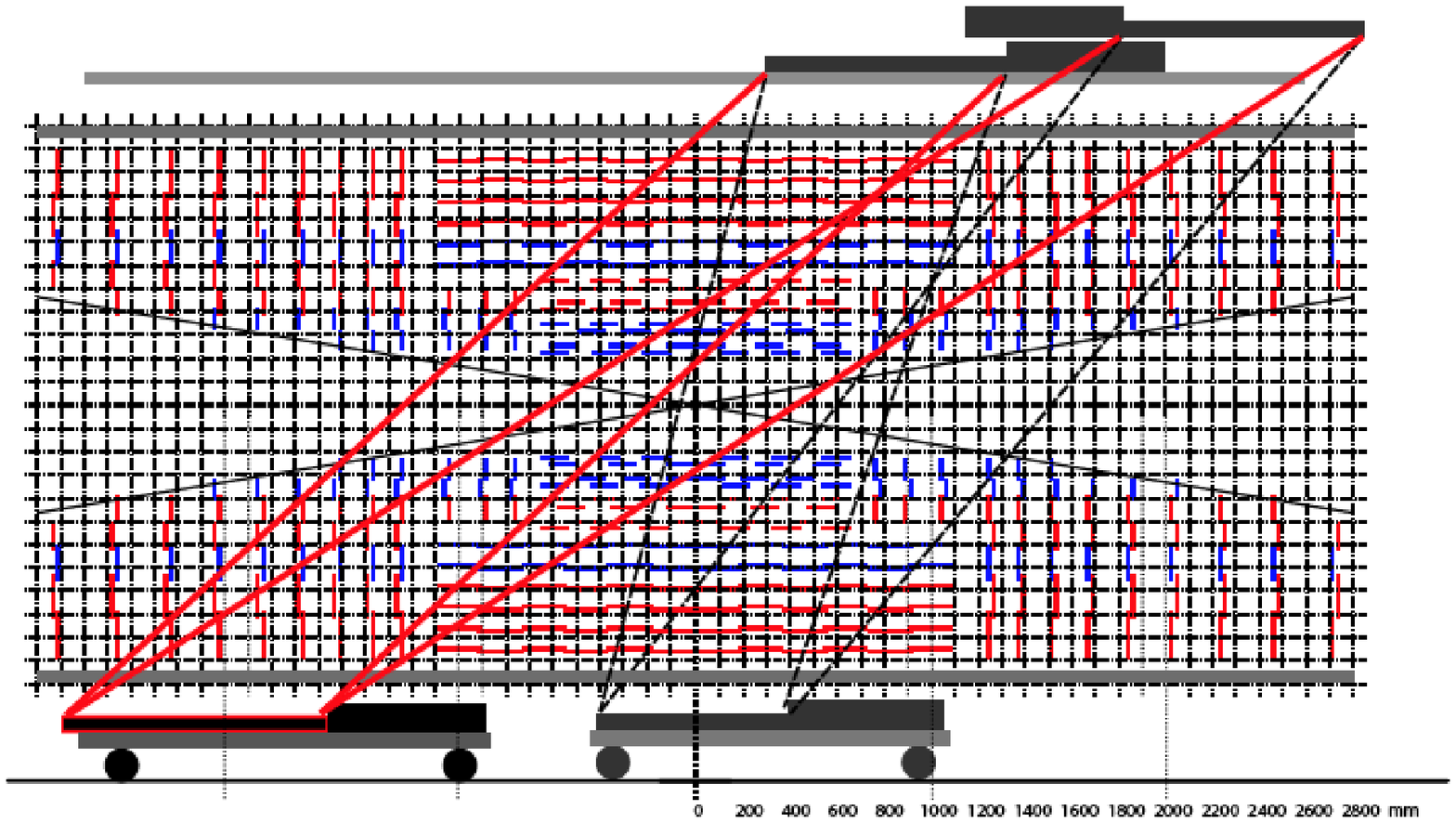,height=5.50cm}
}
\caption{\sl
Layout of the CMS Strip Tracker and of the trigger scintillators at TIF,
front (left) and side view (right).
The acceptance region is indicated by the straight lines connecting the
active areas of the scintillators above and below the tracker.
On the right, 
configuration A corresponds approximately to the acceptance region
defined by the 
right
bottom scintillator;
configuration B corresponds to the left bottom scintillator;
and configuration C combines both.
\label{fig:triggerlayout}
}
\end{center}
\end{figure}

Three different trigger configurations were used in TIF data-taking, 
called A, B and C
and shown in Fig.~\ref{fig:triggerlayout}. 
About $15\%$ of the detector modules, all located at $z>0$, 
were powered and read-out. This includes 
444 modules in TIB ($16\%$), 
720 modules in TOB ($14\%$), 
204 modules in TID ($25\%$), and 
800 modules in TEC ($13\%$).
Lead plates were 
included above the lower trigger scintillators, which enforced a minimum
energy of the cosmic rays of 200~MeV to be triggered. 

The data were collected in trigger configuration A at room
temperature ($+15^\circ$C), both before and after insertion
of the TEC at $z<0$.
All other configurations (B and C) had all strip detector components
integrated. In addition to room temperature, configuration C
was operated at +10~$^\circ$C, -1~$^\circ$C, -10~$^\circ$C, and -15~$^\circ$C.
Due to cooling limitations, a large number of modules had to be turned off
at -15~$^\circ$C.
The variety of different configurations allows us to study alignment
stability with different stress and temperature conditions.
Table~\ref{tab:datasets} gives an overview of the different data sets.
\begin{table}[tb]
  \centering
  \begin{tabular}{|c|c|c|r|c|}
    \hline
    Label & \parbox[c][2.5em][c]{4em}{\centering Trigger\\ Position}& Temperature &
\multicolumn{1}{|c|}{$N_{trig}$} & Comments \\ \hline
    A$_1$ &    A         & 15$^\circ$C   & 665~409 & before TEC- insertion \\
    A$_2$ &    A         & 15$^\circ$C   & 189~925 & after TEC- insertion \\ \hline    \hline
    B     &    B         & 15$^\circ$C   & 177~768 & \\    \hline    \hline
    C$_{15}$&   C        & 15$^\circ$C   & 129~378 & \\    \hline
    C$_{10}$&  C         & 10$^\circ$C   & 534~759 & \\    \hline
    C$_0$  &   C         & -1$^\circ$C   & 886~801 & \\    \hline
    C$_{-10}$& C         & -10$^\circ$C  & 902~881 & \\    \hline
    C$_{-15}$& C         & -15$^\circ$C  & 655~301 & less modules read out \\    \hline
    C$_{14}$  
& C        & 14.5$^\circ$C   & 112~134 &  \\    \hline \hline
    MC        & C        &  --           &3~091~306& simulation \\ \hline
  \end{tabular}
  \caption{\sl
    Overview of different data sets, ordered in time, and their number of triggered
    events $N_{trig}$ taking into account only good running conditions.
  }
  \label{tab:datasets}
\end{table}

We also validate tracking and alignment algorithm performances
with simulation.
A sample of approximately three million cosmic track events was simulated 
using the CMSCGEN simulator~\cite{CMSCGEN}. Only cosmic muon tracks within 
specific geometrical
ranges were selected to simulate the scintillator trigger configuration C. 
To extend CMSCGEN's energy range, events at low muon energy have been
re-weighted to adjust the energy spectrum to the CAPRICE data~\cite{CAPRICE}.

Charged track reconstruction includes three essential steps: seed finding,
pattern recognition, and track fitting. Several pattern recognition algorithms are 
employed on CMS, such as ``Combinatorial Track Finder'' (CTF), ``Road Search'',
and ``Cosmic Track Finder'', the latter being specific to the cosmic track
reconstruction. All three algorithms use the Kalman filter algorithm 
for final track fitting, but the first two steps are different. 
The track model used is a straight line parametrised 
by four parameters where the Kalman filter track fit includes multiple
scattering effects in each crossed layer.
We employ the CTF algorithm for alignment studies in this note. 

In order to recover tracking efficiency which is otherwise lost in the pattern
recognition phase because hits are moved outside the standard search window
defined by the detector resolution, an ``alignment position error''
(APE) is introduced. This APE is added quadratically to the hit resolution,
and the combined value is subsequently used as a search window in the pattern
recognition step. The APE settings used for the TIF data are modelling the assembly
tolerances~\cite{detector-paper}.

There are several important aspects of the TIF configuration which require
special handling with respect to normal data-taking. First of all,
no magnetic field is present. Therefore, the momentum of the tracks cannot be 
measured and estimates of the energy loss and multiple scattering can be done 
only approximately. A track momentum of 1~GeV/$c$ is assumed in the estimates,
which is close to the average cosmic track momentum observed in simulated 
spectra. Other TIF-specific features are due to
the fact that the cosmic muons do not originate from the interaction region.
Therefore the standard seeding mechanism is extended to use also hits in the
TOB and TEC, and no beam spot constraint is applied.
For more details see Ref.~\cite{tiftracking}.

\label{sec:input-tracksel}

Reconstruction of exactly one cosmic muon track in the event is required.
A number of selection criteria is applied on the hits, tracks, and 
detector components subject to alignment, to ensure good quality data.
This is done based on trajectory estimates and the fiducial tracking geometry.
In addition, hits from noisy clusters or from combinatorial background
tracks are suppressed by quality cuts on the clusters.
The detailed track selection is as follows:

\begin{itemize}
\item The direction of the track trajectory satisfies the requirements: 
$-1.5 < \eta_{track} < 0.6$ and $-1.8 < \phi_{track} <-1.2$~rad, 
according to the fiducial scintillator positions. 
\item The $\chi^2$ value of the track fit, normalised to the number of degrees
of freedom, fulfils $\chi^2_{track}/\mbox{ndof} < 4$.
\item The track has 
at least 5 hits associated and among those 
at least 2 matched hits in double-sided modules. 
\end{itemize}
 
A hit is kept for the track fit:

\begin{itemize}
\item If it is associated to a cluster with a total charge of at least 50~ADC
counts. If the hit is matched, both components must satisfy this
requirement. 
\item If it is isolated, i.e. if any other reconstructed hit is found on the same
module within 8.0~mm, the whole track is rejected. This cut helps in
rejecting fake clusters generated by noisy strips and modules.
\item If it is not discarded by the outlier rejection step during the refit
(see below).
\end{itemize}

The remaining tracks and their associated hits are refit in every iteration
of the alignment algorithms. An outlier rejection technique is applied
during the refit. 
Its principle is to iterate the final track fit until no outliers are found. 
An outlier is defined as a hit whose trajectory estimate is larger than a 
given cut value ($e_{cut} = 5$). The trajectory estimate of a hit 
is the quantity:
$e = \mathbf{r}^T \cdot \mathbf{V}^{-1} \cdot \mathbf{r}$, 
where $\mathbf{r}$ is the 1- or 2-dimensional local residual vector and 
$\mathbf{V}$ is the associated covariance matrix. If one or more outliers are
found in the first track fit, they are removed from the hit collection and
the fit is repeated. This procedure is iterated until there are no more 
outliers or the number of surviving hits is less than 4.

Unless otherwise specified, these cuts are common to all alignment algorithms used.
The combined efficiency for all the cuts above is estimated to be 8.3\% on 
TIF data (the $\mathrm{C}_{-10}$ sample is used in this estimate) and 20.5\% 
in the TIF simulation sample.

\subsection{Survey of the CMS Tracker}
\label{sec:input-survey}

Information about the relative position of modules within detector
components and of the larger-level structures within the tracker is
available from the optical survey analysis prior to or during 
the tracker integration. This includes Coordinate Measuring Machine (CMM) 
data and photogrammetry, the former usually used for the
active element measurements and the latter for the larger object
alignment. For the inner strip detectors (TIB and TID), 
survey data at all levels was used in analysis. 
For the outer strip detectors (TOB and TEC), module-level
survey was used only for mounting precision monitoring, 
while survey of high-level structures was used in analysis.

For TIB, survey measurements are available for the module positions
with respect to shells, and of cylinders with respect to the tracker support tube.
Similarly, for TID, survey measurements were done for modules with respect to
the rings, rings with respect to the disks and disks with respect to the tracker support tube.
For TOB, the wheel was measured with respect to the tracker support tube.
For TEC, measurements are stored at the level of disks with respect to the endcaps
and endcaps with respect to the tracker support tube.

\begin{figure}[tb]
\begin{center}
\centerline{
\epsfig{figure=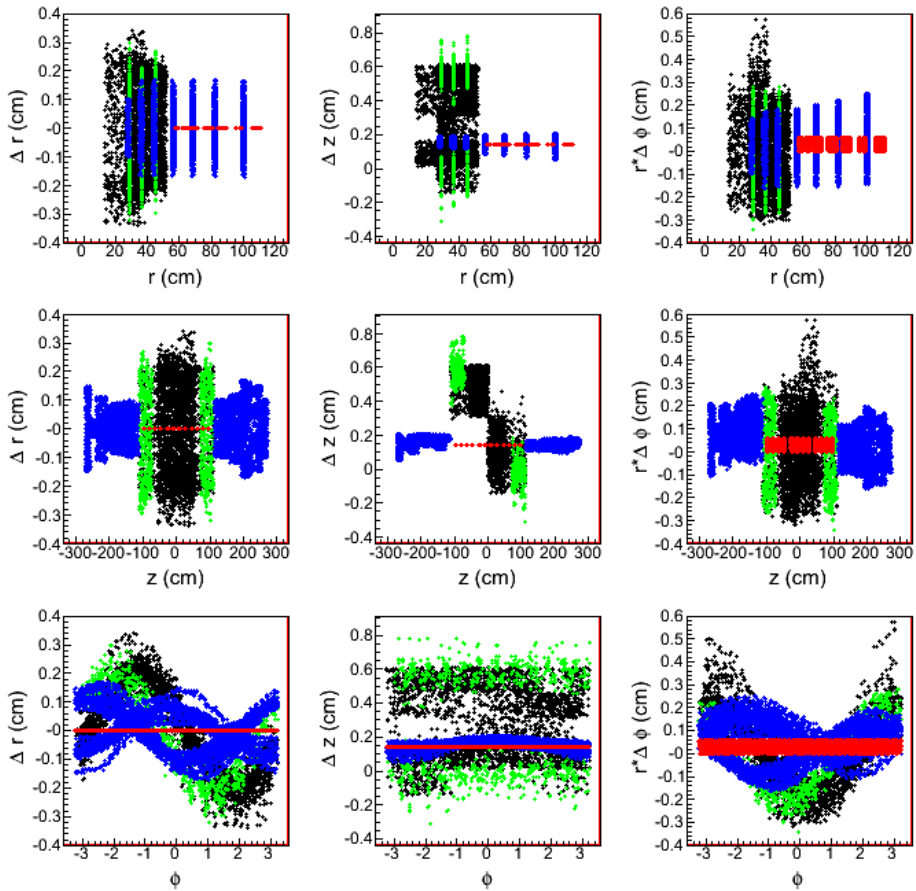,height=12.0cm}
}
\caption{\sl
Displacement of modules in global cylindrical coordinates
as measured in survey with respect to design geometry.
A colour coding is used: black for TIB, green for TID,
red for TOB, and blue for TEC.
}
\label{fig:stripSurveyComp}
\end{center}
\end{figure}

Figure~\ref{fig:stripSurveyComp} illustrates the relative positions of the 
CMS tracker modules with respect to design geometry as measured in 
optical survey: as can be seen,
differences from design geometry as large as several millimetres
are expected.  
Since hierarchical survey measurements were performed and 
TOB and TEC 
have only large-structure information, the corresponding modules appear to be
coherently displaced in the plot.

\subsection{Laser Alignment System of the CMS Tracker}
\label{sec:input-laser}


The Laser Alignment System 
(LAS, see Fig.~\ref{fig:stripLayout})~\cite{trackertdr,detector-paper}
uses infrared laser beams with a wavelength of
$\lambda=1075$\,nm to monitor the position of selected tracker modules. It
operates globally on tracker substructures (TIB, TOB and TEC disks) and cannot
determine the position of individual modules. The goal of the system 
is to
generate alignment information on a continuous basis, providing geometry
reconstruction of the tracker substructures at the level of $100\,\mu$m.
In addition, possible tracker structure movements can be monitored at the level
of $10 \rm\ \mu m$, providing additional input for the track based alignment.

In each TEC, laser beams cross all nine TEC disks in ring 6 and ring 4
on the back petals, equally distributed in $\phi$. Here, special silicon
sensors with a $10$\,mm hole in the backside metallisation and an
anti-reflective coating are mounted. The beams are used for the internal
alignment of the TEC disks. The other eight beams, distributed in
$\phi$, are foreseen to align TIB, TOB, and both TECs with respect to each
other. Finally, there is a link to the muon system, which is
established by 12 laser beams (six on each side) with precise position and
orientation in the tracker coordinate system.

The signal induced by the laser beams on the silicon sensors decreases in
height as the beams penetrate through subsequent silicon layers in the TECs
and through beam splitters in the alignment tubes that partly deflect the
beams onto TIB and TOB sensors. To obtain optimal signals on all sensors, a
sequence of laser pulses with increasing intensities, optimised for each
position, is generated. Several triggers per intensity are taken and the
signals are averaged.  In total, a few hundred triggers are needed to get a
full picture of the alignment of the tracker structure. Since the trigger rate
for the alignment system is around $100$\,Hz, this takes only a few
seconds. 


\section{Statistical Methods and Approaches}
\label{sec:methods}
Alignment analysis with tracks uses
the fact that the hit positions and the measured trajectory impact points
of a track are systematically displaced if the module 
position is not known correctly.
The difference in local module coordinates between these two quantities are the
{\it track-hit residuals} ${\bf r}_{i}$, 
which are 1- (2-dimensional) vectors in the case of a single (double)
sided 
module and which 
one would like to minimise. More precisely, one can
minimise the $\chi^2$ function which includes a covariance
matrix ${\bf V}$ of the measurement uncertainties:
\begin{equation}
\label{eq:chisq}
\chi^2=\sum_i^{\rm hits}{\bf r}_{i}^T({\bf p},{\bf q}){\bf V}^{-1}_{i}{\bf r}_{i}({\bf p},{\bf q})
\end{equation}
where 
${\bf q}$ represents the track parameters
and ${\bf p}$ represents the alignment parameters of the modules.

%

\begin{figure}[b] 
  \centering \includegraphics[width=.6\textwidth]{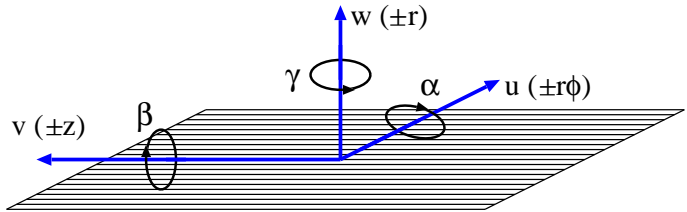} \caption{Schematic
  illustration of the local coordinates of a module as used for
  alignment. Global parameters (in parentheses) are shown for modules in the
  barrel detectors (TIB and TOB).}  \label{fig:localCoordinates}
\end{figure}

A
module is assumed to be a rigid body, so three absolute positions and three
rotations are sufficient to parametrise its degrees of freedom.
These are commonly defined for all methods in the module coordinates 
as illustrated in Fig.~\ref{fig:localCoordinates}. 
The local positions are called $u$, $v$ and $w$, where $u$ is along the
sensitive coordinate (i.e.\ across the strips),
$v$ is perpendicular to $u$ in the sensor plane and $w$ is
perpendicular to the $uv$-plane, completing the right-handed coordinate
system.
The rotations around the $u$, $v$ and $w$ axes are called $\alpha$,
$\beta$ and $\gamma$, respectively.
In the case of alignment of intermediate structures like rods, strings
or petals, we follow the convention that $u$ and $v$ are
parallel and perpendicular to the precisely measured coordinate,
while for the large structures like layers and disks, the local
coordinates coincide with the global ones.


The different alignment methods used to minimise Eq.~(\ref{eq:chisq}) 
are described in the following.

\subsection{HIP algorithm}
\label{sec:methods-concepts-hip}

The HIP (Hits and Impact Points) algorithm is described in detail in 
Ref.~\cite{hipalignment}. 
Neglecting the track parameters in Eq.~(\ref{eq:chisq}), the alignment
parameters ${\bf p}_m$ of each module can be found independently from
each other. The general formalism of the $\chi^2$ minimisation in the
linear approximation leads to
%
\begin{eqnarray}
\label{eq:chisqsolution}
{\bf p}_m =
\left[
\sum_{i}^{\rm hits}{\bf J}_i^T{\bf V}_i^{-1}{\bf J}_i 
\right]^{-1}
\left[
\sum_{i}^{\rm hits}{\bf J}_i^T{\bf V}_i^{-1}{\bf r}_i 
\right]
\end{eqnarray}
where the Jacobian ${\bf J}_{i}$ is defined as the derivative of the residual 
with respect to the sensor position parameters and can be found
analytically with the small angle approximation~\cite{KarimakiDerivs}
(used by the other algorithms as well).
Correlations between different modules and effects on the track parameters
are 
accounted for by iterating the minimisation process and by
refitting the tracks with new alignment constants after each iteration.


\subsection{Kalman filter algorithm}
\label{sec:methods-concepts-kalman}
{
\newcommand{\bm}[1]{\mbox{\boldmath$#1$}}
\newcommand{\noT}{\phantom{T}}
The Kalman alignment algorithm~\cite{kalmanAlignment} is a sequential 
method, derived using the Kalman filter formalism. It is sequential 
in the sense that the alignment parameters are updated after each 
processed track. 
%
The 
algorithm is based on the track model 
$\bm{m}=\bm{f}(\bm{q}_t,\bm{p}_t)+\bm{\varepsilon}$. 
This model relates the observations $\bm{m}$ to the true 
track parameters $\bm{q}_t$ and the true alignment constants 
$\bm{p}_t$ via the deterministic function $\bm f$. Energy 
loss is considered to be deterministic and is dealt with 
in the track model. The stochastic vector $\bm{\varepsilon}$ 
as well as its variance-covariance matrix $\bm{V}$ contain 
the effects of the observation error and of multiple scattering. Therefore the
matrix $\bm{V}$ contains correlations between hits such that
equation~(\ref{eq:chisq}) is a sum over tracks, with residuals being of 
higher dimension according to the number of hits along the track
trajectory. Linearised around an expansion point $(\bm{q}_0,\bm{p}_0)$, 
i.e. track parameters from a preliminary track fit and an 
initial guess for the alignment constants, the track model reads:
\begin{equation}
  \bm{m}=\bm{c}+\bm{D}_q\bm{q}_t+\bm{D}_p\bm{p}_t+\bm{\varepsilon},
\end{equation}
with
\begin{equation}
  \bm{D}_q=\bigl.\partial\bm{f}/\partial\bm{q}_t\bigr|_{\bm{q}_0}, \quad\bm{D}_p=\bigl.\partial\bm{f}/\partial\bm{p}_t\bigr|_{\bm{p}_0}, \quad \bm{c}=\bm{f}(\bm{q}_0,\bm{p}_0)-\bm{D}_q\bm{q}_0-\bm{D}_p\bm{p}_0
\end{equation}
By applying 
the Kalman filter formalism to this relation, updated equations for the alignment parameters ${\bm p}$ and their variance-covariance matrix ${\bm C}_p$ can be extracted.
}

\subsection{Millepede algorithm}
\label{sec:methods-concepts-millepede}

Millepede~II~\cite{MillepedeII}
is an upgraded version of the Millepede
program~\cite{MillepedeI}. Its principle is a global fit to minimise
the $\chi^2$ function,
simultaneously taking into account track and alignment parameters.
Since angular corrections are small, the linearised problem is a good approximation for alignment. 
Being interested only in the $n$ alignment parameters, the problem is
reduced to the solution of a matrix equation of size $n$. 


The $\chi^2$ function, Eq.~(\ref{eq:chisq}), depends on
track (local, ${\bf q}$) and alignment (global, $\bf p$) parameters.
For uncorrelated hit measurements $y_{ji}$ of the track $j$,
with uncertainties $\sigma_{ji}$, it can be rewritten as

\begin{equation}
\label{eq:objFuncRes}
\chi^2({\bf p},{\bf q}) = 
\sum_j^{\rm tracks}
\sum_i^{\rm hits} \frac{\left(y_{ji}-f_{ji}({\bf p},{\bf q}_j)\right)^2}
                      {\sigma_{ji}^2}
\end{equation}
where ${\bf q}_j$ denotes the parameters of track $j$.

Given reasonable start values ${\bf p}_0$ and ${\bf q}_{j0}$ as expected
in alignment, the track model prediction $f_{ji}({\bf p},{\bf q}_j)$ can be linearised.
%
Applying the least squares method to minimize $\chi^2$,
results in a large linear system with one equation for each alignment parameter
and all the track parameters of each track.
The particular structure of the system of equations allows a reduction of
its size,
leading to the matrix equation 
\begin{equation}
\label{eq:matrix}
{\bf C} \, {\bf a} = {\bf b} 
\end{equation}
for the small corrections ${\bf a}$ to the alignment parameter start values
${\bf p}_0$.


\subsection{Limitations of alignment algorithms}
\label{sec:methods-concepts-limitations}

We should note that Eq.~(\ref{eq:chisq}) may be invariant under certain
coherent transformations of assumed module positions, 
the so-called 
``weak'' modes. 
The trivial transformation which is $\chi^2$-invariant is a global
translation and rotation of the whole tracker. 
This transformation has no effect in internal alignment, and is easily 
resolved by a suitable convention for defining the global reference frame.
Different algorithms employ different 
approaches and conventions here, so we will discuss this in more detail 
as it applies to each algorithm. 

The non-trivial $\chi^2$-invariant transformations which preserve
Eq.~(\ref{eq:chisq}) are of larger concern. For the full CMS tracker with
cylindrical symmetry one could define certain ``weak'' modes, such
as elliptical distortion, twist, etc., depending on the track sample 
used. However, since we use only a partial CMS tracker without the 
full azimuthal coverage, different ``weak'' modes may show up.
For example, since we have predominantly vertical cosmic
tracks (along the global $y$ axis), a simple shift of all modules in the
$y$ direction approximately constitutes a ``weak'' mode, this transformation 
preserving the size of the track residuals for a vertical track. 
However, since we still have tracks with some angle to vertical
axis, some sensitivity to the $y$ coordinate remains. 

In general, any particular track sample would have its own
``weak'' modes and the goal of an unbiased alignment procedure is
to remove all $\chi^2$-invariant transformations with a balanced
input of different kinds of tracks. In this study we are limited
to only predominantly vertical single cosmic tracks and this limits
our ability to constrain $\chi^2$-invariant transformations, 
or the ``weak'' modes. This is discussed more in the validation
section.

\subsection{Application of Alignment Algorithms to the TIF Analysis}
\label{sec:commonalisel}

Accurate studies have been performed with all algorithms in order to determine
the maximal set of detectors that can be aligned and 
the aligned coordinates that are sensitive to the peculiar
track pattern and limited statistics of TIF cosmic track events.

For the tracker barrels (TIB and TOB), the collected statistics is sufficient to
align at the level of single modules if restricting to a geometrical subset 
corresponding to the positions of the scintillators used for triggering.
The detectors aligned are those whose centres lie inside the geometrical 
ranges 
$z >$ 0, $x <$ 75~cm and 0.5 $< \phi <$ 1.7~rad
where all the coordinates are in the global CMS frame.

The local coordinates aligned for each module are
\begin{itemize}
\item $u$, $v$, $\gamma$ for TOB double-sided modules,
\item $u$, $\gamma$ for TOB single-sided modules,
\item $u$, $v$, $w$, $\gamma$ for TIB double-sided modules and
\item $u$, $w$, $\gamma$ for TIB single-sided modules.
\end{itemize}

Due to the rapidly decreasing cosmic track rate $\sim\cos^2\psi$ (with $\psi$
measured from zenith) only a small fraction of tracks cross the endcap
detector modules at an angle suitable for alignment. Therefore, the
$z^{+}$-side Tracker endcap (TEC) could only be aligned at the level of
disks. All nine disks are considered in TEC alignment, and the only aligned
coordinate is the angle $\Delta\phi$ 
around the CMS $z$-axis.  
Because there are only data in two sectors of the TEC, the track-based alignment
is not sensitive to the $x$ and $y$ coordinates of the disks. 

\begin{figure}[t]
\begin{center}
  \includegraphics[width=0.6\textwidth]{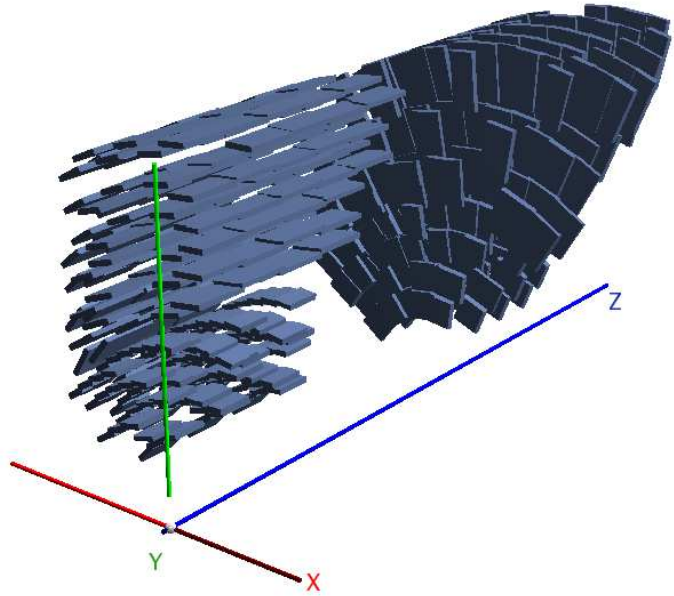}
  \caption{\sl
Visualisation of the modules used in the track-based alignment procedure.
Selected modules based on the common geometrical and track-based selection for the algorithms.
  }
  \label{fig:visualization}
\end{center}
\end{figure}

The Tracker Inner Disks (TID) are not aligned due to
lack of statistics.
Figure~\ref{fig:visualization} visualises the modules selected for the
track-based alignment procedure.

\subsubsection{HIP algorithm}
\label{sec:methods-hip}



Preliminary residual studies show that, in real data, the misalignment of 
the TIB is larger than in TOB, and TEC alignment is quite independent from
that of other structures. For this reason, the overall alignment result
is obtained in three steps:
\begin{enumerate}
\item In the first step, the TIB is excluded from the analysis and the tracks 
are refit using only reconstructed hits in the TOB. Alignment parameters
are obtained for this subdetector only. No constraints are applied on the
global coordinates of the TOB as a whole.
\item In the second step, the tracks are refit using all their hits; the
TOB is fixed to the positions found after step 1 providing the global reference 
frame; and alignment parameters are obtained for TIB only. 
\item The alignment of the TEC is then performed as a final step starting 
from the aligned barrel geometry found after steps 1 and 2.
\end{enumerate} 

Selection of aligned objects and coordinates is done according to the common 
criteria 
described in Secs.~\ref{sec:input-tracking} and ~\ref{sec:commonalisel}. 

The Alignment Position Error (APE) for the aligned detectors is set at the 
first iteration to a value 
compatible with the expected positioning uncertainties after assembly, 
then decreased linearly with the iteration number, reaching zero at iteration
$n$ ($n$ varies for different alignment steps). Further iterations
are then run using zero APE.

In order to avoid a bias in track 
refitting from parts of the TIF tracker that are not aligned in this
procedure (e.g. low-$\phi$ barrel detectors), an arbitrarily large APE is 
assigned for all iterations to trajectory measurements whose corresponding
hits lie in these detectors, de-weighting them in the $\chi^2$ 
calculation. 

\begin{figure}[htp]
\begin{center}
\epsfig{figure=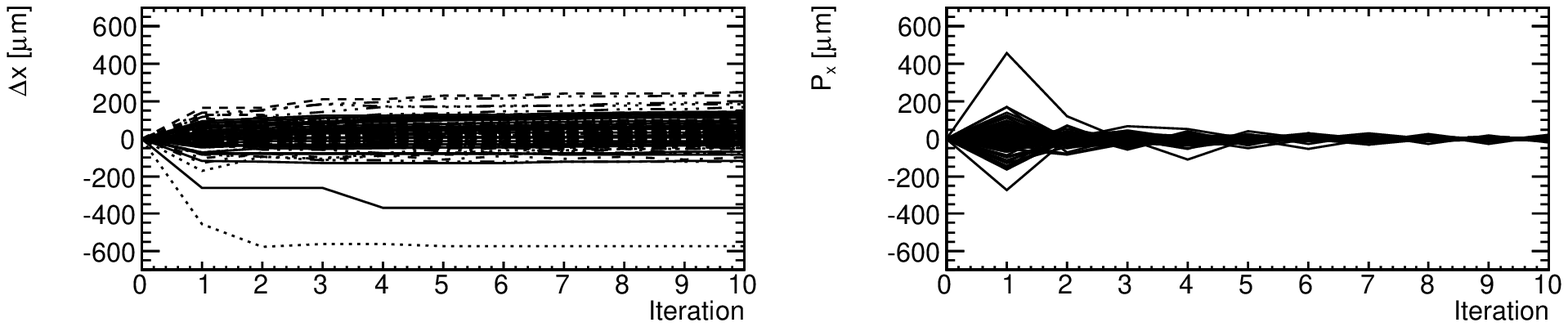,height=3.5cm} \\
\epsfig{figure=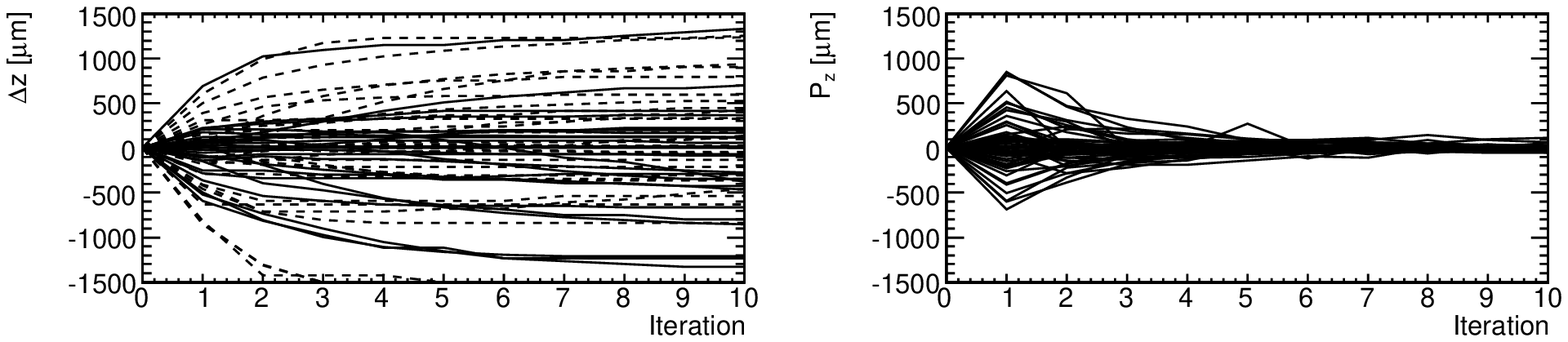,height=3.6cm}
 \caption{\sl
 Results of the first HIP alignment step (TOB modules only) on the C$_{-10}$
 TIF data sample. From top to bottom the
 plots show respectively the quantities $\Delta x$ for all modules and
 $\Delta z$ for double-sided modules 
 where $\Delta$ stands for the difference between the aligned local position of
 a module at a given iteration of the algorithm and the nominal 
 position of the same
 module. On the left column the evolution of the object position is plotted vs.
 the iteration number (different line styles correspond to the 6 TOB layers), 
 while on the right the parameter increment for each iteration
 of the corresponding
 alignment parameters is shown. 
 }
 \label{fig:hiptob}
 \end{center}
 \end{figure}

For illustrative purposes, we show here the results of HIP alignment on the
C$_{-10}$ TIF data sample after event selection. 
Figure~\ref{fig:hiptob} 
shows examples of the evolution of the aligned positions and the
alignment parameters calculated by the HIP algorithm after every iteration.
We observe reasonable
convergence for the coordinates that are expected to be most precisely
determined (see Sec.~\ref{sec:track-based}) and a stable result in
subsequent iterations using zero APE.


\subsubsection{Kalman filter algorithm}
\label{sec:methods-kalman}


In the barrel, the alignment is carried out 
starting from the module survey geometry. The alignment parameters 
are calculated for all modules in the TIB and the TOB at once, 
using the common alignable selection described in 
Sec.~\ref{sec:commonalisel}. No additional alignable 
selection criteria, for instance a minimum number of hits per module, 
is used. Due to the lack of any external aligned reference 
system, some global distortions in the final alignment can 
show up, e.g. shearing or rotation with respect to the true 
geometry. 

The tracking is adapted to the needs of the algorithm, 
especially to include the current estimate of the alignment 
parameters. Since for every module the position error can 
be calculated from the up-to-date parameter errors, no 
additional fixed Alignment Position Error (APE) is used. 
The material effects are crudely taken into account by 
assuming a momentum of 1.5~GeV/$c$, which is larger than the one used in
standard track reconstruction.

TEC alignment is determined on disk level. Outlying tracks, which would
cause unreasonably large changes of the alignment parameters if used by the
algorithm, are discarded. 
Due to the experimental setup, the total number of hits per disk decreases such that the
error on the calculated parameter increases from disk one to disk nine.
%
During the alignment process, disk 1 is used as reference. After that, the
alignment parameters are transformed into the coordinate system defined by
fixing the mean and slope of $\phi(z)$ to zero. This is done because there is
no sensitivity to a linear torsion, which, in a linear approximation,
corresponds to a slope in $\phi(z)$, expected for the TEC. Due to differences
in the second order approximation between a track inclination and a torsion of
the TEC, the algorithm basically has a small sensitivity to a torsion of the
endcap. Here, the linear component is expected to be superimposed into
movements of the disks in $x$ and $y$, which are converted by the algorithm
into rotations because these are the only free parameters.

The alignment parameters do not seem to depend strongly on the temperature
(see section \ref{sec:validation-tec-stability}), so all data except for the
runs at -15~$^\circ$C were merged to increase the statistics. 
%
%

\subsubsection{Millepede algorithm}
\label{sec:methods-millepede}

%
%


Millepede alignment is performed at module level in both TIB and TOB, and 
at disk level in the TEC, in one step only. 
To fix the six degrees of freedom from global translation and rotation,
equality constraints are used on the parameters in the
TOB: These inhibit overall shifts and rotations of the TOB, while the TIB 
parameters are free to adjust to the fixed TOB position. In addition,
TEC disk one is kept as fixed.

The requirements to select a track useful for alignment are 
described in Sec.~\ref{sec:input-tracksel}. All these criteria are
applied, except for the hit outlier rejection since outlier down-weighting
is applied within the minimisation process.
Since Millepede internally refits the tracks,
it is additionally required that a track hits at least five of those modules
which are subject to the alignment procedure.
Multiple scattering and energy loss effects are treated, as in the
Kalman filter alignment algorithm, by increasing and correlating the hit uncertainties,
assuming a track momentum of 1.5 GeV/$c$. This limits the accuracy
of the assumption of uncorrelated measured hit positions in Eq.~(\ref{eq:objFuncRes}).

The alignment parameters are calculated for all modules using the
common alignable selection described in Sec.~\ref{sec:commonalisel}.
Due to the fact that barrel and endcap are aligned together in one
step, no request on the minimum number of hits in the subdetector for
a selected track is done.

The required minimum number of hits for a module to be
aligned is set to 50.
Due to the modest number of parameters, the matrix equation~(\ref{eq:matrix})
is solved by
inversion with five Millepede global iterations.
In each global iteration,
the track fits are repeated four times
with alignment parameters updated from the previous global iteration.
Except for the first track fit iteration,
down-weighting factors are assigned
for each hit depending on its normalised residuum of the previous fit
(details see~\cite{MillepedeII}).
About 0.5\% of the tracks with an average hit weight below 0.8 are rejected
completely.

\begin{figure}[tb]
\includegraphics[width=.47\textwidth]{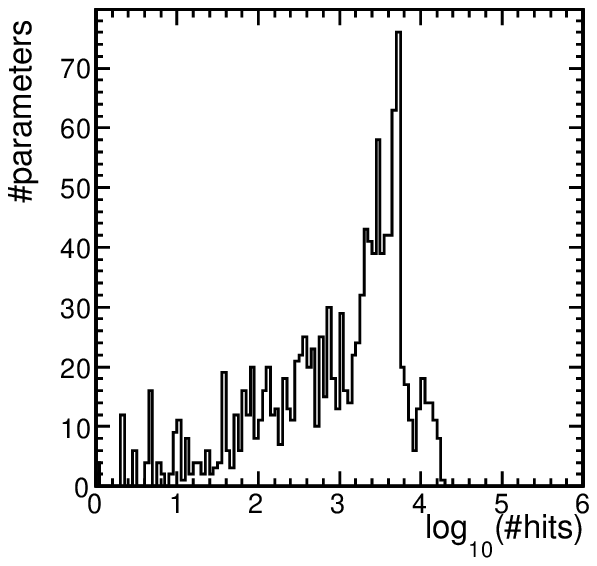} \hfill
\includegraphics[width=.47\textwidth]{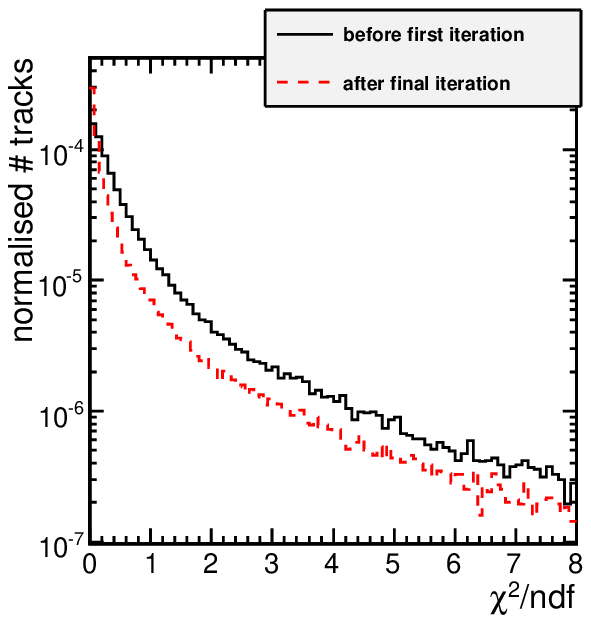}
\caption{\sl
Number of hits for the parameters aligned with Millepede (left) and
improvement of the normalised $\chi^2$ distribution as seen by Millepede
(right).
\label{fig:mpResults}
}
\end{figure}

Fig.~\ref{fig:mpResults} shows, on the left, the number of hits per alignment
parameter used for the global minimisation; 58 modules fail the cut of 50 hits.
On the right, the normalised $\chi^2$ distributions of the Millepede
internal track fits before and after minimisation are shown.
The distributions do not have a peak close to one, indicating that the
hit uncertainties are overestimated. Nevertheless, the effect of minimisation
can clearly be seen.

\section{Validation of Alignment of the CMS Tracker at TIF}
\label{sec:validation-pass4}

In this section we present validation of the alignment results. 
Despite the limited precision of alignment that prevents
detailed systematic distortion studies, 
the available results from TIF provide important 
validation of tracker alignment for the 
set of modules used in this study. 

The evolution of the 
module positions is shown starting from the design geometry,
moving to survey measurements, 
and finally comparing to the results from the track-based algorithms. 
Both the overall track quality and individual hit residuals improve between 
the three steps. 
All three track-based algorithms produce similar results when 
the same input and similar approaches are taken. We show that the
residual misalignments are consistent with statistical uncertainties in
the procedure. Therefore, we pick just one alignment geometry from
the track-based algorithms for illustration of results when comparison 
between different algorithms is not relevant.

\subsection{Validation Methods}
\label{sec:vali}

We use two methods in validation and illustration of the alignment results.
One approach is track-based and the other approach directly compares
geometries resulting from different sets of alignment constants.

In the track-based approach, we refit the tracks with all Alignment Position
Errors (APE) set to zero.
A loose track selection is applied, requiring at least six hits where more
than one of them must be two-dimensional.
Hit residuals will be shown as the difference between the measured hit position
and the track position on the module plane.
To avoid a bias, the latter is predicted without using
the information of the considered hit. In the barrel part of the tracker,
the residuals in local $x'$ and
$y'$ direction, parallel to $u$ and $v$, will be shown. The sign is chosen such that
positive values always point into the same $r\phi$ and $z$ directions,
irrespective of the orientation of the local coordinate system.
For the wedge-shaped sensors as in TID and TEC, the residuals have a 
correlation
depending on the local $x$- and $y$-coordinates of the track impact point.
The residuals in global $r\phi$- and $r$-coordinates therefore are
used for these modules.


In addition to misalignment, hit residual distributions depend on the intrinsic
hit resolution and the track prediction uncertainty. For low-momentum tracks
(as expected to dominate the TIF data) in the CMS tracker, the latter is large.
For a momentum of 1~GeV/$c$ and an extrapolation as between two adjacent TOB layers
between two consecutive hits, the mean multiple scattering displacement
is about 250~$\mu$m. 
So even with perfect alignment one expects a width of the
residual distribution that is significantly larger than the intrinsic hit resolution
of up to $23-53$~$\mu$m in the strip tracker barrel~\cite{detector-paper}.


Another way of validating alignment results is provided by direct 
comparison of the obtained tracker geometries. This is done by showing
differences between the same module coordinate in two geometries
(e.g. ideal and aligned) vs. their geometrical position (e.g. $r$, $\phi$ or
$z$) or correlating these differences as seen by two different 
alignment methods.
%
%
Since not all alignment algorithms fix the position and orientation
of the full tracker, comparison between two geometries is done after
making the centre of gravity and the overall orientation of the considered modules
coincide.

\subsection{Validation of the Assembly and Survey Precision}
\label{sec:assembly}

\begin{figure}[tb]
\begin{center}
  {\includegraphics[width=.45\textwidth]{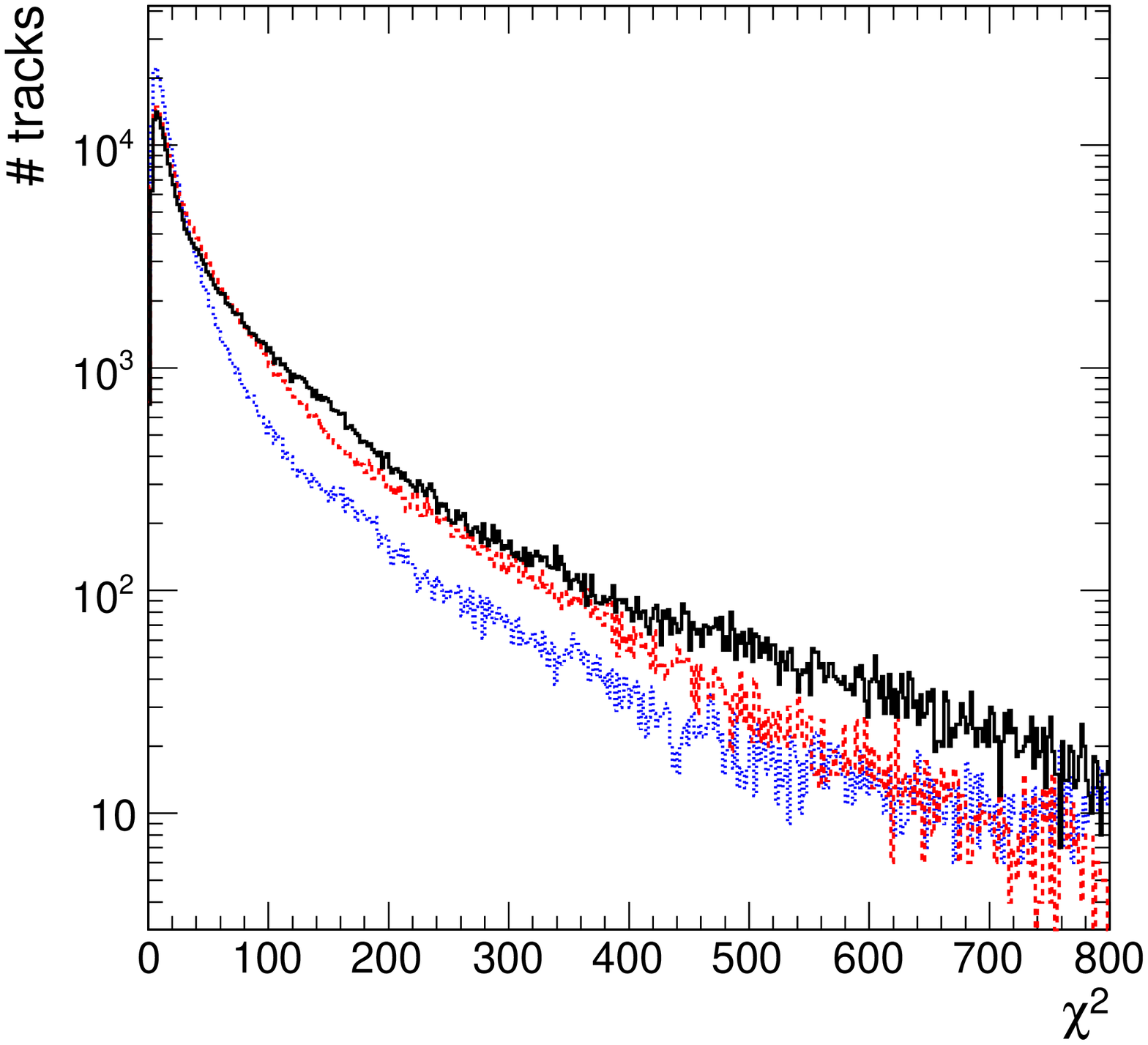}}
  {\includegraphics[width=.45\textwidth]{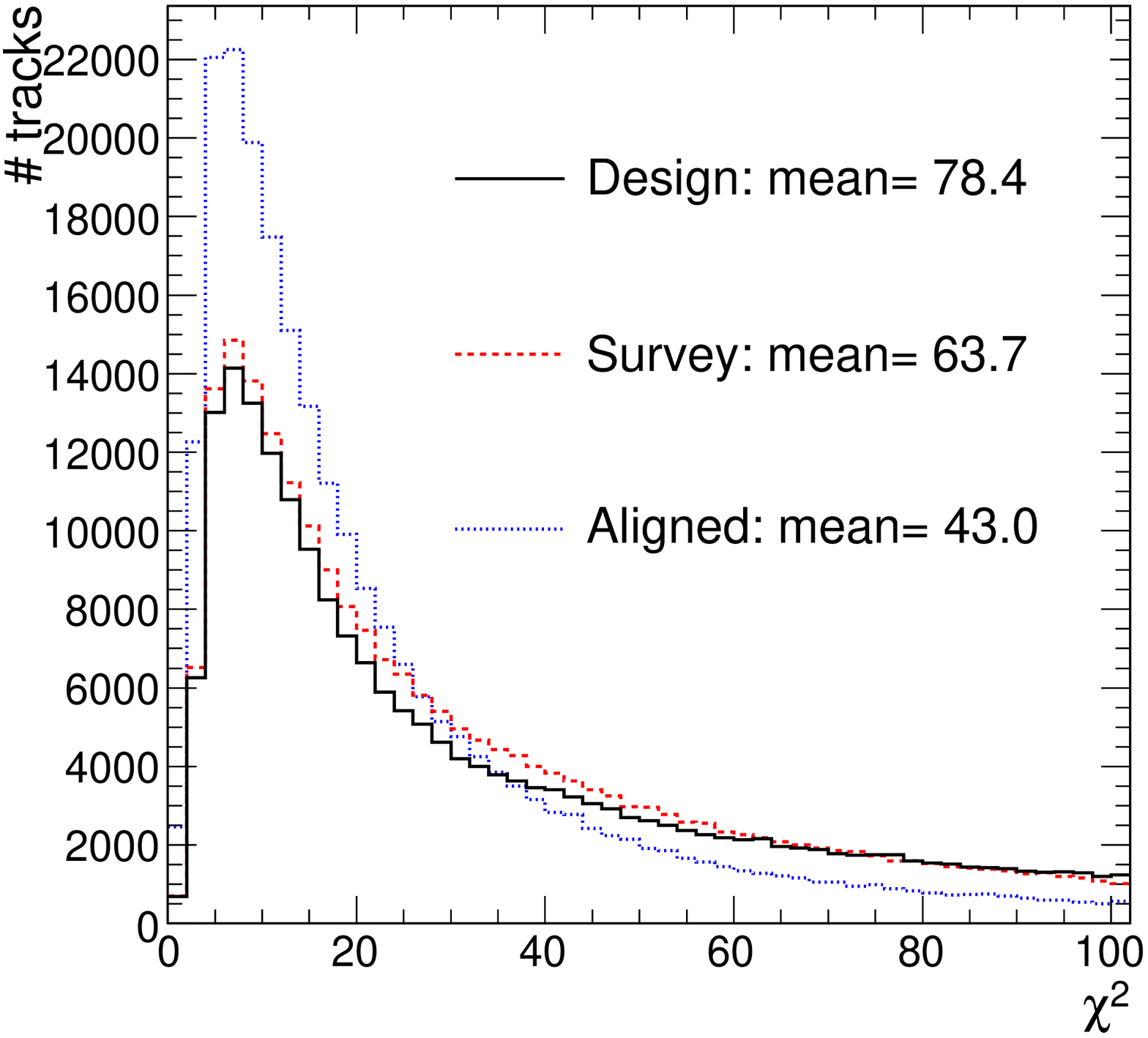}}
  \caption{\sl
Distributions of the absolute $\chi^2$-values of the track fits
for the design and survey geometries as well as the one from HIP track-based alignment.
  }
  \label{fig:absolute_Chi2_Ideal}
\end{center}
\end{figure}

\begin{figure}[ptb]
\begin{center}
 {\includegraphics[width=\textwidth]{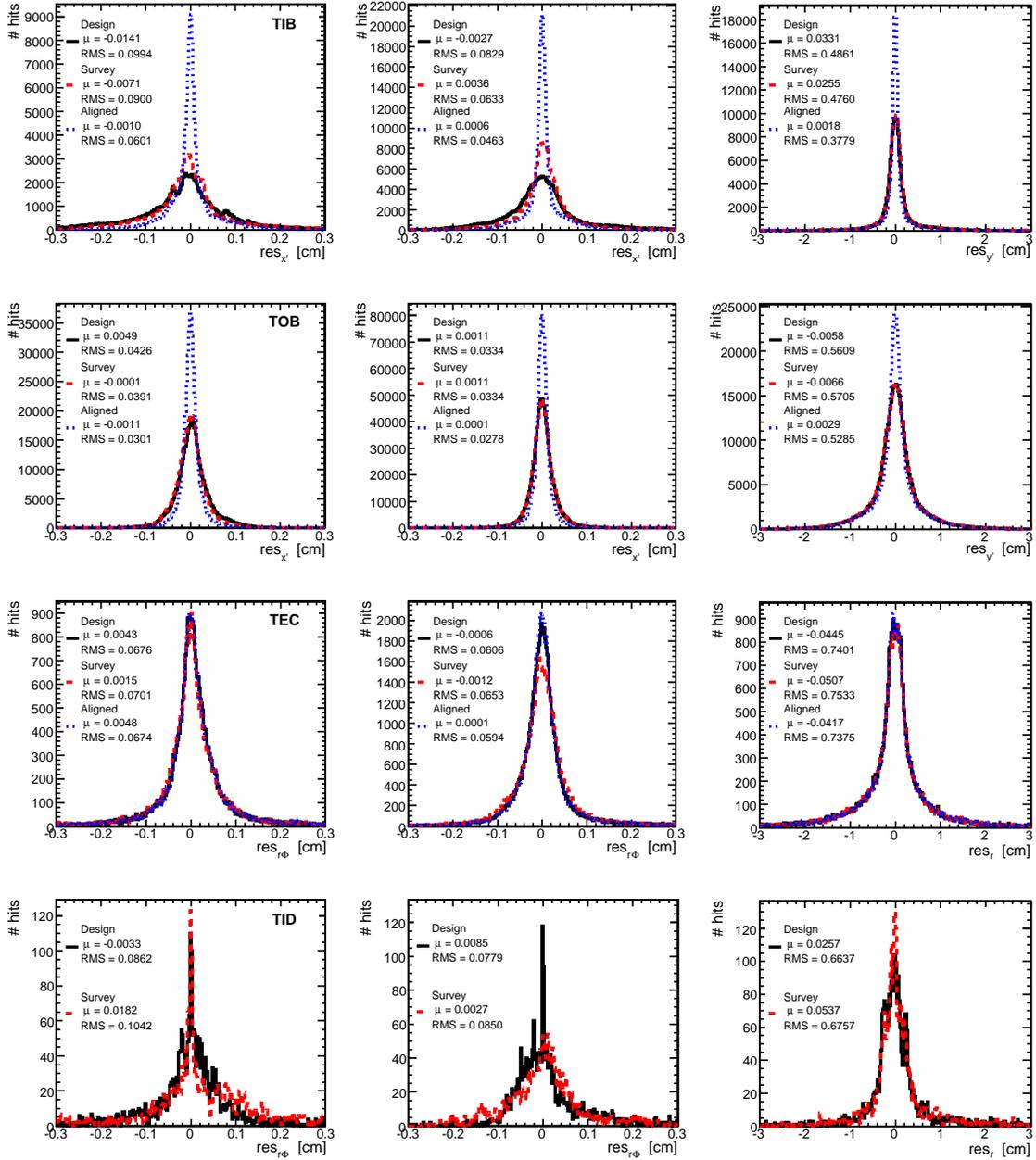}}
  \caption{\sl
Hit residuals for different geometries: ideal (solid/black), 
survey (dashed/red), and track-based alignment (dotted/blue, HIP).
Four Tracker sub-detectors are shown in the
top row (TIB), second row (TOB), third row (TEC), and bottom row (TID).
The absolute local $x'$-residuals are shown for single-sided modules (left)
and double-sided modules (middle), while local $y'$-residuals are shown for
the double-sided modules only (right).
For the endcap modules (in TEC and TID) transformation to the
$r\phi$ and $r$ residuals is made.
  }
  \label{fig:residuals_Ideal}
\end{center}
\end{figure}

Improvements of the absolute track fit $\chi^2$ are observed when
design geometry, survey measurements, and track-based alignment 
results are compared, as shown in Fig.~\ref{fig:absolute_Chi2_Ideal}.
The average $\chi^2$ changes from $78\to64\to43$ between the
three geometries, respectively. 
This is also visible in the absolute hit residuals shown 
in Fig.~\ref{fig:residuals_Ideal}.
In general, an improvement can be observed by comparing the survey information
to the design geometry, and comparing the track-based
alignment to survey results.
The residual mean values are closer to zero, and the standard deviations are
smaller.  

\begin{figure}[t]
\begin{center}
  \includegraphics[width=\textwidth]{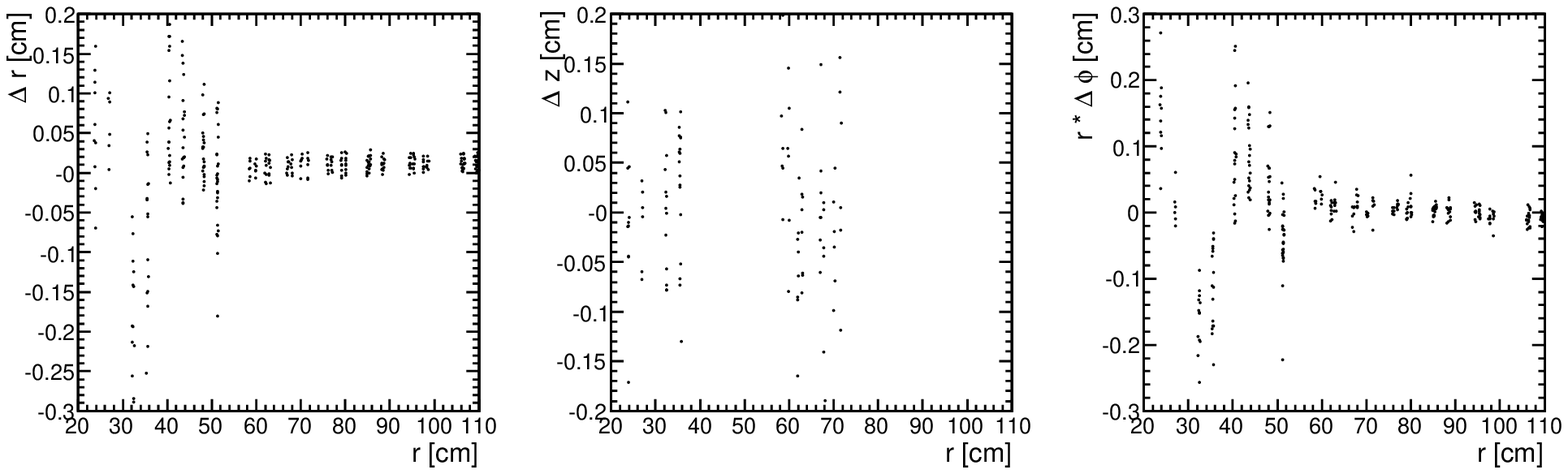}
  \caption{\sl
Difference of the module positions between the measured (in HIP track-based alignment)
and design geometries for TIB (radius $r<55$ cm) and TOB ($r>55$ cm). Projection on
the $r$ (left), $z$ (middle), and $\phi$ (right) directions are shown. 
Only double-sided modules are considered in the $z$ comparison.
  }
  \label{fig:vIdeal_tif3x1}
\end{center}
\end{figure}

In Fig.~\ref{fig:vIdeal_tif3x1}, the differences of the module positions
between the design geometry and the geometry aligned with the HIP algorithm
are shown for TIB and TOB.
There is a clear coherent movement of the four layers of the TIB
in both radial ($r$) and azimuthal ($\phi$) directions. The scale
of the effect is rather large, $1-2$~mm. 
At the same time, mounting placement uncertainty 
of modules in TOB is much smaller for both layers within the TOB 
and for modules within layers. No obvious systematic deviations are 
observed apart from statistical scatter due to mounting precision.

Given good assembly precision of the TOB discussed above, 
the ideal geometry is a sufficiently good starting geometry for TOB.
Therefore, only high-level structure survey is considered for TOB
and no detailed comparison can be discussed.
As a result, TOB residuals in Fig.~\ref{fig:residuals_Ideal} 
do not change much between survey and ideal geometries,
the two differing only in the overall TOB global position
as shown in Fig.~\ref{fig:stripSurveyComp}.

However, the situation is different for TIB and optical survey is
necessary to improve the initial understanding of the module positions
in this detector. From Figs.~\ref{fig:stripSurveyComp} and
\ref{fig:vIdeal_tif3x1} it is evident that survey of the layer
positions in TIB does not reflect the situation in data (displacement
appears to be even in the opposite direction). Therefore, we do not
consider layer-level survey of TIB in our further analysis
and do not include it in the track-based validation.
However, the position of modules within a layer is reflected well in
the optical survey. This is evident by significant improvement 
of the TIB residuals between the ideal and survey geometries
shown in Fig.~\ref{fig:residuals_Ideal}, and in the track $\chi^2$ 
in Fig.~\ref{fig:absolute_Chi2_Ideal}.



\subsection{Validation of the Track-Based Alignment}
\label{sec:track-based}


The three track-based alignment algorithms used in this study
employ somewhat different
statistical methods to minimise hit residuals and overall 
track $\chi^2$. Therefore, comparison of their results is an important
validation of the systematic consistency of the methods.
 
\begin{figure}[htbp]
  \begin{center}
        \hspace{.0666\textwidth}
        \includegraphics[width=0.4\textwidth]{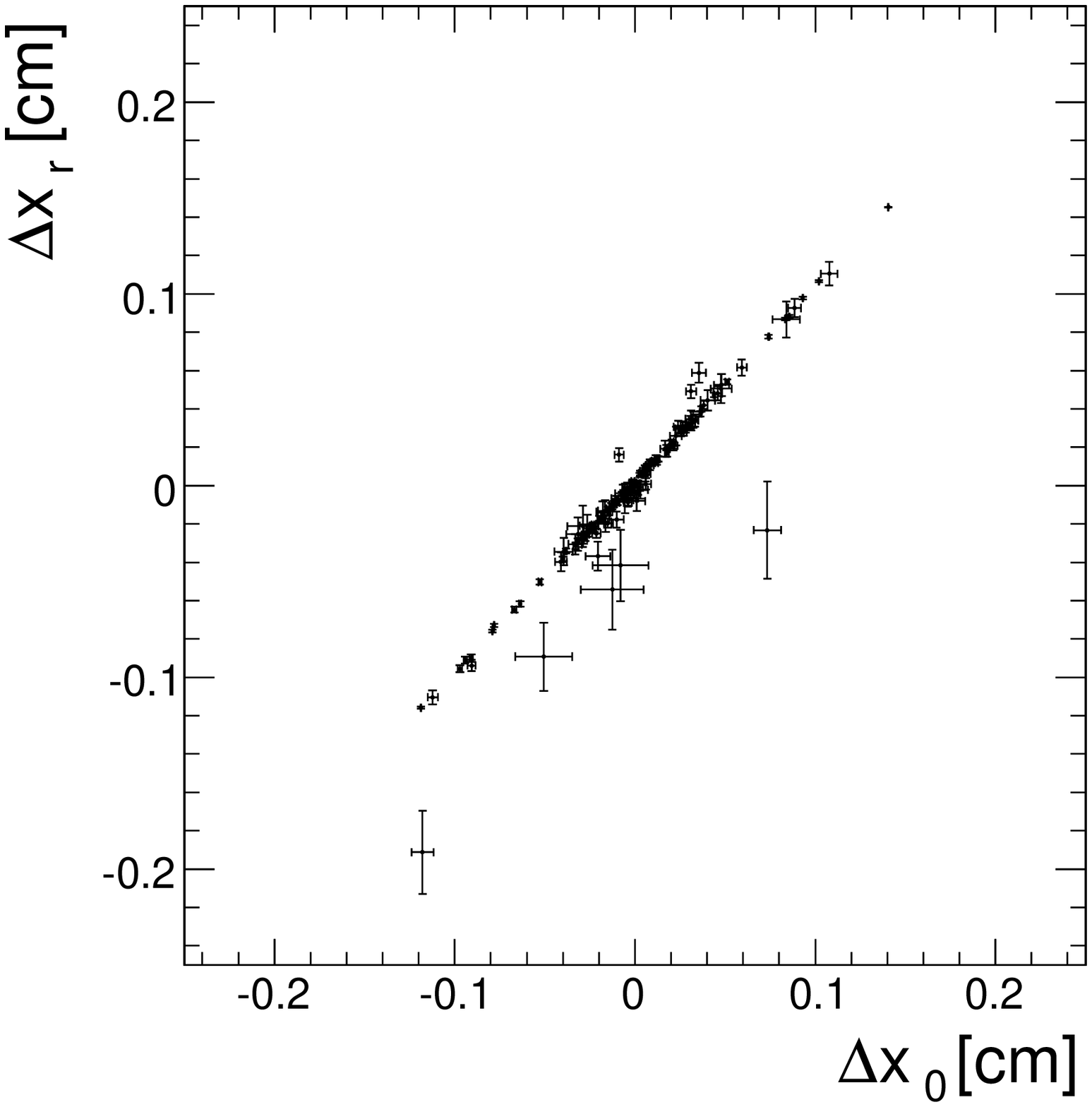}
         \hspace{0.06\textwidth}
      \includegraphics[width=0.4\textwidth]{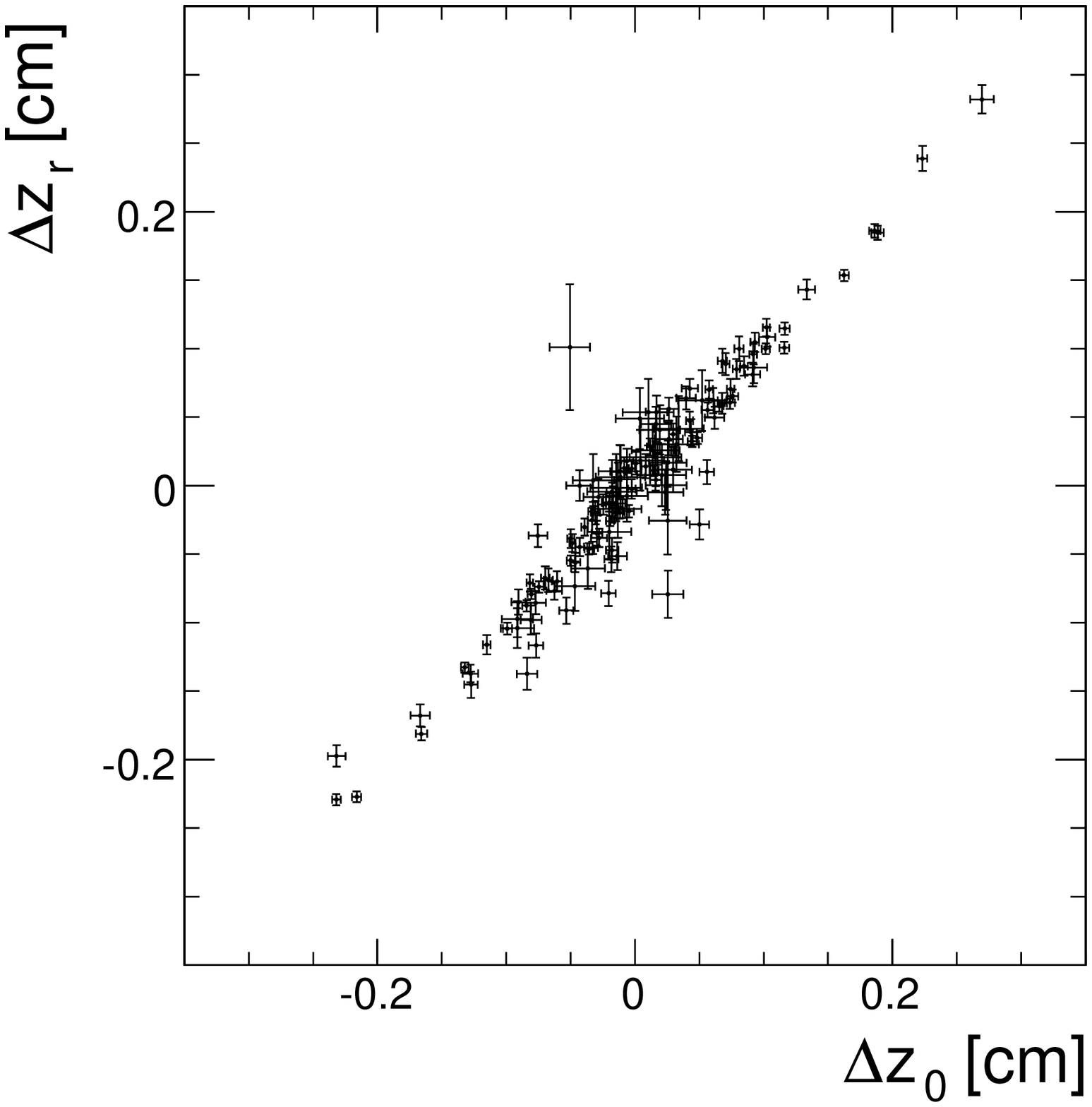}
        \hspace*{0.0666\textwidth}

        \hspace{0.0666\textwidth}
        \includegraphics[width=0.4\textwidth]{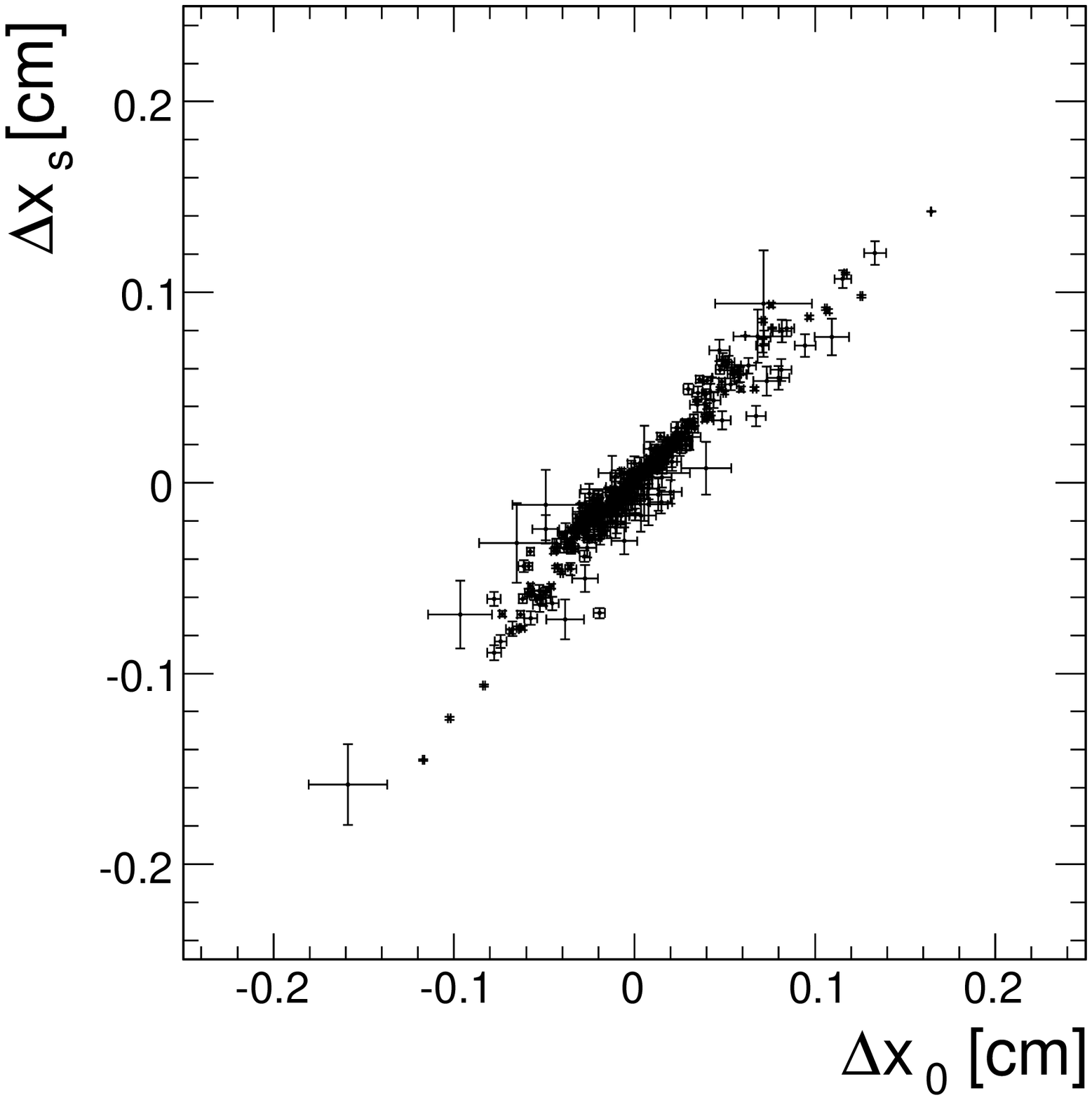} 
        \hspace{0.0666\textwidth}
        \includegraphics[width=0.4\textwidth]{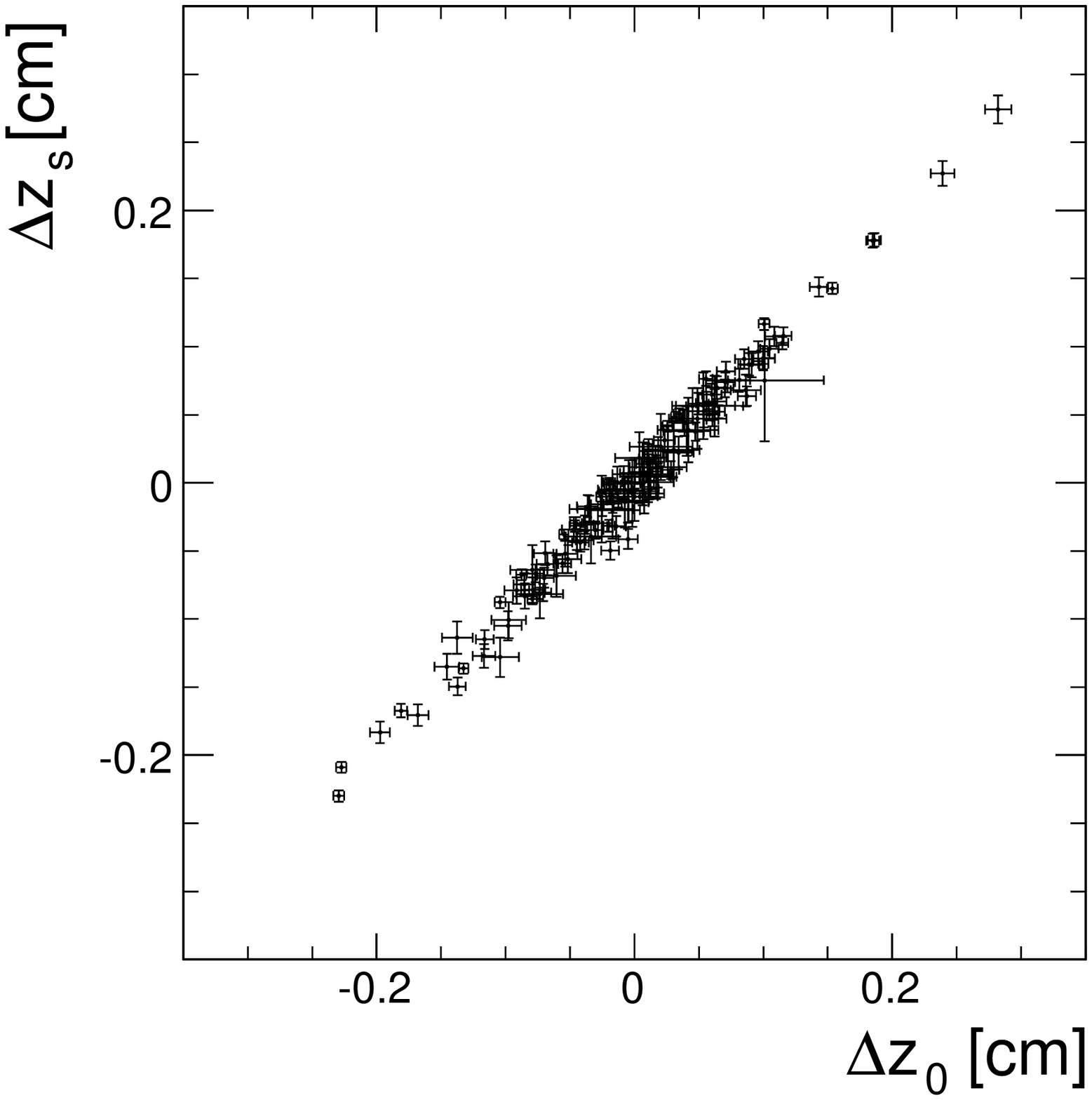}
        \hspace*{0.0666\textwidth}
    \caption{Comparison of the global shifts computed with different starting values,
        using the Kalman alignment algorithm. 
        For the computation of $\Delta x_{0}$ and $\Delta z_{0}$ the starting values 
        for parameters were set to 0, for $\Delta x_{\rm r}$ and $\Delta z_{\rm r}$
        they were drawn from a Gaussian distribution and for $\Delta x_{\rm S}$
        and $\Delta z_{\rm S}$ they are taken from the module survey geometry.}
    \label{fig:methods-kalman:shifts-compare}
  \end{center}
\end{figure}

To exclude the possibility of bad convergence of the track-based alignment, 
the alignment constants have been computed with random starting values. 
As an example, the 
starting values for the local shifts were drawn from a Gaussian distribution 
with a variance of $\sigma=200~\mu$m. The corresponding results for the Kalman algorithm
can be seen in Fig.~\ref{fig:methods-kalman:shifts-compare}, where in the upper two
plots the computed
global shifts for the sensitive coordinates are compared to ones from the standard
approach. Also, starting from the survey geometry rather than the ideal geometry was
attempted, as shown in the lower two plots. 
The results are
compatible within their uncertainties as they are calculated inside the 
Kalman algorithm.

\begin{figure}[tbp]
\begin{center}
  {\includegraphics[width=0.45\textwidth]{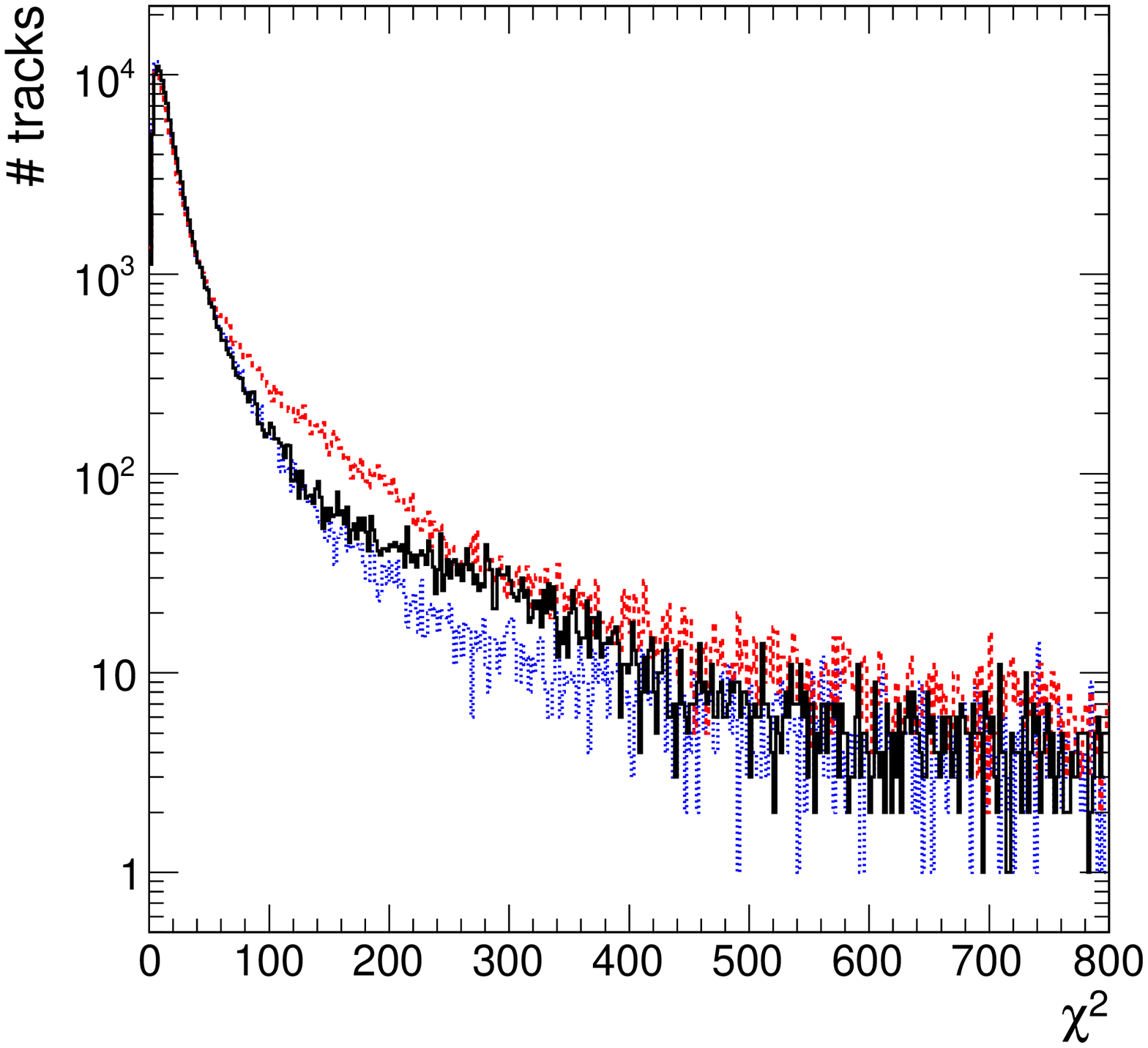}}
  {\includegraphics[width=0.45\textwidth]{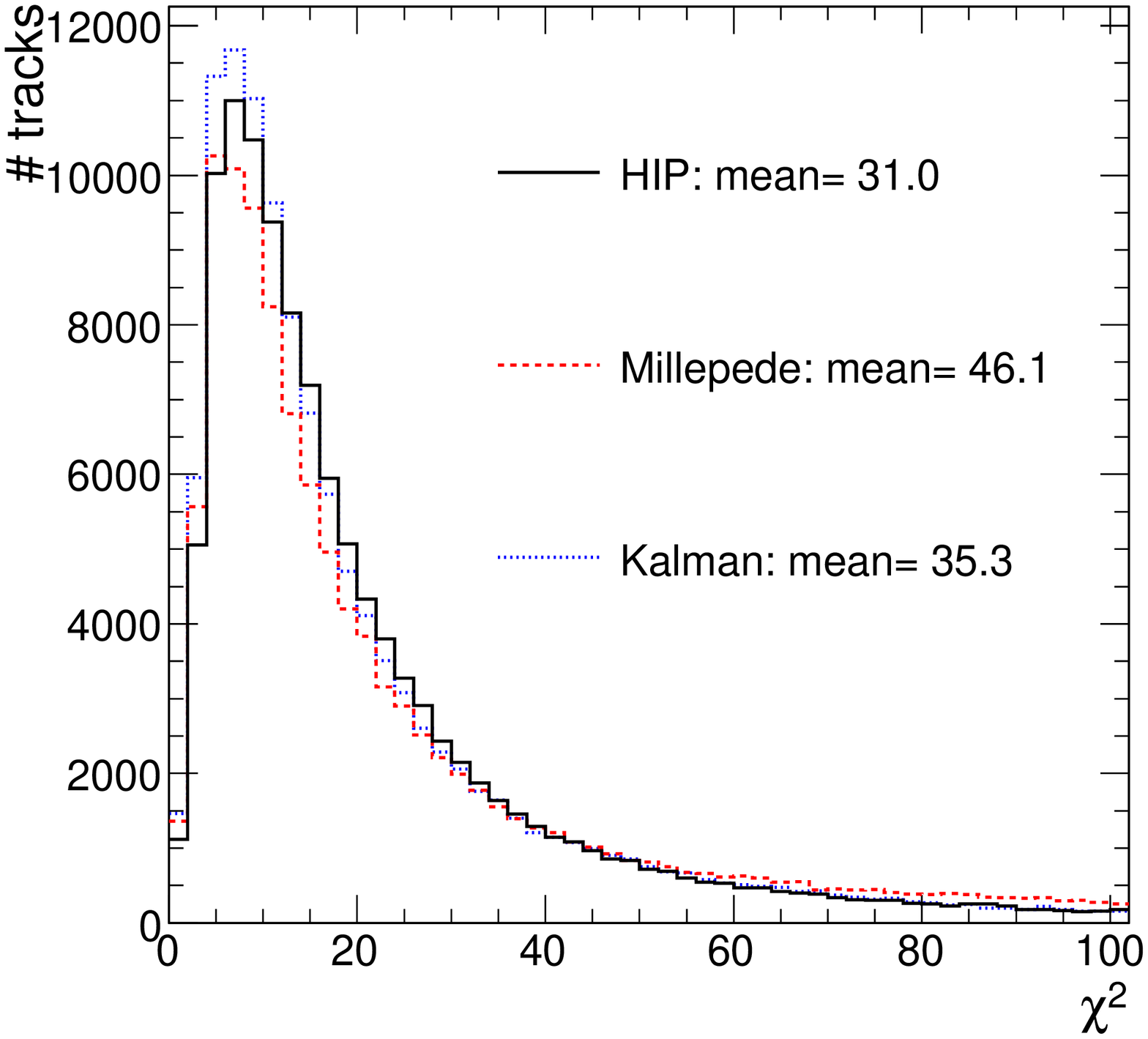}}
  \caption{\sl
Distributions of the absolute $\chi^2$-values of the track fits
for the geometries resulting from HIP, Kalman, and Millepede alignment.
The track fit is restricted to modules aligned by all three algorithms.
  }
  \label{fig:absolute_Chi2}
\end{center}
\end{figure}

\begin{figure}[tbp]
\begin{center}
  {\includegraphics[width=1.0\textwidth]{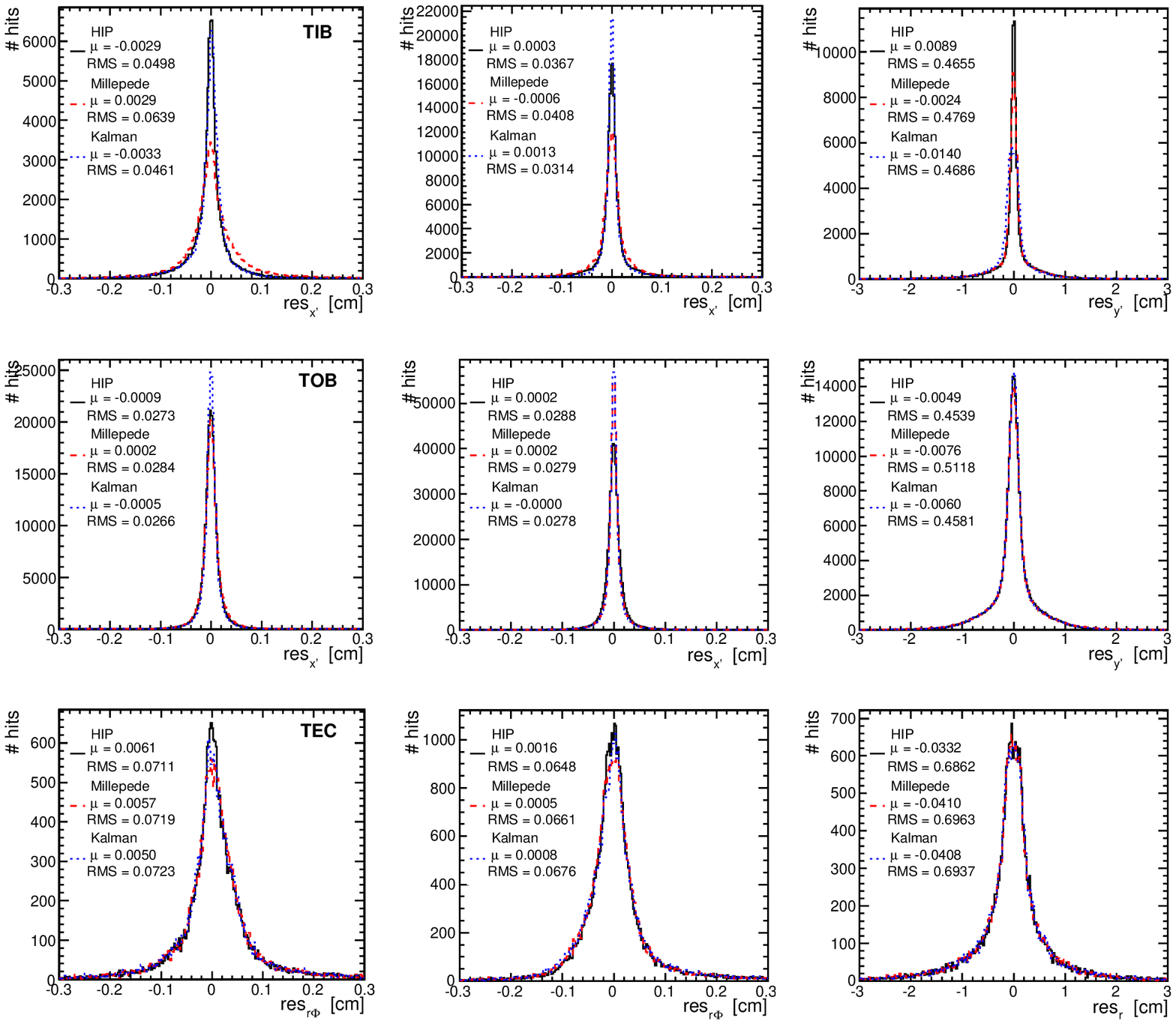}}
  \caption{\sl 
Hit residuals for different geometries
from three track-based algorithms: 
HIP (solid/black), 
Millepede (dashed/red), 
and Kalman (dotted/blue) based alignment.
Three Tracker sub-detectors are shown in the
top row (TIB), second row (TOB), and bottom row (TEC).
The absolute local $x'$-residuals are shown for single-sided modules (left)
and double-sided modules (middle), while local $y'$-residuals are shown for
the double-sided modules only (right).
For the endcap modules (TEC) transformation to the
$r\phi$ and $r$ residuals is made.
The track fit is restricted to modules aligned by all three algorithms.
  }
  \label{fig:residuals_alg}
\end{center}
\end{figure}

The three alignment algorithms show similar distributions of
the track $\chi^2$ shown in Fig.~\ref{fig:absolute_Chi2}.
HIP constants give the smallest mean value whereas
Kalman and Millepede have more tracks at low $\chi^2$ values 
than the HIP constants.
The three algorithms also have consistent residuals in all Tracker sub-detectors
as shown in Fig.~\ref{fig:residuals_alg}, though the most relevant
comparison is in the barrel region (TIB and TOB) since the endcaps
were not aligned at the module level.
For both Figs.~\ref{fig:absolute_Chi2} and~\ref{fig:residuals_alg}, only 
modules selected for alignment have been taken into account in the refit and
in the residual distributions.

\begin{figure}[tbp]
\begin{center}
 \includegraphics[width=0.48\textwidth]{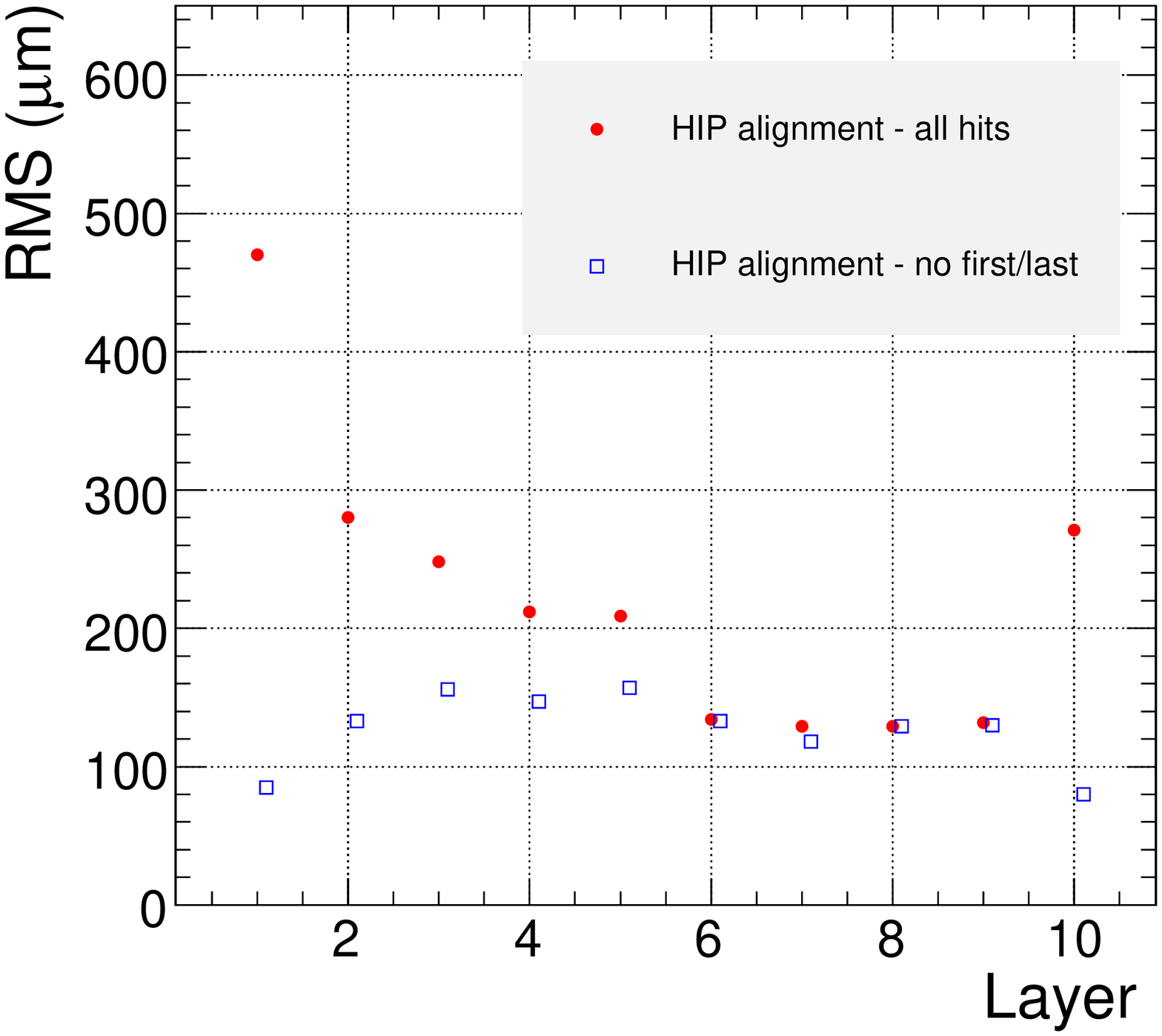}
 \caption{\sl
 Hit residual RMS in local $x'$ coordinate in ten layers of the barrel 
 tracker, i.e. four layers of TIB and six layers of TOB, after 
 track-based alignment with HIP. In contrast to the blue squares, the red circles
 are obtained including residuals from the first and last hits of the track.
Hits on modules not aligned are not considered
 in the track fit.
}       
  \label{fig:residuals_layers2}
\end{center}
\end{figure}

A more quantitative view of the residual distributions and their improvement
with alignment can be gained
by looking at 
their widths.
To avoid influence
of modules not selected for alignment in the following, these are excluded
from the residual
distributions and from the track refits. Furthermore, taking the
pure RMS of the distributions gives a high weight to outliers e.g. from
wrong hit assignments in data or artificially large misaligned modules
in simulations (see Sec.~\ref{sec:track-based2}).
For this reason truncated mean and RMS values are calculated 
from the central 99.87\% interval of each distribution, 
corresponding to 2.5$\sigma$ for a Gaussian-distributed variable.
The resulting widths of the residual distributions in $x'$ 
after alignment (HIP constants) are shown in Fig.~\ref{fig:residuals_layers2}
for the ten barrel layers.
They are about 120~$\mu$m in TOB layers 2-5, between 200 and 300~$\mu$m
in TIB layers 2-3 and much larger in TIB layer~1 and TOB layer~6.
This is due to the much larger track pointing uncertainty if the
track prediction is an extrapolation to the first and last hit of a track
compared to interpolations for the hits in between, as can be seen
from the second curve in Fig.~\ref{fig:residuals_layers2}. Here
residuals from the first and last hits of the tracks are not considered.
Residual widths in TIB decrease clearly to about 150~$\mu$m, making
it evident that many tracks end within the TIB. TIB layer~1 and TOB layer~6 
now show especially small values since all remaining residuals
come from sensor overlap and have short track interpolation distances.

\begin{figure}[tbp]
\begin{center}


{\includegraphics[width=0.48\textwidth]{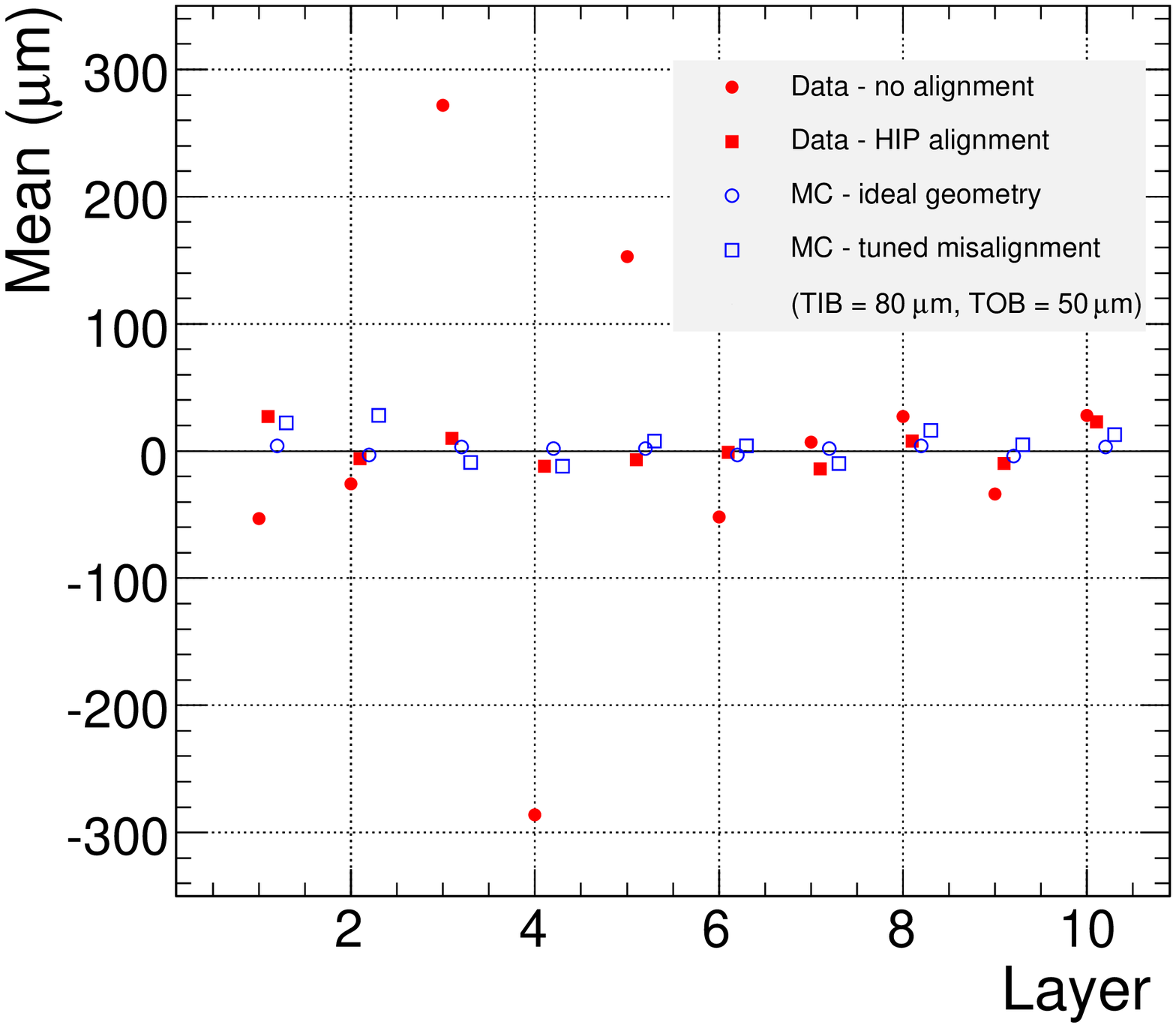}}
{\includegraphics[width=0.48\textwidth]{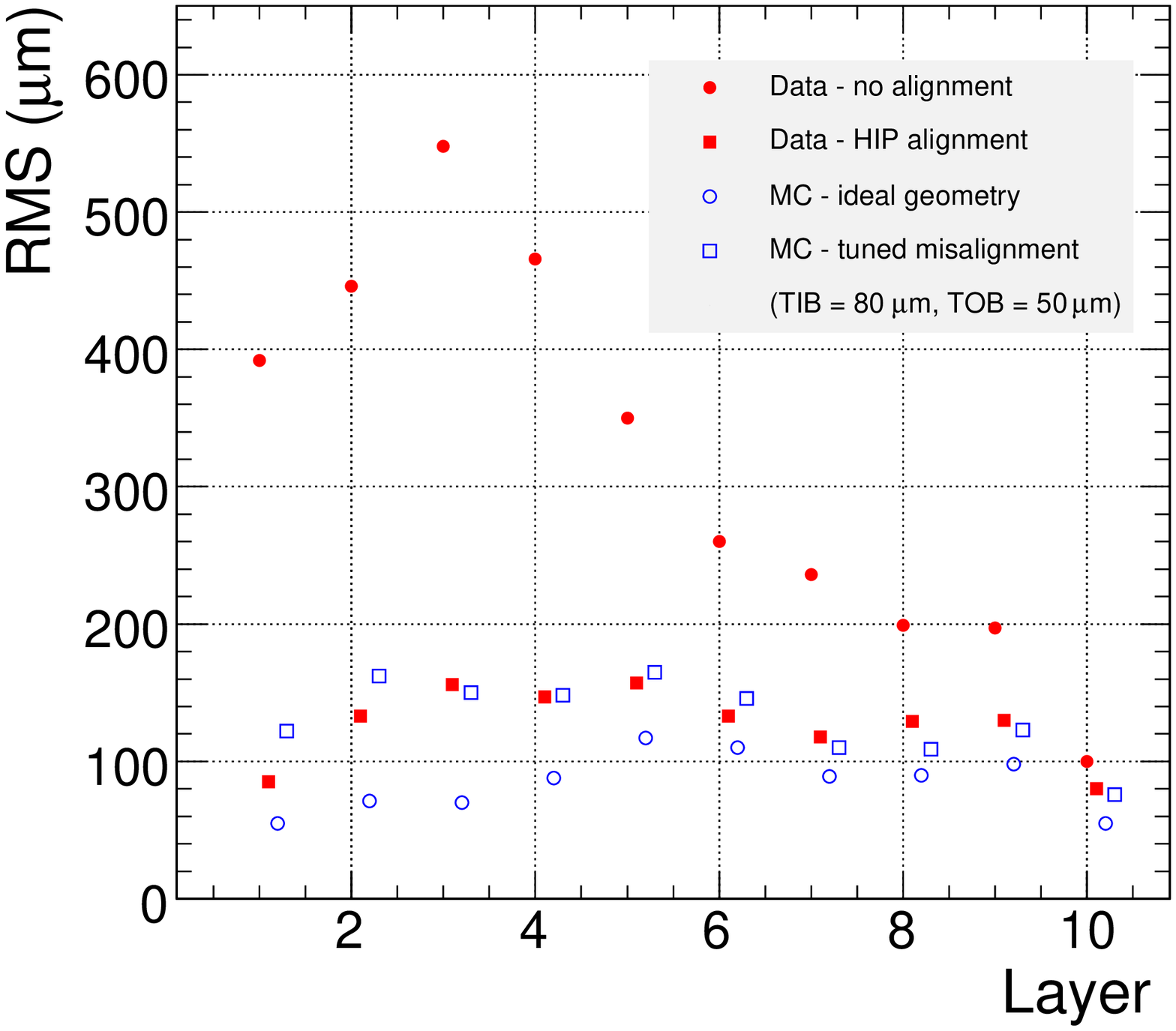}}

  \caption{\sl
Hit residual means in local $x'$ coordinate (left) and RMS (right) 
in ten layers of the barrel tracker, i.e. four layers of TIB and six layers 
of TOB, shown in data before track-based alignment (red full circles),
after track-based alignment (HIP, red full squares), in simulation with 
ideal geometry (blue open circles) and in simulation after tuning of
misalignment according to data (blue open squares).
}

  \label{fig:residuals_layers}
\end{center}
\end{figure}

The truncated mean and RMS values of these residual distributions are
shown in Fig.~\ref{fig:residuals_layers} for the HIP alignment result 
compared to the results before alignment, showing clearly the improvements.
The mean values are now close to zero and the RMS decreases by at least
almost a factor of two.



\subsection{Geometry comparisons}
\label{sec:track-based3}

\begin{figure}[tbp]
\begin{center}
  \includegraphics[width=\textwidth]{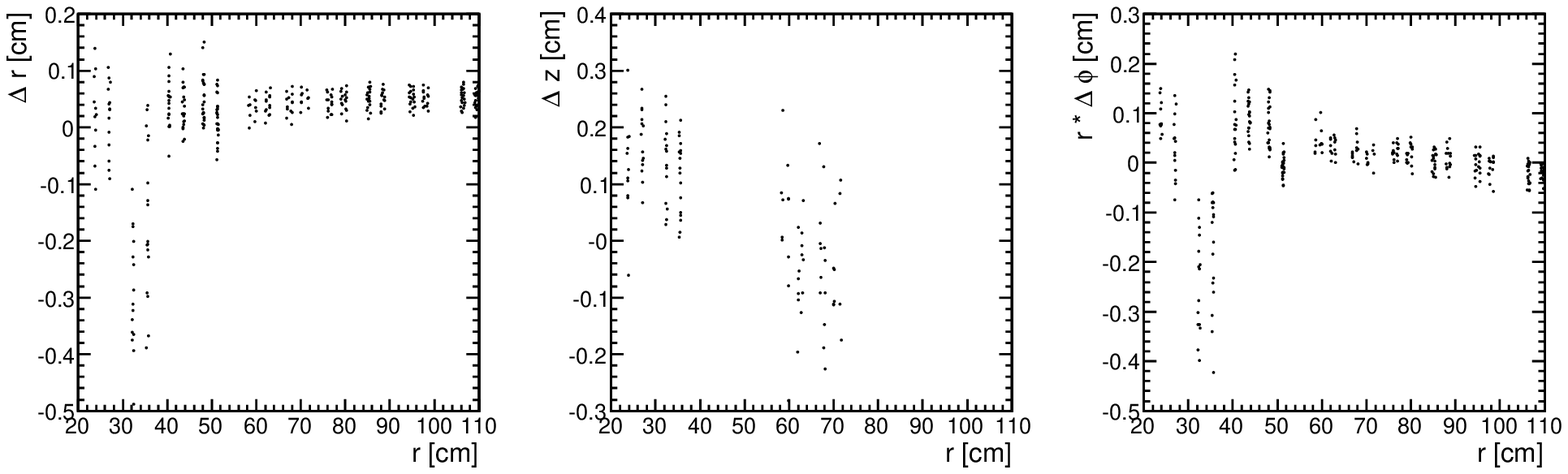}
  \includegraphics[width=\textwidth]{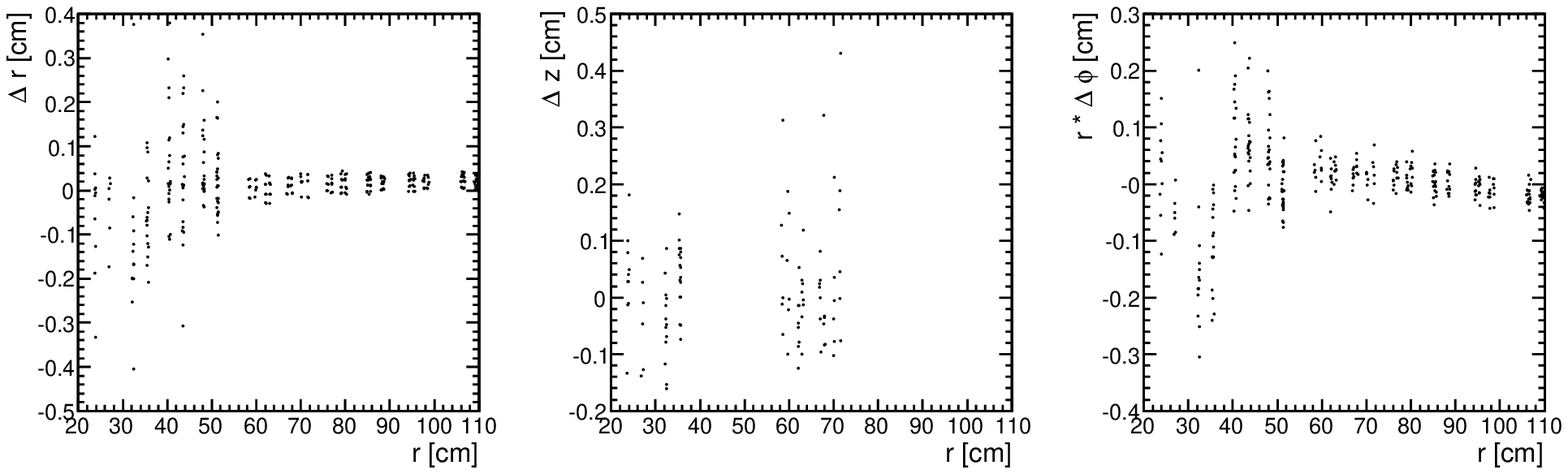}
  \caption{\sl
Difference of the module positions between the measured (in track-based alignment)
and design geometries shown for Kalman (top) and Millepede (bottom) algorithms
for TIB (radius $r<55$ cm) and TOB ($r>55$ cm). Projection on
the $r$ (left), $z$ (middle), and $\phi$ (right) directions are shown. 
Only double-sided modules are considered in the $z$ comparison.
  }
  \label{fig:KAAvIdeal_tif3x1}
\end{center}
\end{figure}

Overall, a very consistent picture is observed when the same comparison to 
design geometry, as shown in Fig.~\ref{fig:vIdeal_tif3x1} for the HIP constants,
is done
with the other two algorithms in Fig.~\ref{fig:KAAvIdeal_tif3x1}.
In all cases, the same coherent movement of TIB layers is found,
while TOB mounting precision is consistently better.


\begin{figure}[tbp]
\begin{center}
  \includegraphics[width=1.0\textwidth]{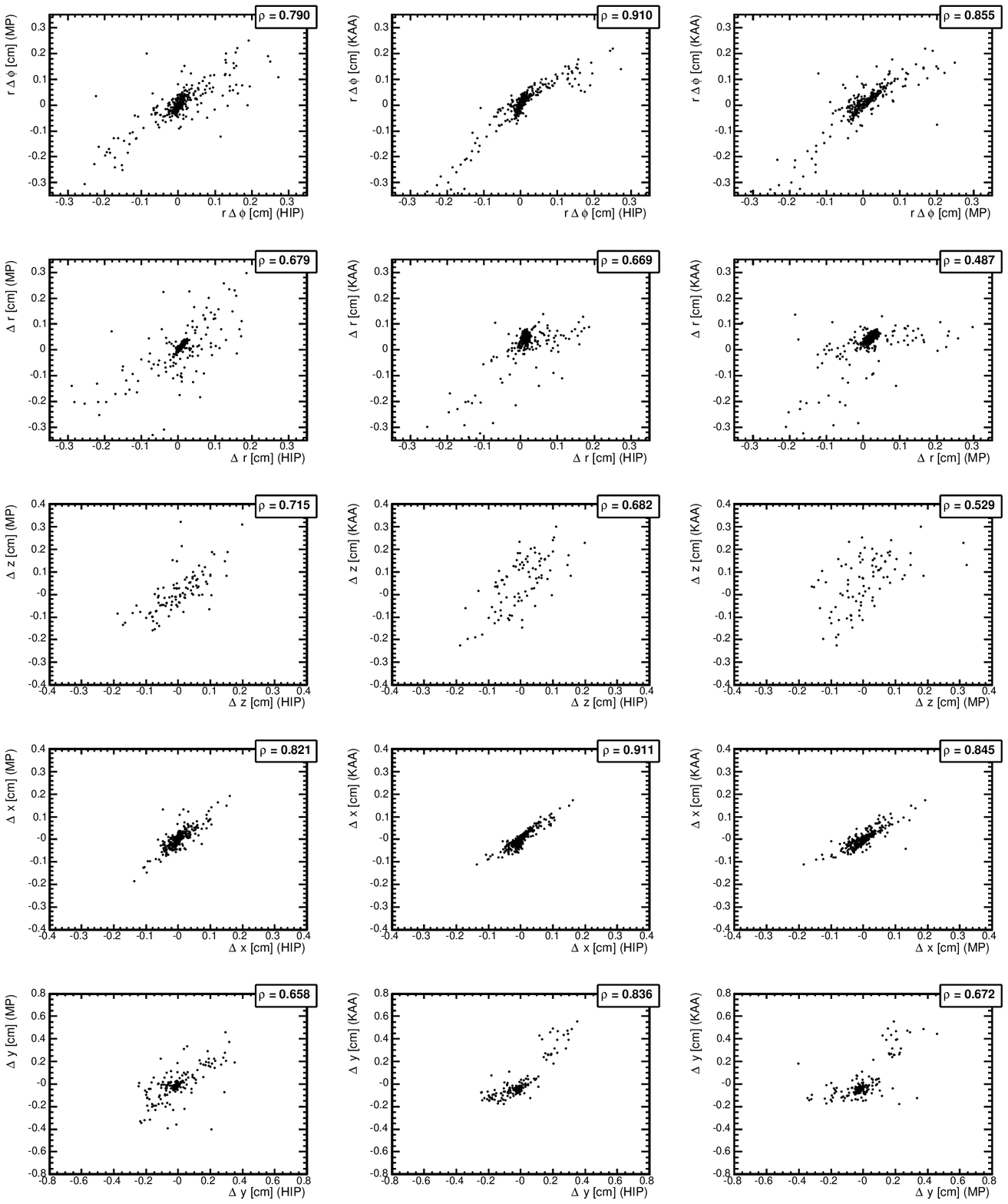}
  \caption{\sl
Direct comparison of differences from ideal geometry 
in $r\phi$ (top), $z$ (second row), $r$ (third row), $x$ (fourth row), and $y$ (bottom)
between Millepede and HIP (left), Kalman and HIP (middle), 
and Kalman and Millepede (right).
The correlation coefficient $\rho$ is stated.
  }
  \label{fig:correlation3x3}
\end{center}
\end{figure}

Consistency of the three algorithms is shown in 
Fig.~\ref{fig:correlation3x3} where the 
$r\phi$, $z$, $r$, $x$ and $y$ 
differences from ideal geometry of the result of the three algorithms
is compared to each other. 
A good correlation of the results is observed, especially in the
$x$ displacement, which is the most sensitive coordinate
with vertical tracks.
The main residual deviation from the diagonal $100\%$ correlation 
is due to statistical and systematic differences in the approaches,
therefore reflecting the achieved precision of the methods.
The numerical results of comparison of different geometries
are shown in Table~\ref{tab:difference}.
The RMS of agreement between algorithms in the $x$ coordinate is
as good as $150~\mu$m,
except for comparison of the Millepede constants in the TIB.

 \begin{table}[t]
 \caption{\sl
 Comparison of the global $x$
 RMS difference (in $\mu$m) of module
 positions between different geometries indicated in the first 
 two columns for TOB and TIB. Single-sided (SS) and double-sided (DS)
modules are shown together and separately.
 }
 \label{tab:difference}
\begin{center}
\begin{tabular}{|c|c|c|ccc|ccc|}
\hline
Geom 1 & Geom 2 & difference & TIB  & TIB\,(SS) & TIB\,(DS)  &
                 TOB & TOB\,(SS) & TOB\,(DS) \cr
\hline
HIP & Design & $\Delta x$ & 526 & 438 & 581 & 130 & 108 & 142 \cr 
MP & Design & $\Delta x$ & 623 & 500 & 653 & 236 & 206 & 208 \cr 
KAA & Design & $\Delta x$ & 543 & 500 & 519 & 237 & 215 & 173 \cr 
KAA & HIP & $\Delta x$ & 165 & 138 & 193 & 159 & 156 & 74 \cr 
MP & HIP & $\Delta x$ & 341 & 297 & 383 & 162 & 148 & 151 \cr 
KAA & MP & $\Delta x$ & 304 & 226 & 396 & 123 & 132 & 97 \cr 
\hline
\end{tabular}
\end{center}
\end{table}

\begin{figure}[tbp]
\begin{center}
  \includegraphics[width=0.5\textwidth]{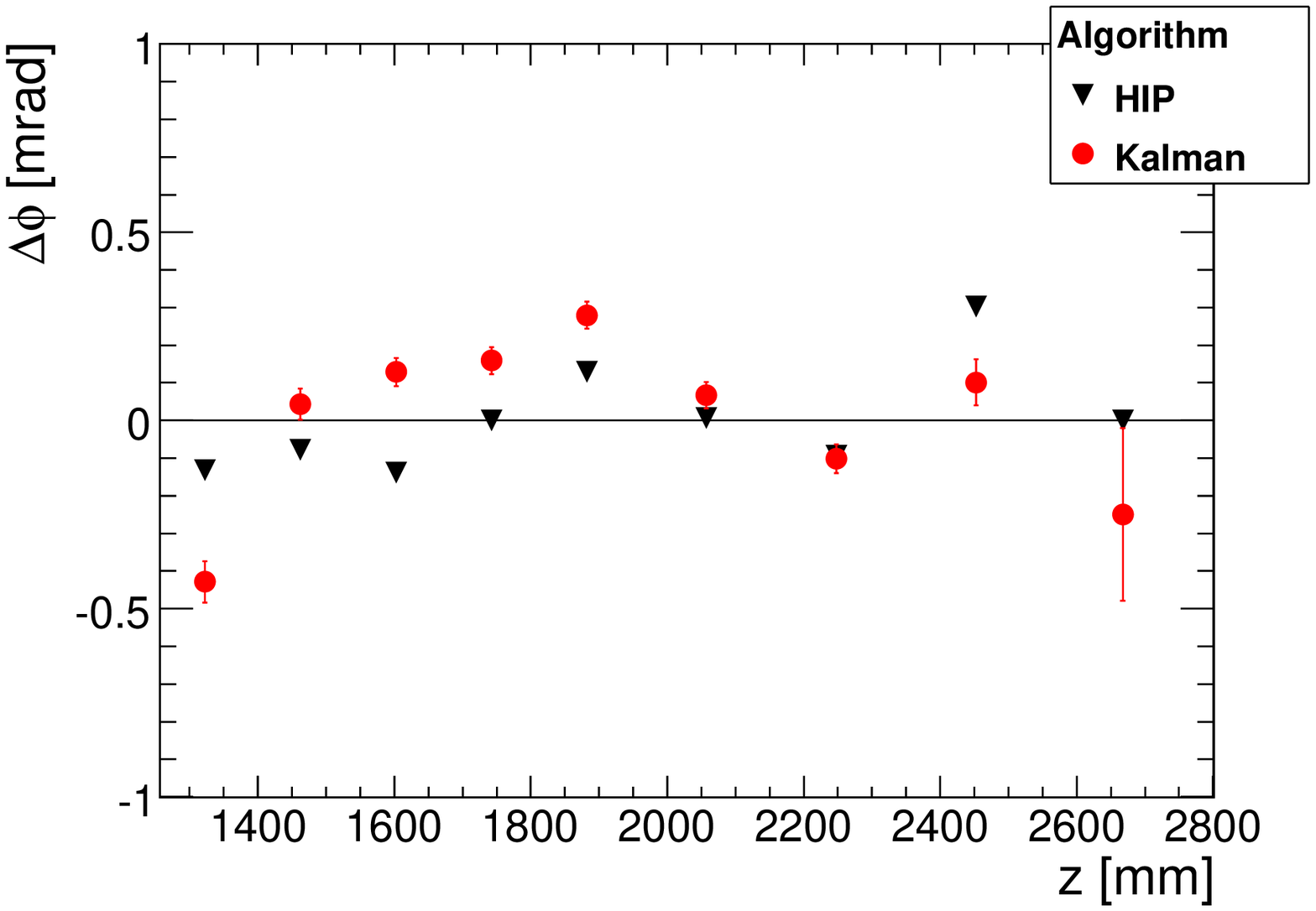}
  \caption{\sl
Rotations of the TEC disks around the global $z$
in comparison of the measured (in track-based alignment)
and design geometries for TEC. Two track-based results are shown: 
HIP (triangles) and Kalman filter (circles) algorithms.
  }
  \label{fig:TECcomparison}
\end{center}
\end{figure}

As discussed in Sec.~\ref{sec:commonalisel}, no attempt to align the TID 
was made and for the TEC, only rotation
of the nine disks around the global $z$ coordinate was studied, due to limited 
track statistics in the endcaps. Comparison of
the resulting geometry in two algorithms (HIP and Kalman) is shown in 
Fig.~\ref{fig:TECcomparison}. 
The results exhibit slight differences, but they clearly show the same trend.

\subsection{Track-Based Alignment with Simulated Data and Estimation of Alignment Precision}
\label{sec:mc}

\begin{figure}[htp]
  \begin{center}
    \vspace*{2mm}
    \begin{minipage}{4.5cm}
      \centerline{\includegraphics[height=4.5cm]{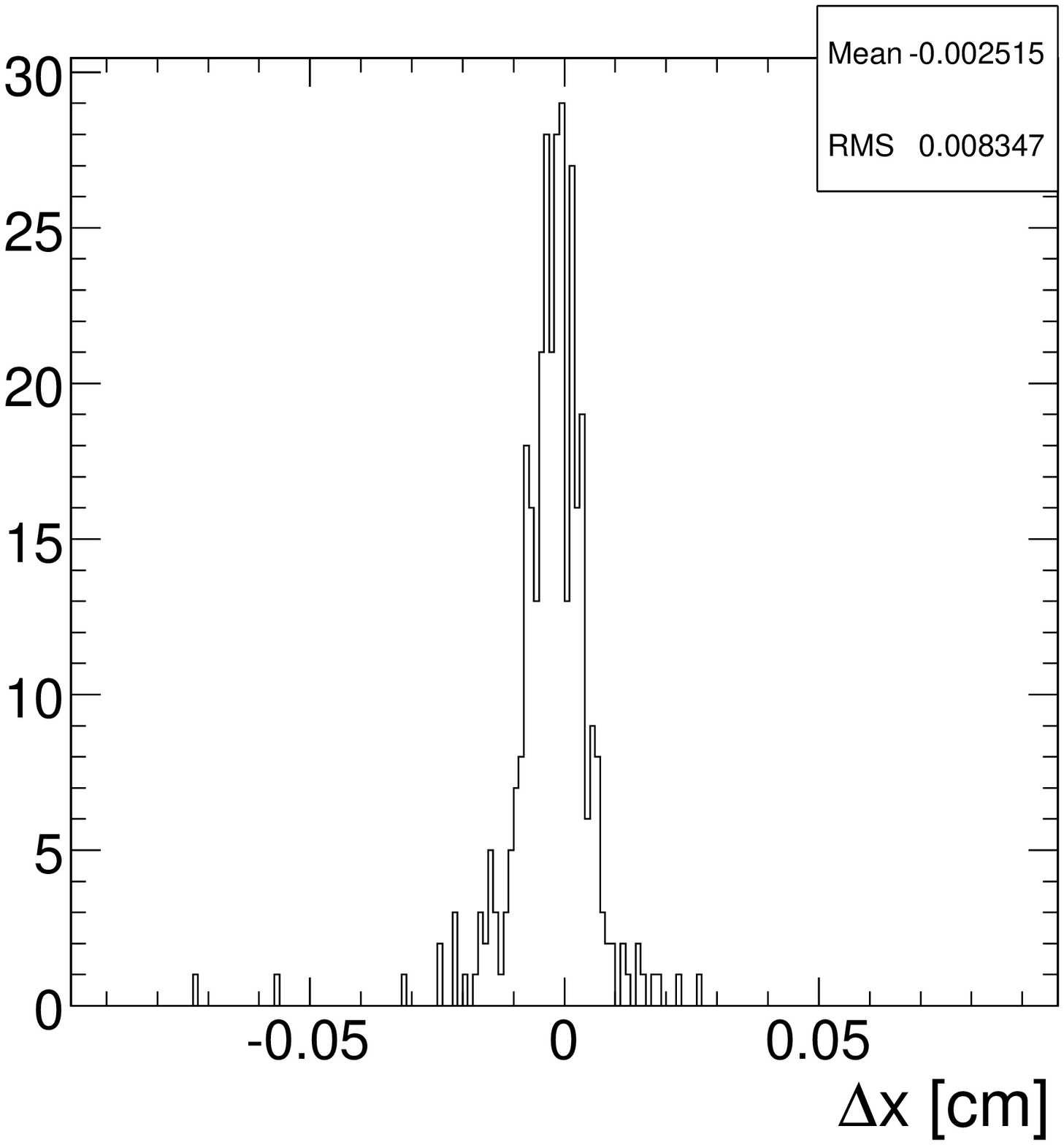}}
    \end{minipage}
    \hspace{0.6cm}
    \begin{minipage}{4.5cm}
      \centerline{\includegraphics[height=4.5cm]{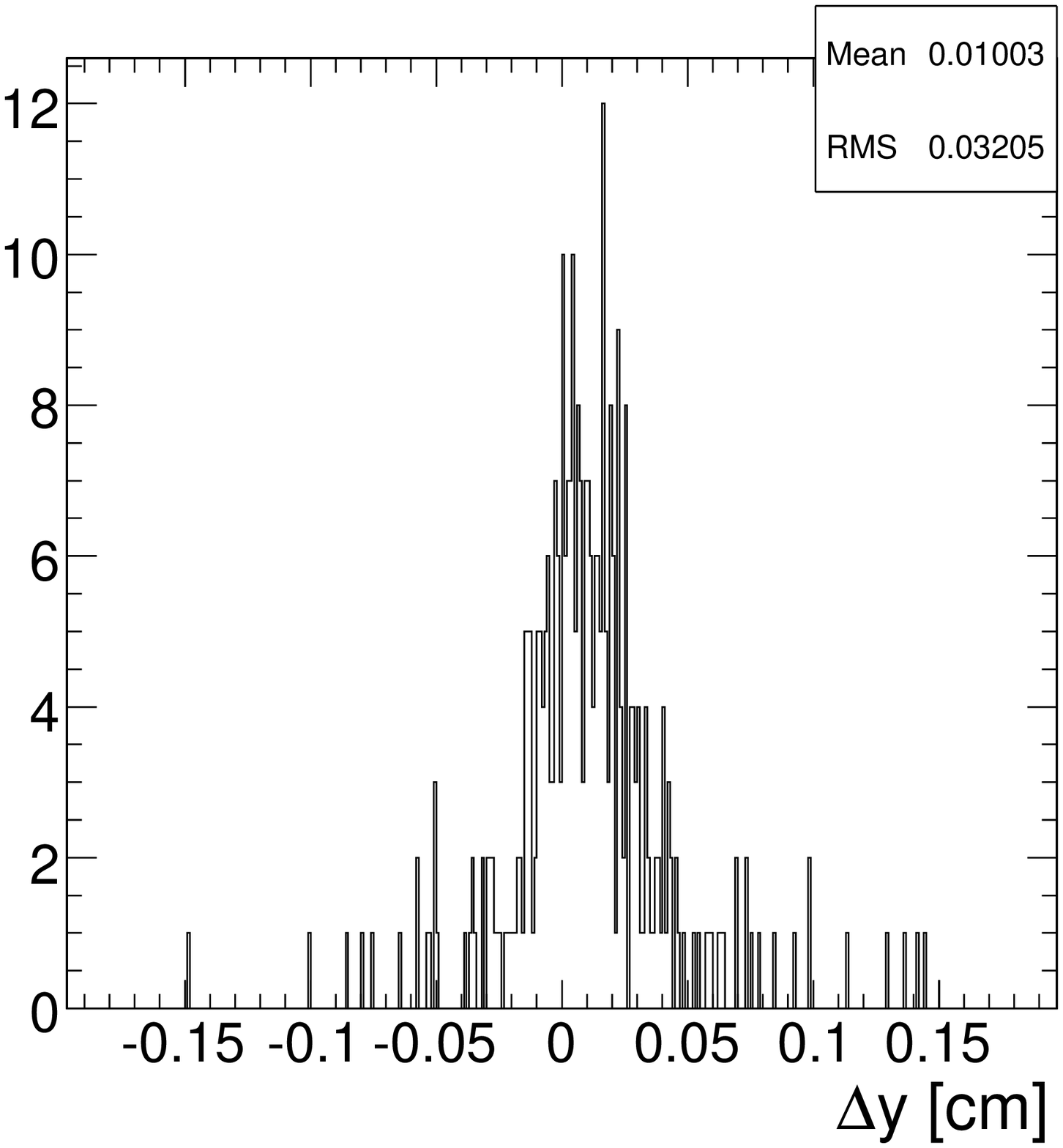}}
    \end{minipage}
    \hspace{0.6cm}
    \begin{minipage}{4.5cm}
      \centerline{\includegraphics[height=4.5cm]{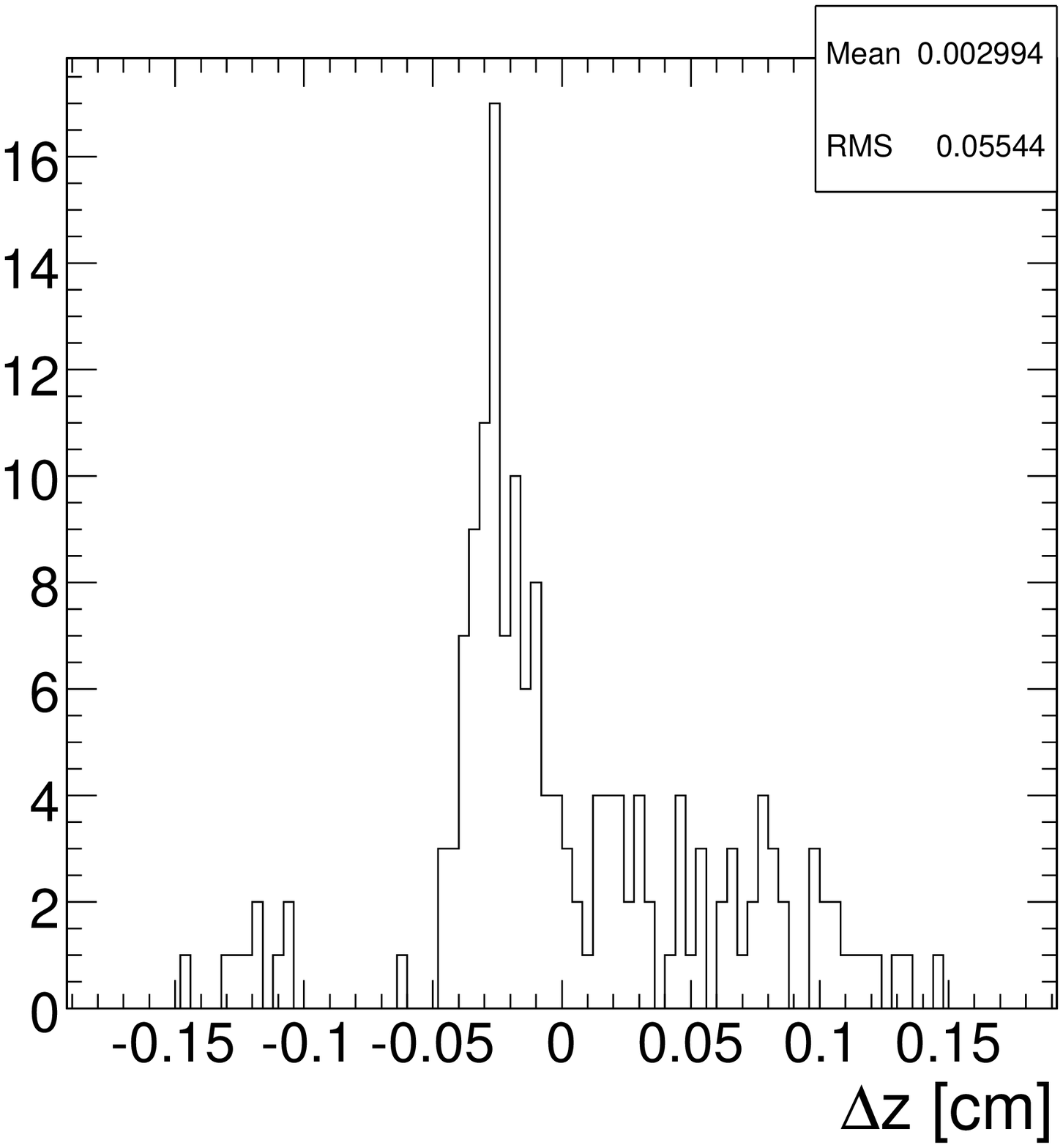}}
    \end{minipage}
    \caption{Alignment resolution in global coordinates achieved with the Kalman alignment
      algorithm on simulated data.
      \label{fig:methods-kalman:mc-results}
    }
  \end{center}
\end{figure}

%
Alignment tests on simulated data have been performed 
with the Kalman algorithm on approximately
40k events from a sample that mimics the situation at the TIF. In
order to reproduce our knowledge of the real tracker geometry after
survey measurements only, movements and errors to the tracker elements
are applied according to the expected starting misalignment~\cite{scenarios}.
The alignment strategy and track selection discussed
above are applied to obtain the results shown in
Fig.~\ref{fig:methods-kalman:mc-results},
resulting in a precision of $80$~$\mu$m in global $x$ position.

\begin{figure}[tbp]
\begin{center}
  {\includegraphics[width=\textwidth]{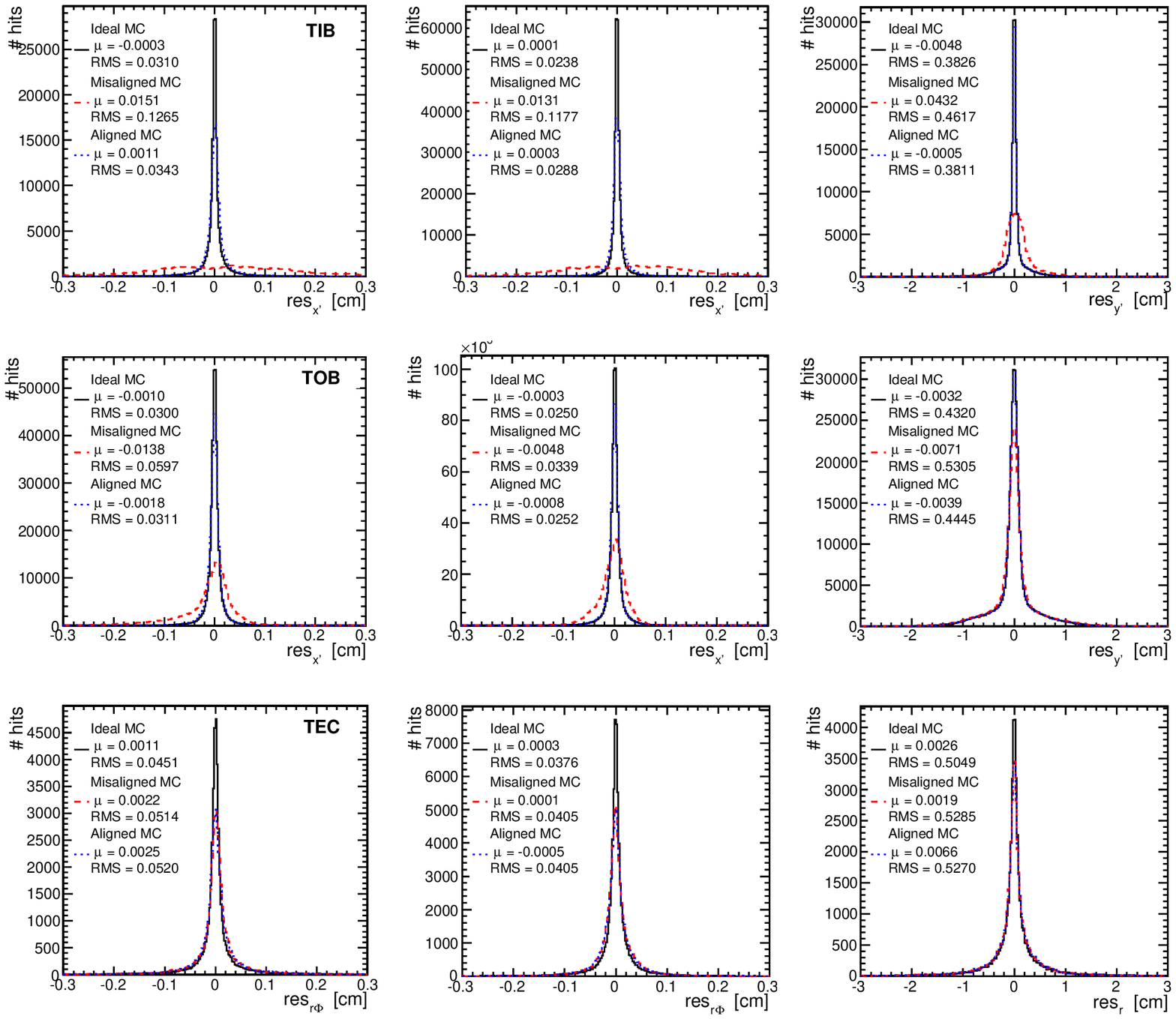}}
  \caption{\sl 
Hit residuals for different geometries
in different conditions for the simulated data sample:
ideal geometry (solid/black),
misaligned geometry according to expected starting misalignment (dashed/red), 
and geometry after alignment (dotted/blue).
Three Tracker sub-detectors are shown in the
top row (TIB), second row (TOB), and bottom row (TEC).
The absolute local $x$-residuals are shown for single-sided modules (left)
and double-sided modules (middle), while local $y$-residuals are shown for
the double-sided modules only (right).
For the endcap modules (TEC) transformation to the
$r\phi$ and $r$ residuals is made.
  }
  \label{fig:residuals_mc}
\end{center}
\end{figure}

An alignment study on the full MC data set has been performed with
the Millepede algorithm with the same settings as for the data, i.e.
alignment of a subset of the barrel part at module level and of the
TEC at disk level. 
The resulting residual distributions in TIB, TOB and TEC are
shown in Fig.~\ref{fig:residuals_mc} and compared with the startup 
misalignment~\cite{scenarios} and the ideal geometry.
Comparison with the distributions obtained from data using the design
geometry (Fig.~\ref{fig:residuals_alg}) reveals that in TIB and TOB the
starting misalignment is overestimated while in TEC it is slightly underestimated.
The residual widths after alignment are generally much smaller than those
obtained from the aligned data, especially in the TIB.
This could be due to the larger statistics of the simulation data sample, but 
also due to effects not properly simulated, e.g. relative misalignment of the
two components of a double-sided module or possible differences in the 
momentum spectrum of Monte Carlo.

\label{sec:track-based2}


The results of the truncated RMS of the layerwise residual
distributions in Fig.~\ref{fig:residuals_layers}
are used to estimate alignment precision in the aligned barrel region
via comparison with simulations.
Different misalignment scenarios have been applied to the
ideal (``true'') Tracker geometry used in reconstructing the simulated data
until truncated RMS values are found to be similar to the ones
in data in all layers.
The modules in TIB and TOB have been randomly shifted in three
dimensions by Gaussian distributions. 
The influence of possibly large misalignments from the tails
of these Gaussians
is reduced by truncating the distributions as stated above.

Besides the truncated mean and RMS values from data before and after
alignment, Fig.~\ref{fig:residuals_layers} shows also the results
from the simulation reconstructed with the ideal geometry and 
reconstructed with a random 
misalignment according to Gaussian distributions with standard deviations
of 50~$\mu$m and 80~$\mu$m in the TOB and the TIB, respectively.
It can be clearly seen that the simulation with the ideal, i.e. true,
geometry has smaller widths than the data, especially in the TIB.
On the other hand, the geometry with a simulated misalignment of
50~$\mu$m and 80~$\mu$m, respectively, resembles rather well the 
data after alignment, such that these numbers can well be taken 
as an estimate of the size of the remaining misalignment.


\section{Stability of the Tracker Geometry with Temperature and Time}
\label{sec:validation-stability}
\subsection{Stability of the Tracker Barrels}

In order to investigate the stability of the tracker components with
respect to the cooling temperature and stress due to TEC insertion,
full alignment
of the Tracker in different periods 
has been performed
and the positions
of modules in space 
are compared.
The advantage of this approach is that
we can see module movements directly, but the potential problem
is that we may be misled by a systematic effect or a weakly
constrained misalignment. 
%
%
Statistical scatter of up to 100~$\mu$m limits the resolution of the method.
These tests have been done with the HIP algorithm.

\begin{enumerate}
\item {\bf +15~$^\circ$C (A$_{1}$,
before TEC- insertion) vs. +10~$^\circ$C (C$_{10}$, after TEC- insertion).}

This test is intended to show the effect of the insertion of a mechanical
object between two data-taking conditions. 
Fig.~\ref{fig:newtimecomp} shows the shifts between the two sets of aligned 
positions in global $x$, $y$ and $z$
as a function of the radial coordinate and projected separately for TIB and 
TOB.
\begin{figure}[tbp]
\begin{center}
\centerline{
\epsfig{figure=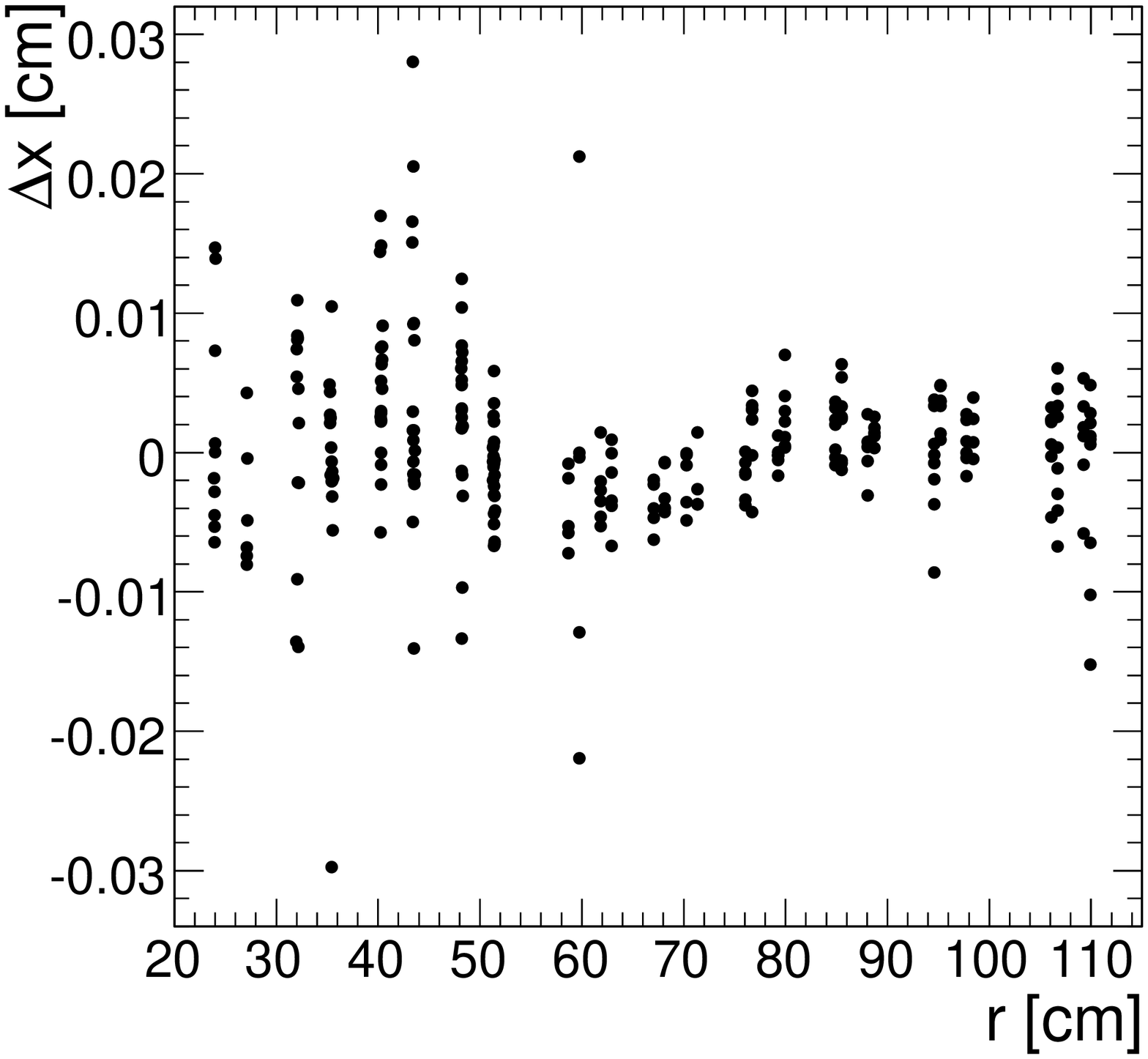,height=3.82cm}
\epsfig{figure=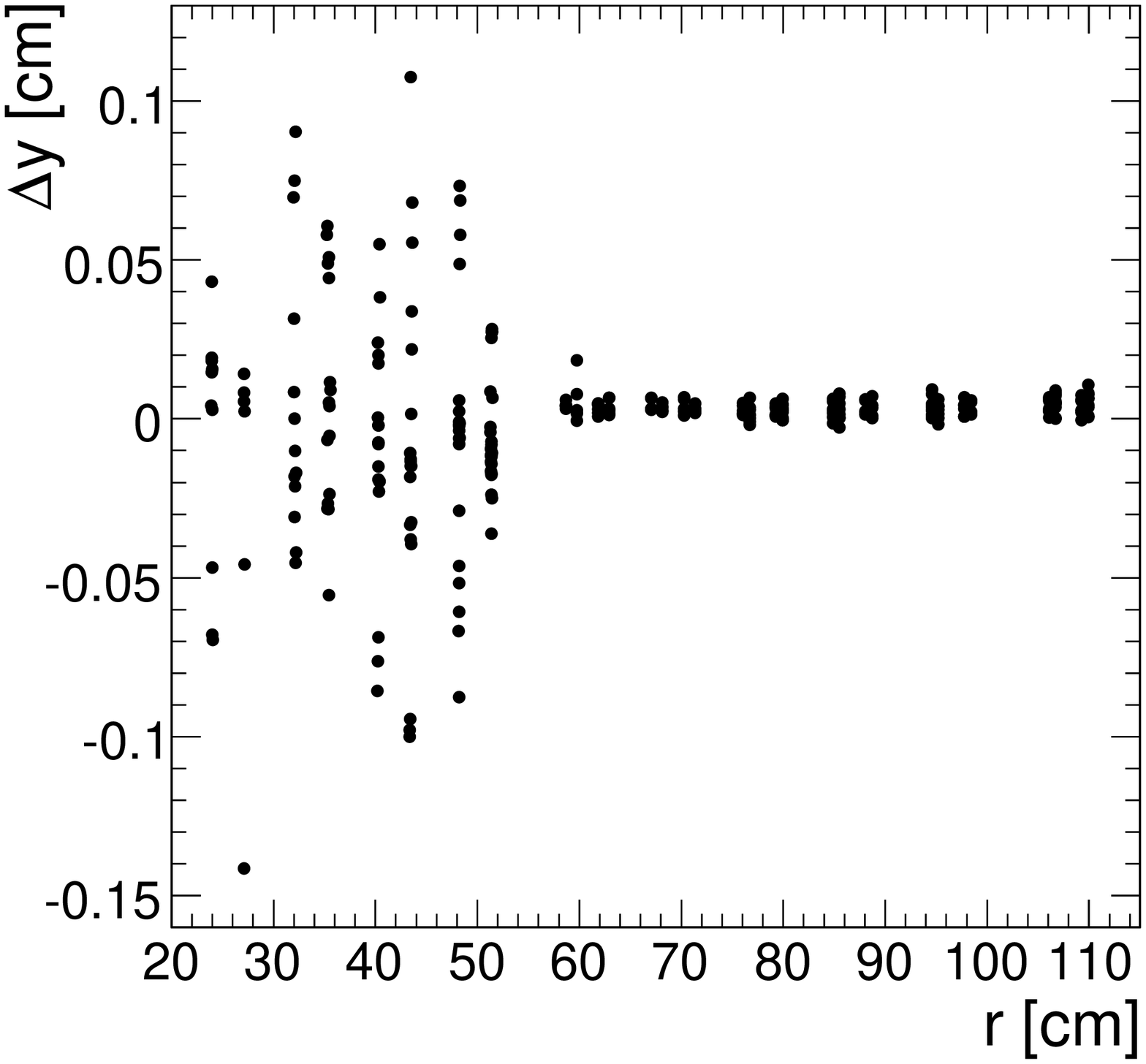,height=3.82cm}
\epsfig{figure=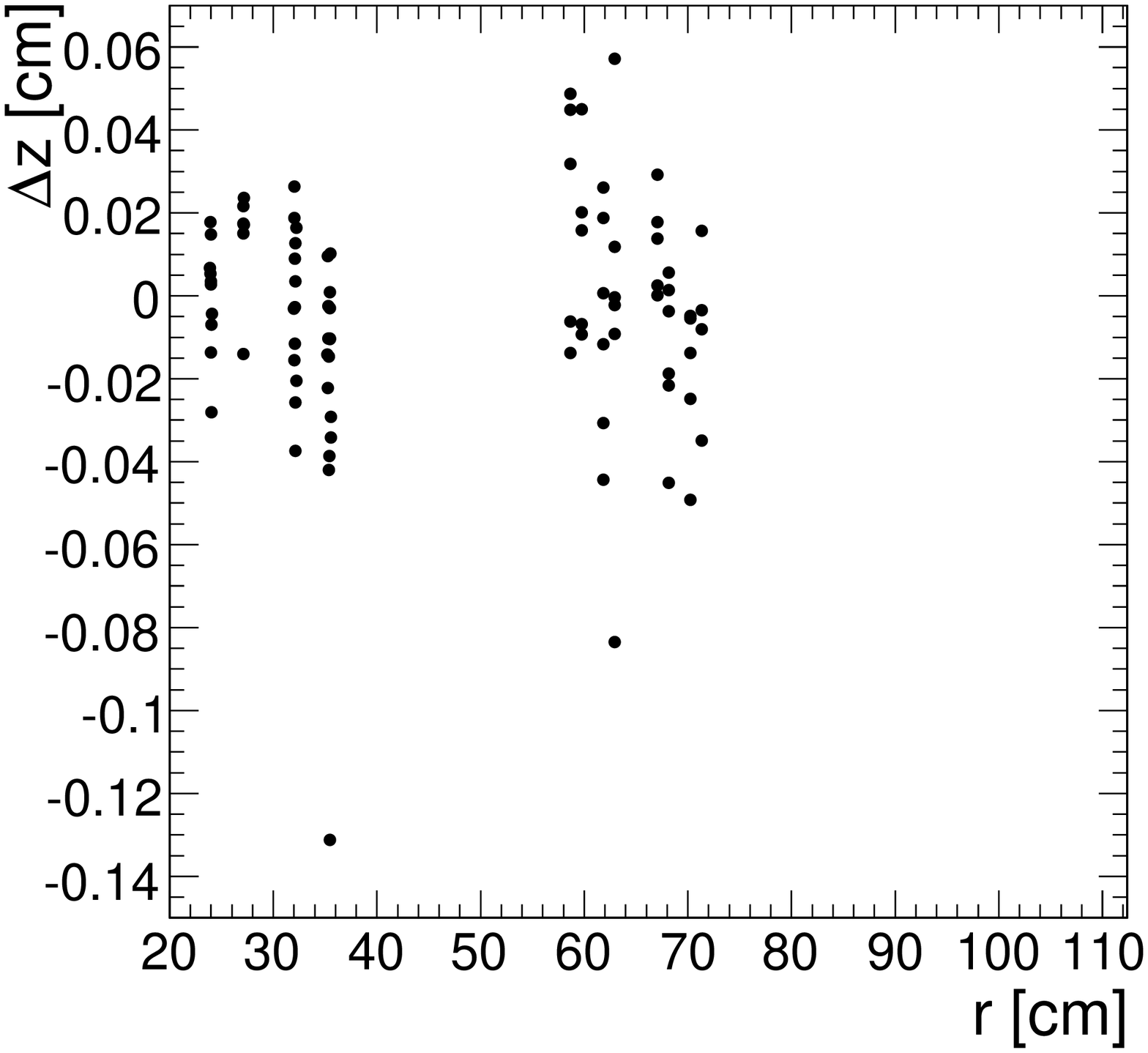,height=3.82cm}
}
\centerline{
\epsfig{figure=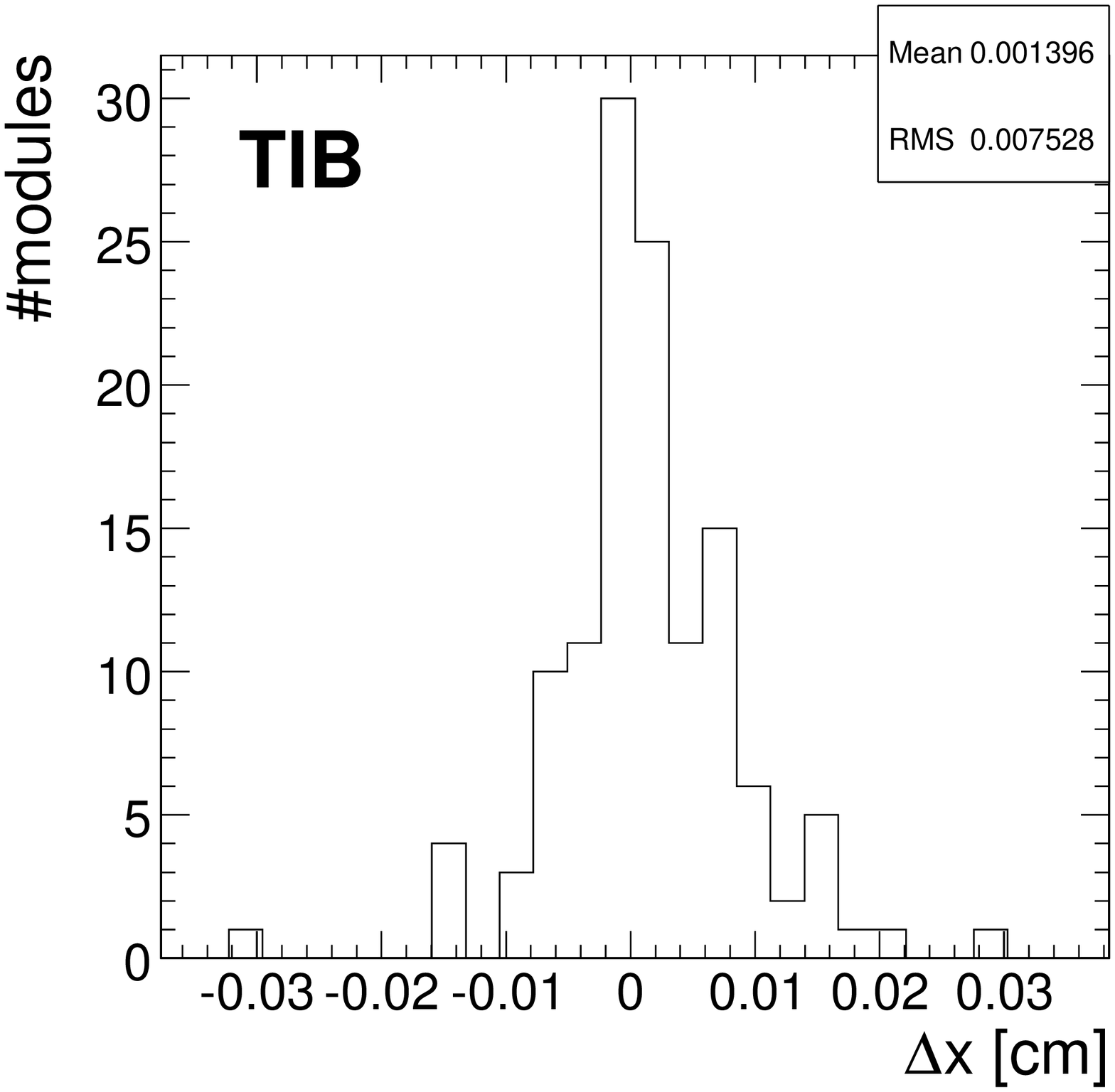,height=3.82cm}
\epsfig{figure=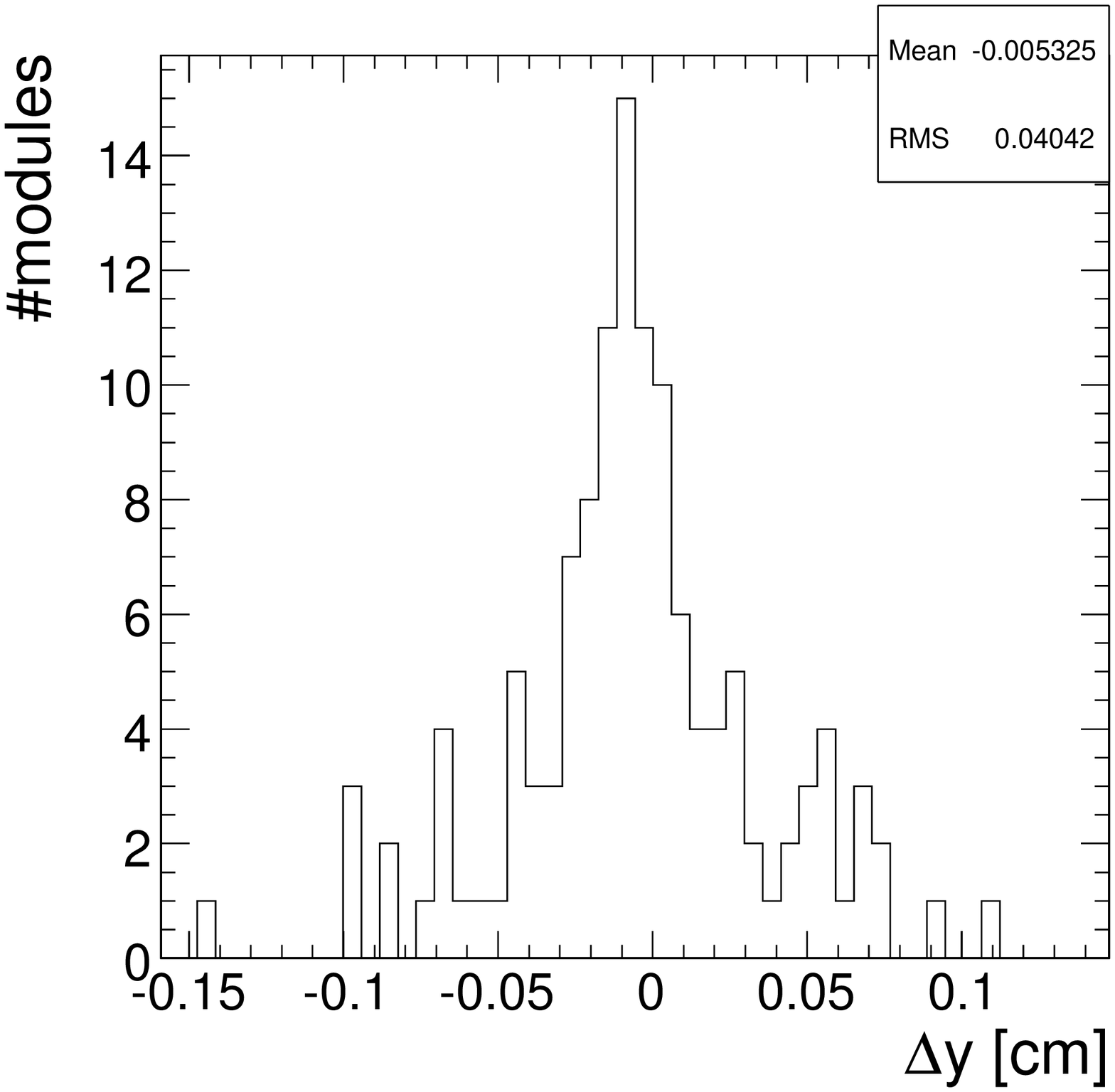,height=3.82cm}
\epsfig{figure=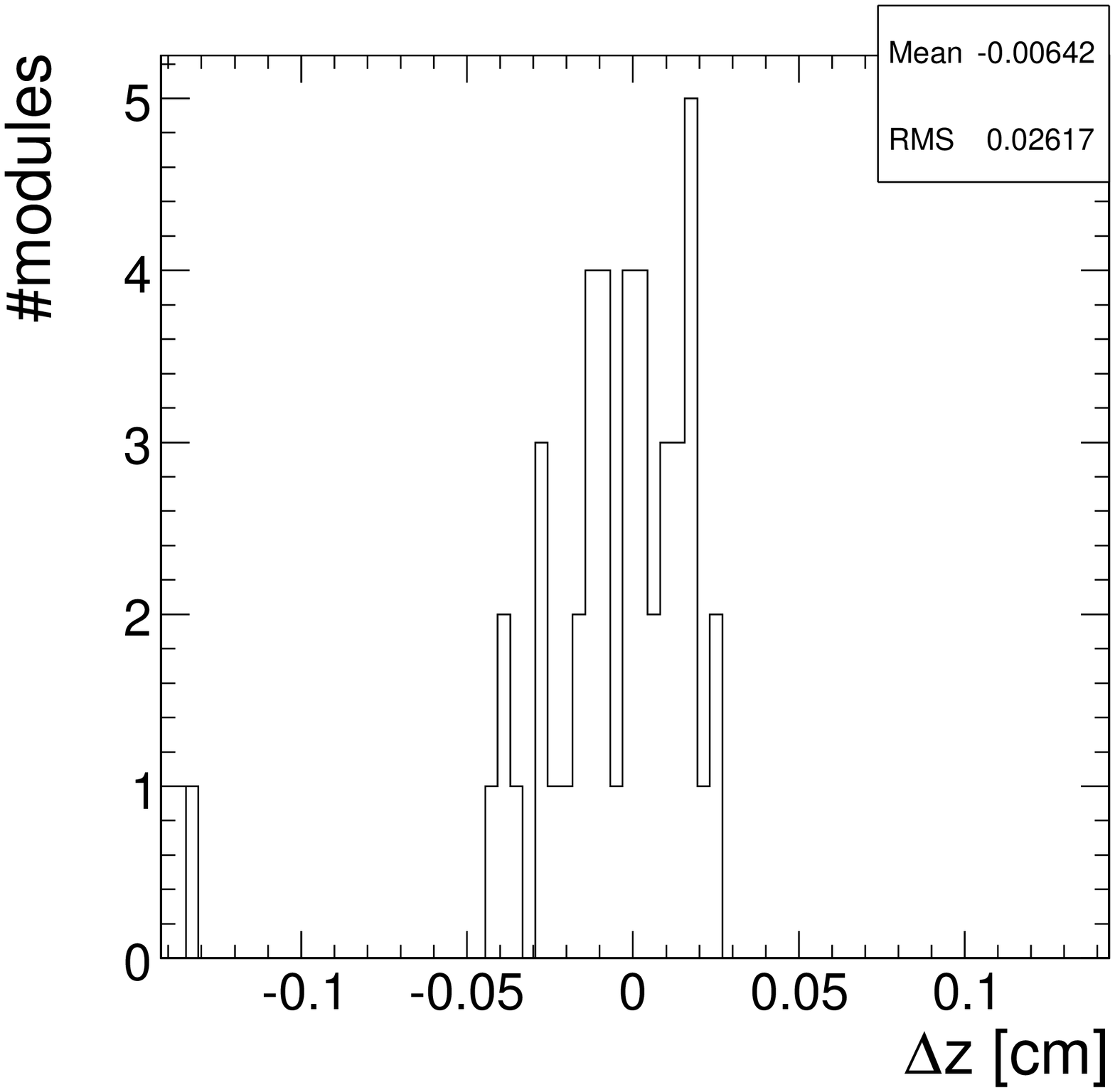,height=3.82cm}
} 
\centerline{
\epsfig{figure=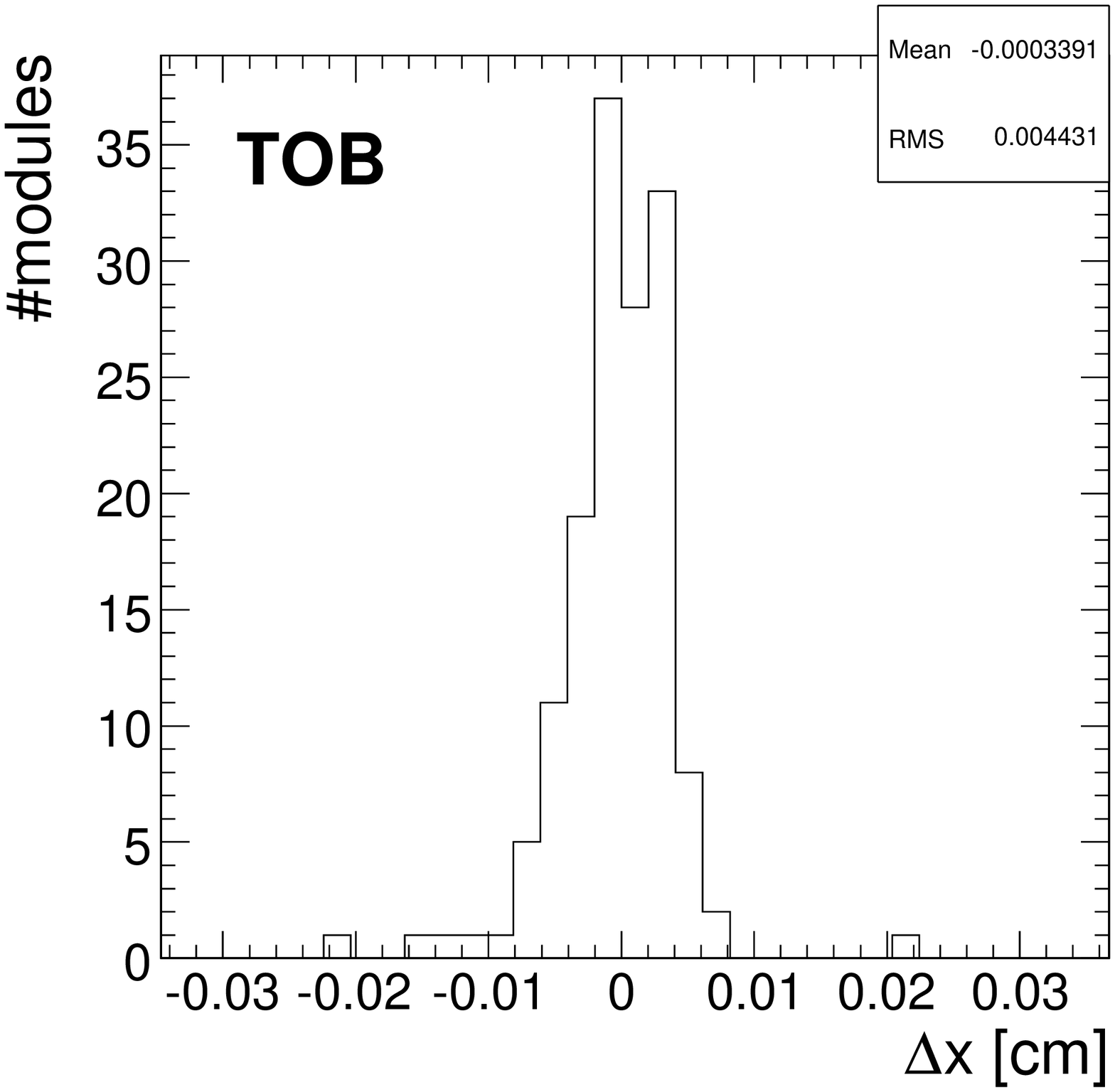,height=3.82cm}
\epsfig{figure=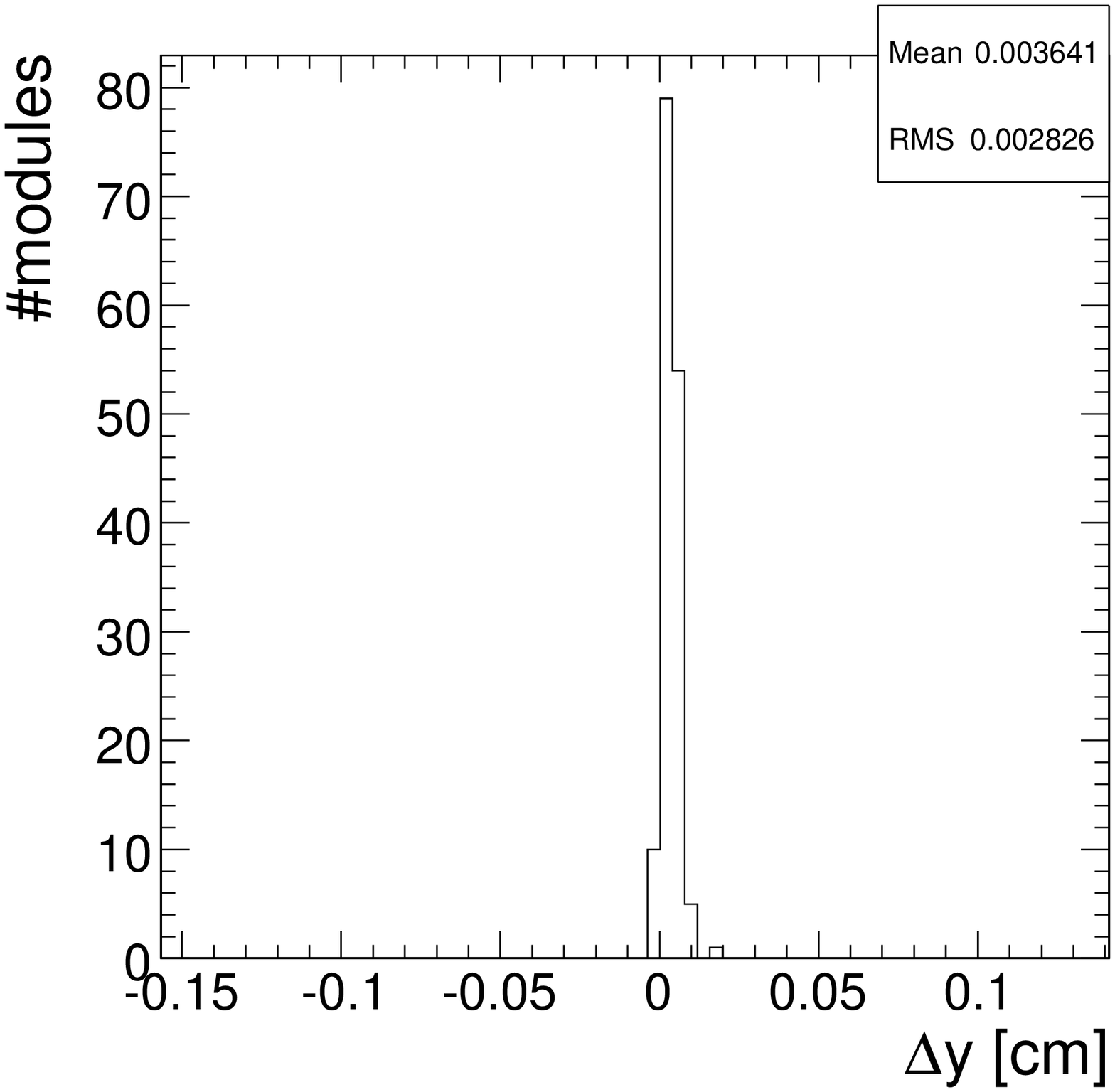,height=3.82cm}
\epsfig{figure=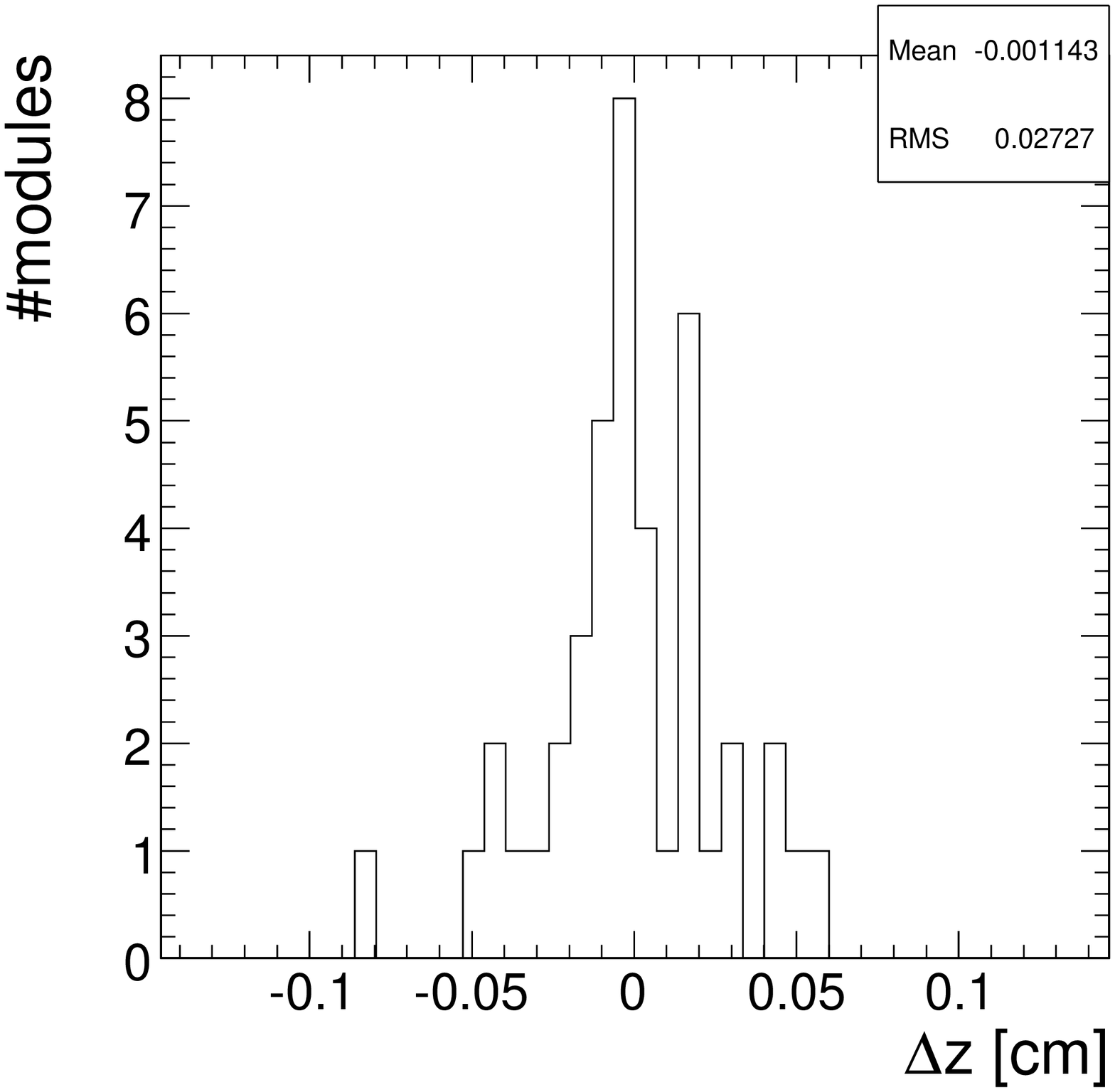,height=3.82cm}
} 
\caption{\sl
Differences in determined  $x$- (left), $y$- (centre) and $z$-positions
(right, only double-sided) of active modules comparing the
configurations before and after TEC- insertion.
The differences are stated as a function of
the module radius $r$ (top row) and for modules in TIB (middle row) and TOB (bottom row)
separately. 
\label{fig:newtimecomp}
}
\end{center}
\end{figure}

\begin{figure}[tbp]
\begin{center}
\centerline{
\epsfig{figure=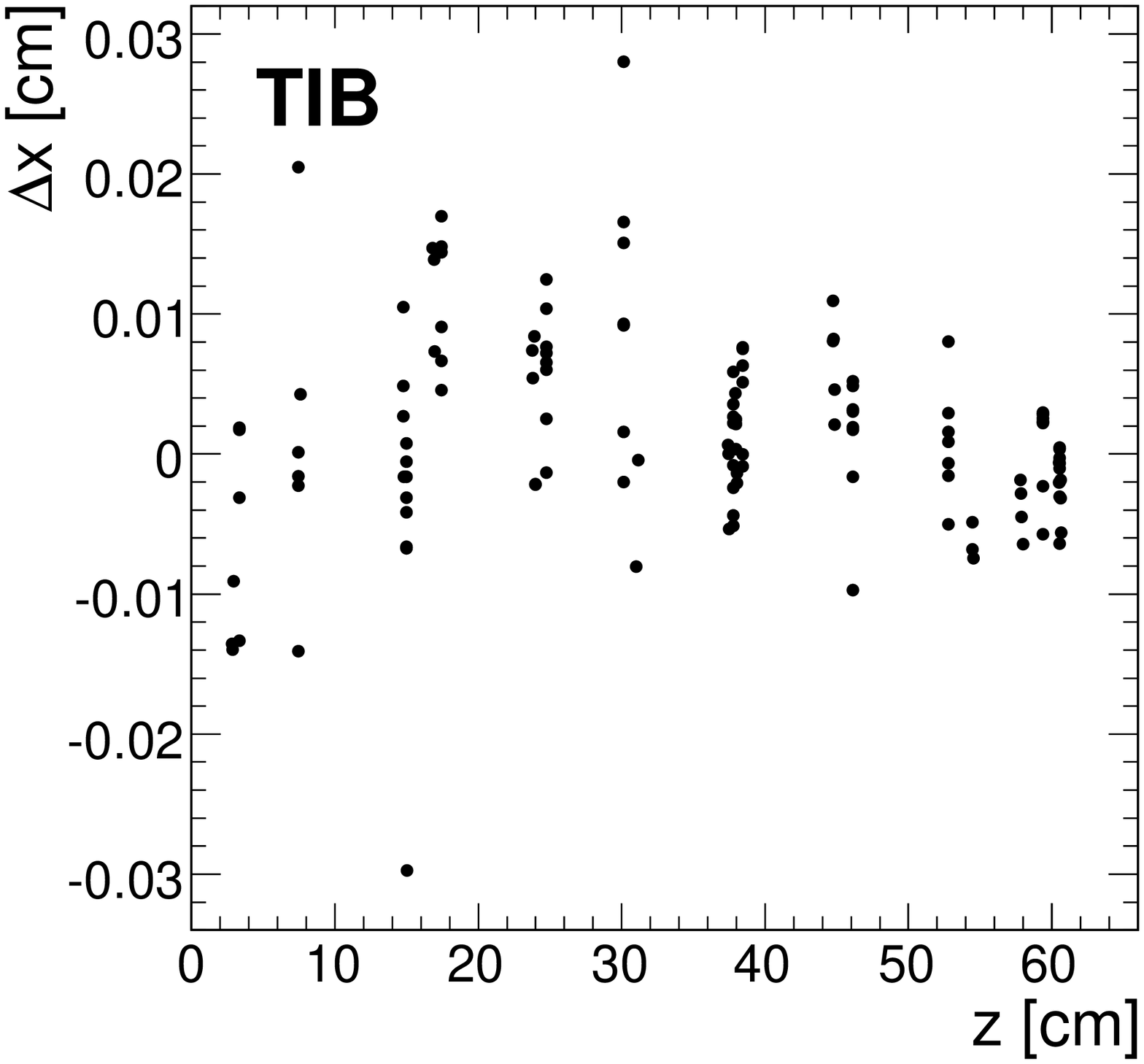,height=3.82cm}
\epsfig{figure=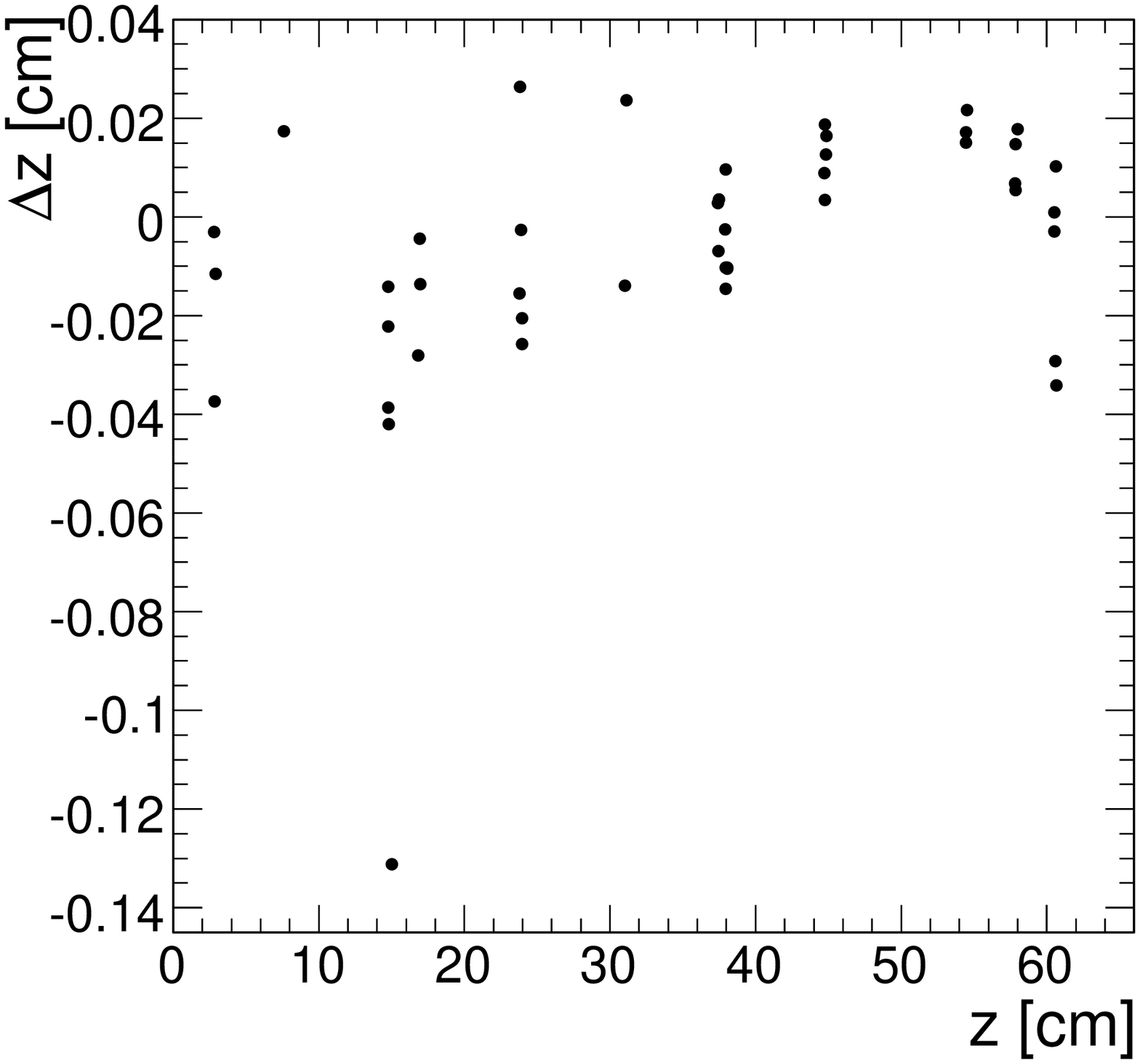,height=3.82cm}
}
\centerline{
\epsfig{figure=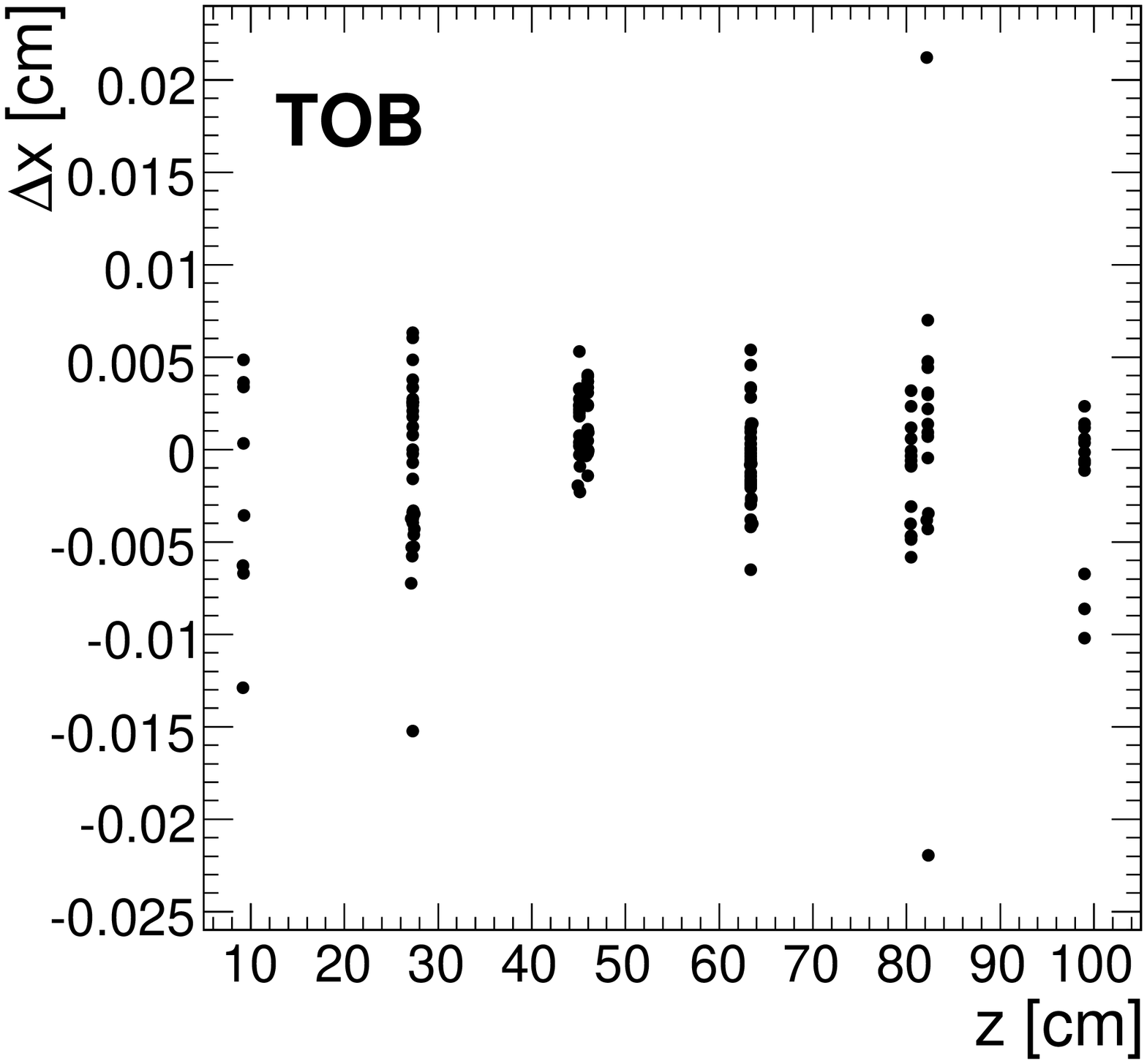,height=3.82cm}
\epsfig{figure=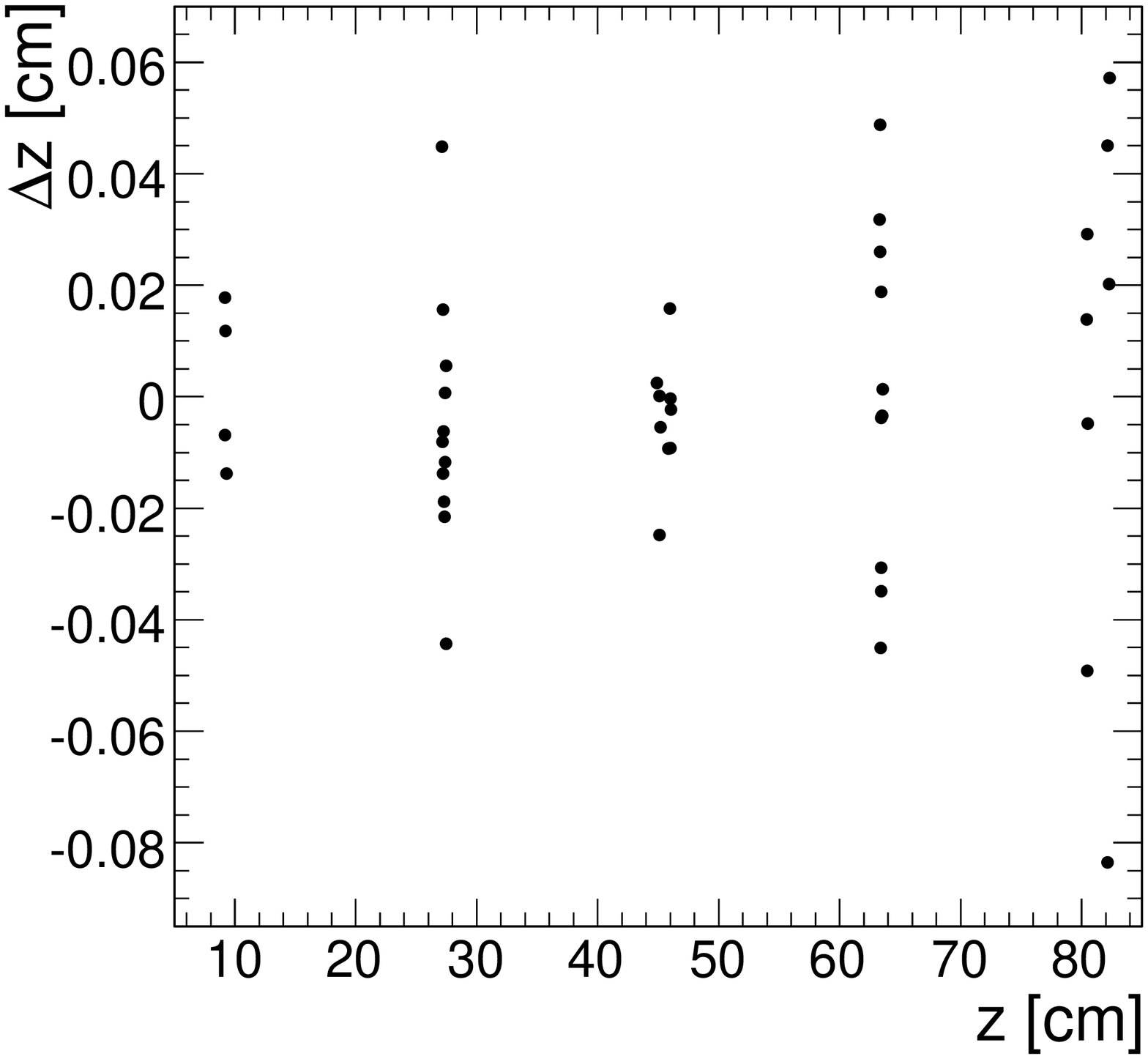,height=3.82cm}
}
\caption{\sl
Differences in determined  $x$- (left) and $z$-positions
(right, only double-sided) of active modules comparing the
configurations before and after TEC- insertion as a function of
the module $r$-position for modules in TIB (upper row) and TOB (bottom row).
\label{fig:newzcomp}
}
\end{center}
\end{figure}

In the TOB, a very small layer-wise shift is visible, 
especially in layers one and two. 

As can be seen from Fig.~\ref{fig:newzcomp}, there is no further 
structure as a function of the $z$ coordinate.
This could be a hint of a small layer-wise rotation around the $z$ axis.
In the TIB, coherent movements are larger in the azimuthal direction and are also 
layer-dependent; but here, they are reflected in the corresponding structures in 
the longitudinal direction:
the movement is largest
closer to $z=0$ and is reduced to small values at large $z$ 
(see Fig.~\ref{fig:newzcomp}).
We interpret it as a layer- and side-dependent twist where the outer
edges in $z$ are better constrained due to the mechanical mounting
technique.
However, it is also possible that there is not enough information to
constrain the ``weak'' degrees of freedom, or this could
be an artificial effect due to different modules being
aligned in different configurations and different track samples
due to different trigger configurations.

\item {\bf -10~$^\circ$C (C$_{-10}$,
default sample) vs. +10~$^\circ$C (C$_{10}$).}

This test is intended to show the effect of a large temperature gap between
two data-taking conditions. 
Figure~\ref{fig:newtempcomp} shows the shifts between the two sets of aligned 
positions in global $x$, $y$ and $z$
as a function of the radial coordinate and projected separately for TIB and 
TOB.
\begin{figure}[htbp]
\begin{center}
\centerline{
\epsfig{figure=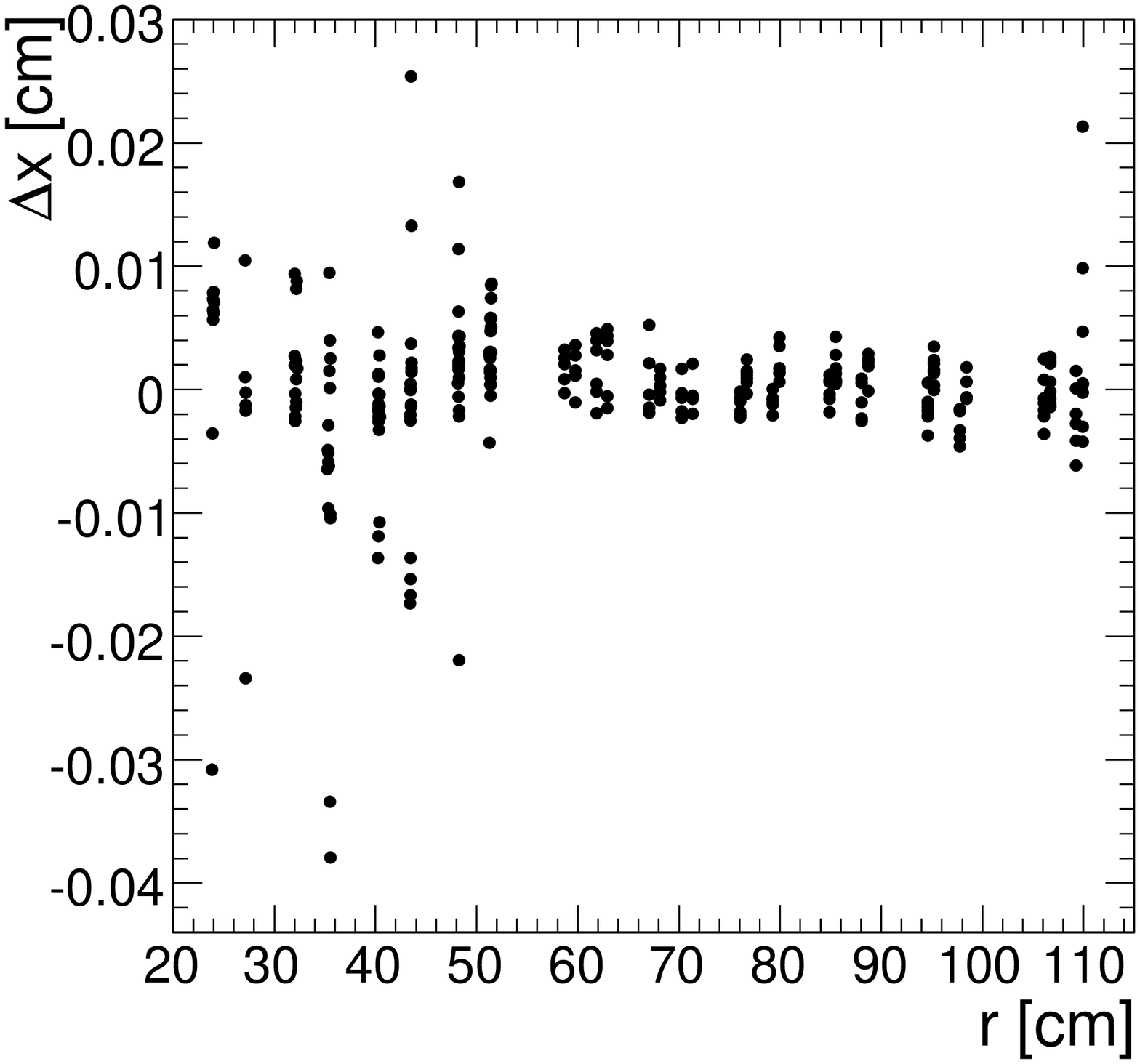,height=3.82cm}
\epsfig{figure=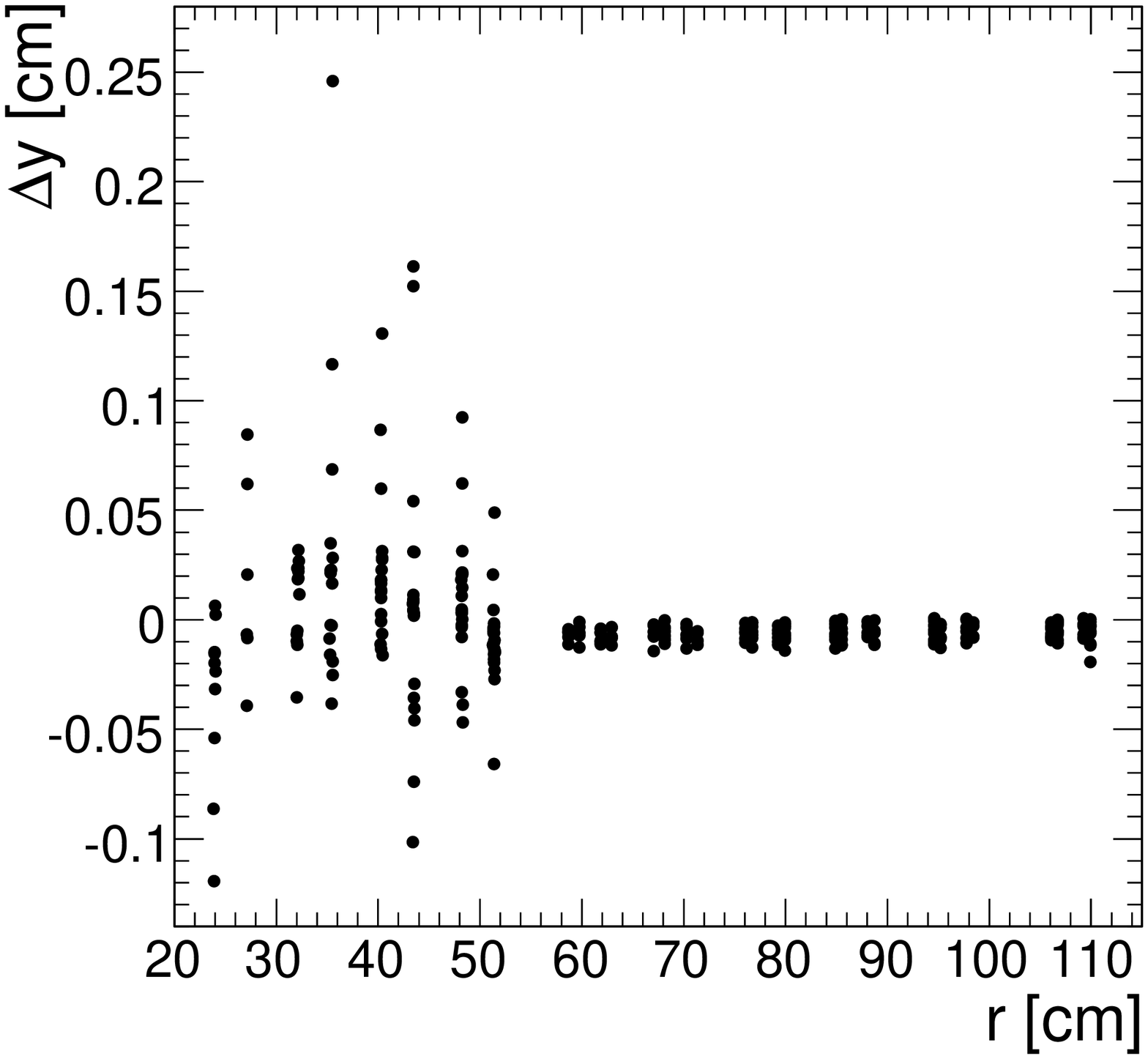,height=3.82cm}
\epsfig{figure=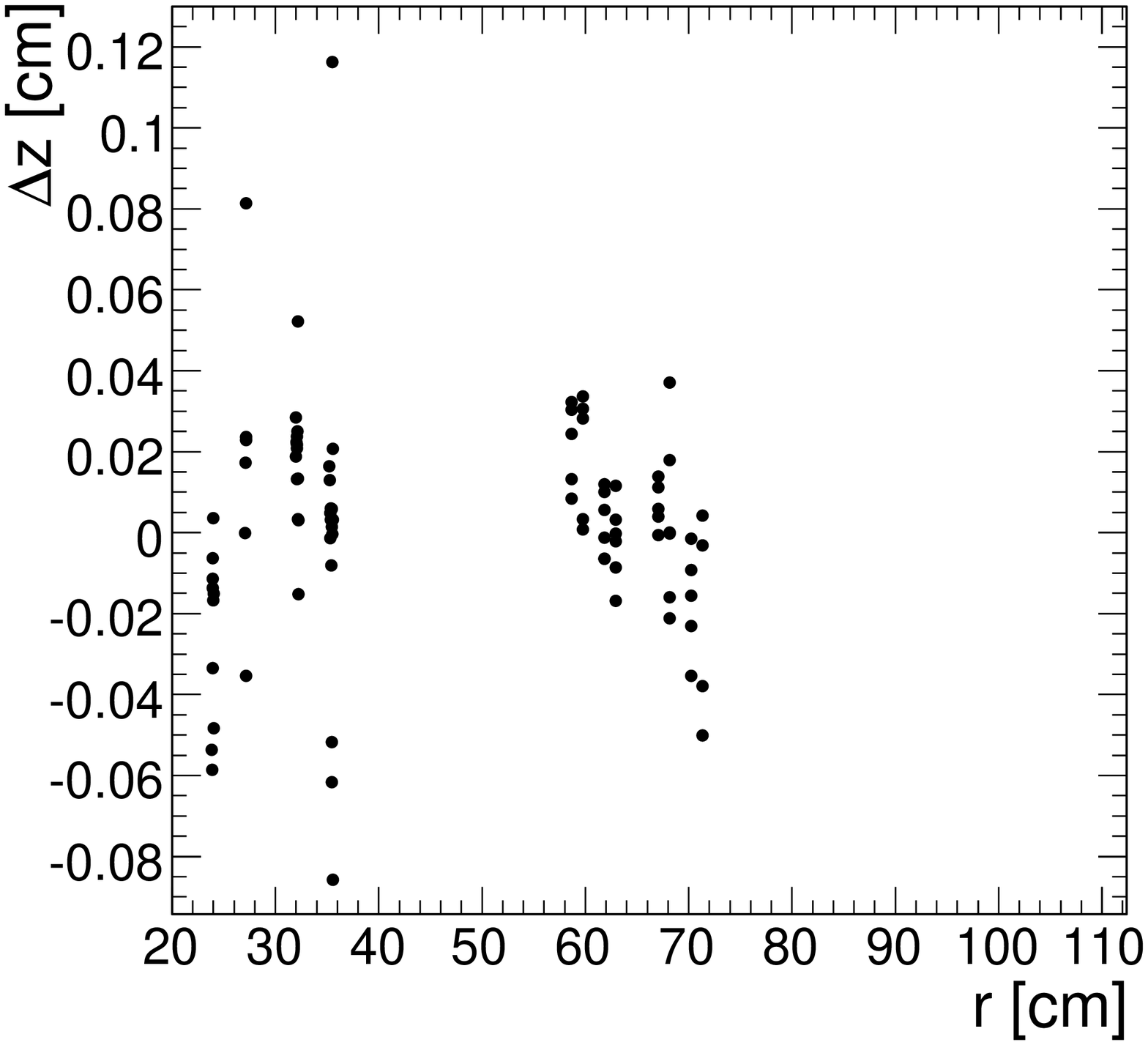,height=3.82cm}
}
\centerline{
\epsfig{figure=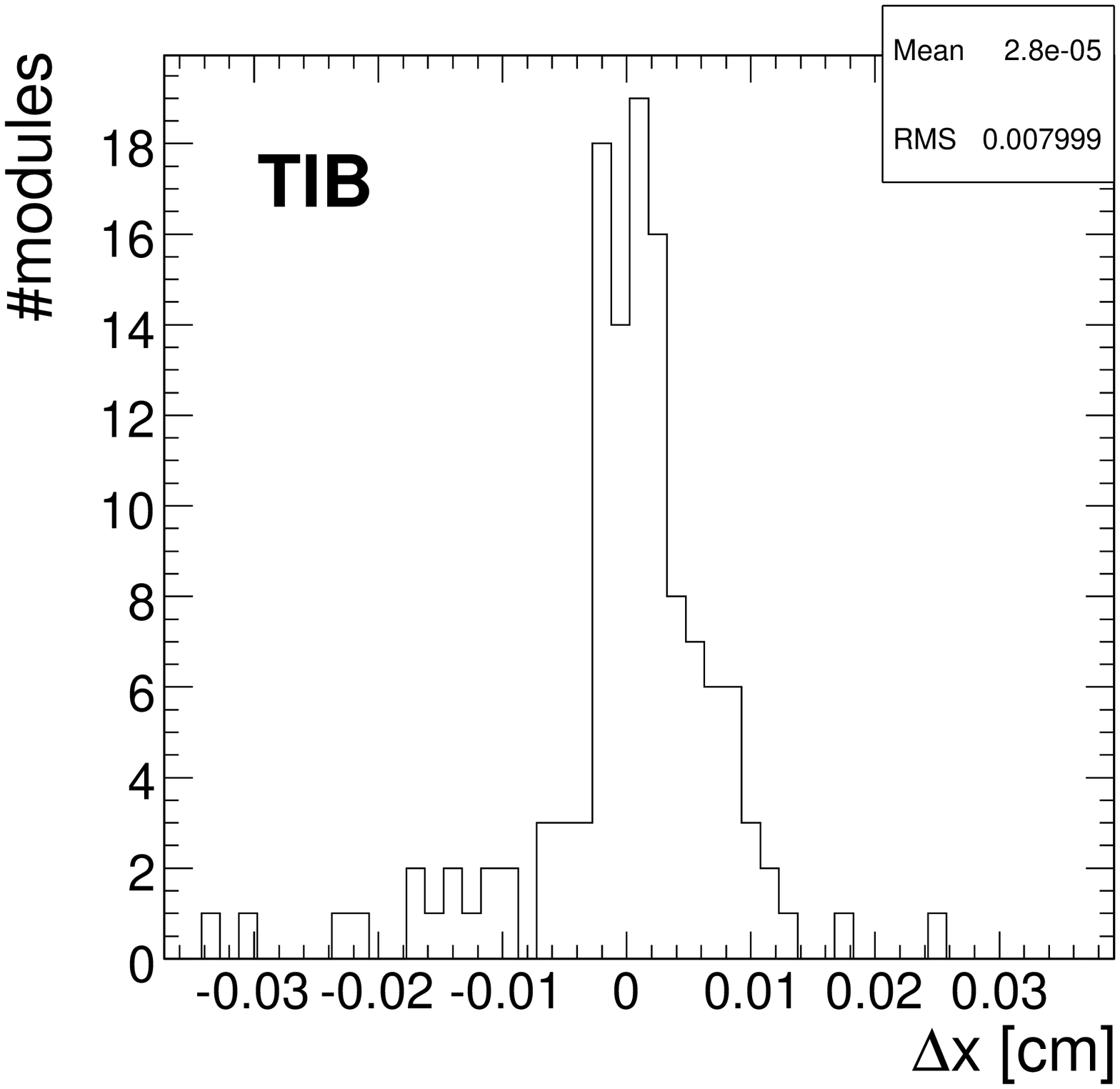,height=3.82cm}
\epsfig{figure=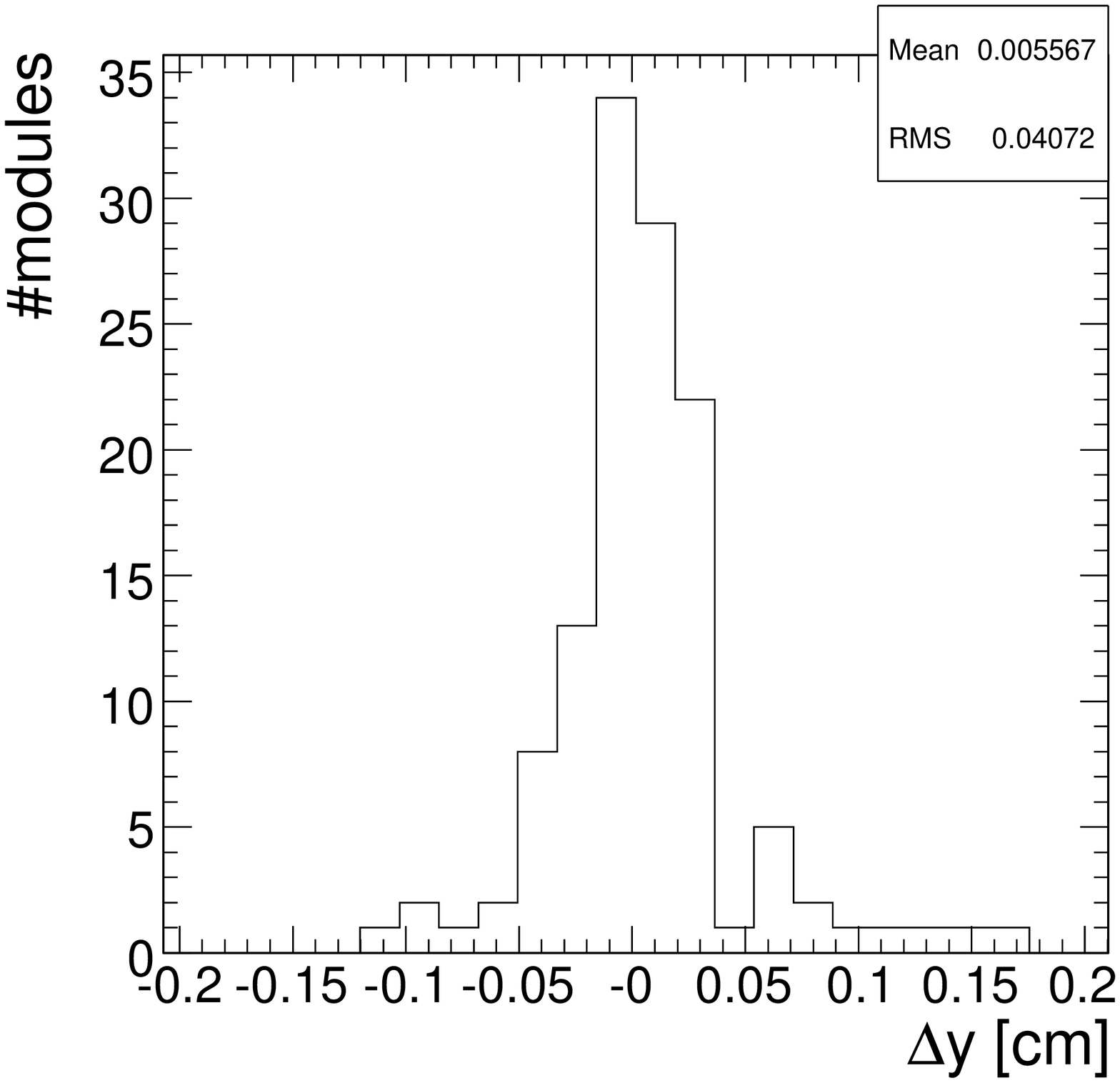,height=3.82cm}
\epsfig{figure=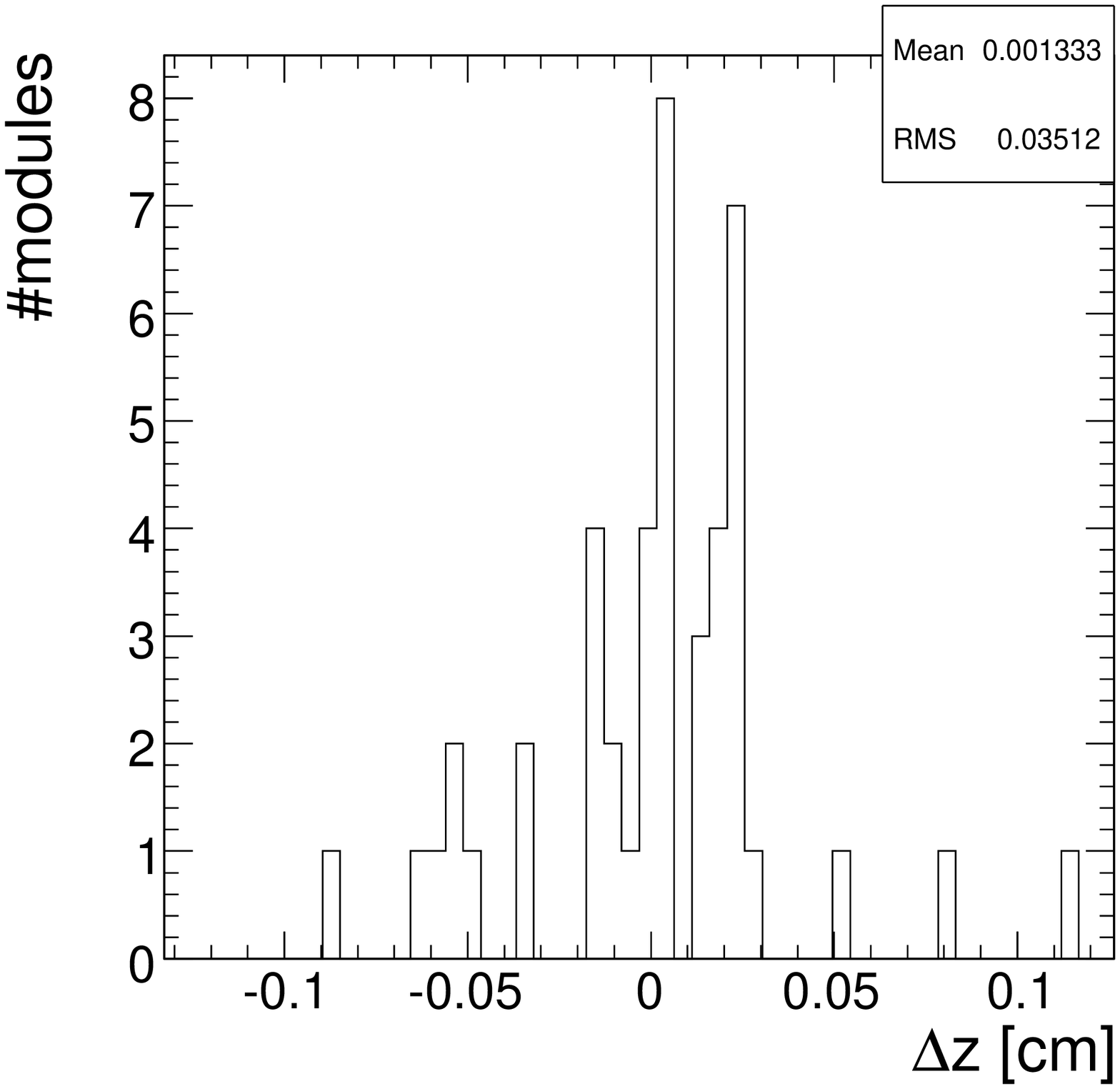,height=3.82cm}
} 
\centerline{
\epsfig{figure=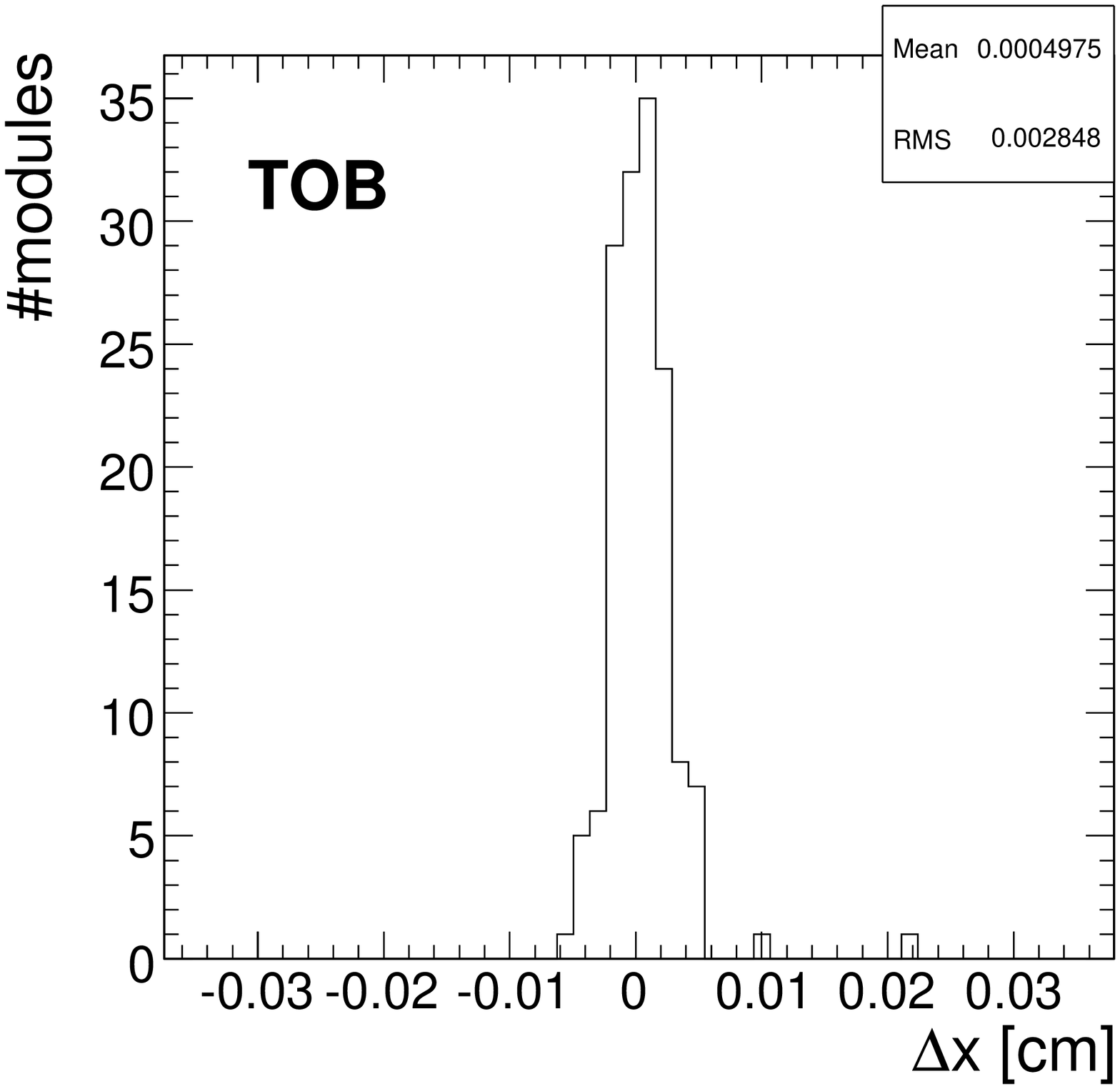,height=3.82cm}
\epsfig{figure=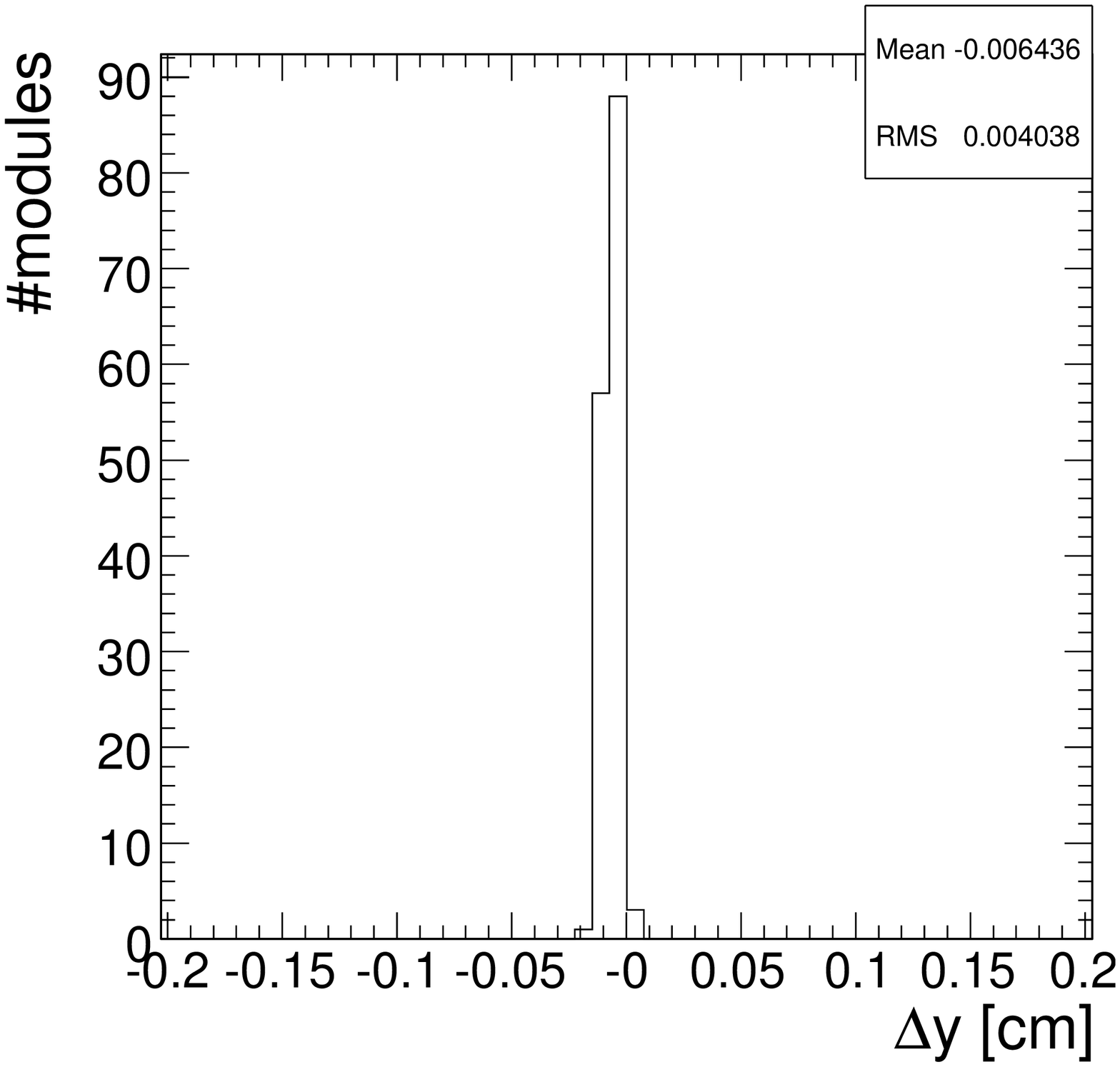,height=3.82cm}
\epsfig{figure=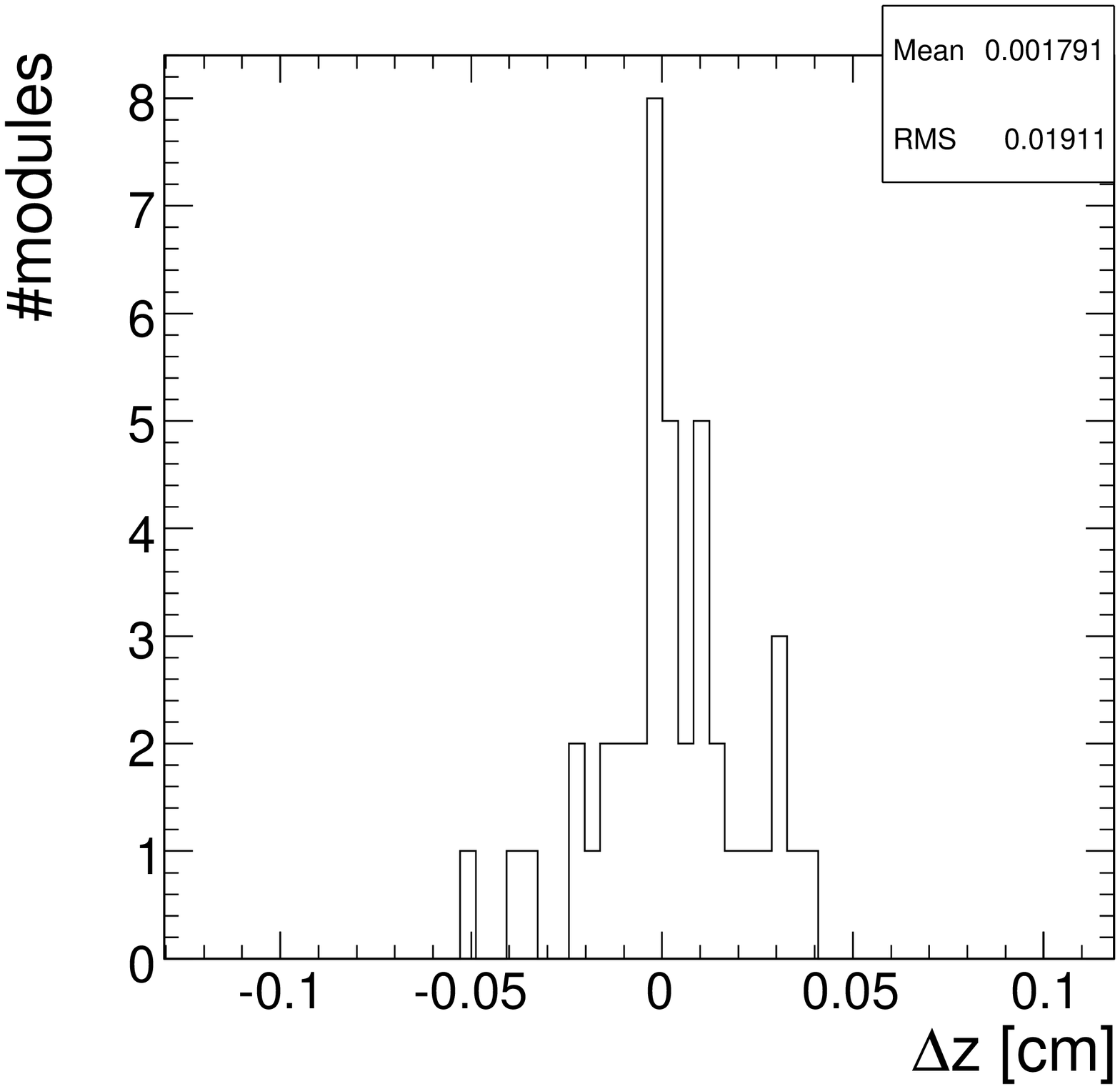,height=3.82cm}
} 
\caption{\sl
Differences in determined  $x$- (left), $y$- (centre) and $z$-positions
(right, only double-sided) of active modules comparing the
+10~$^\circ$C and -10~$^\circ$C configurations.
The differences are stated as a function of
the module radius $r$ (top row) and for modules in TIB (middle row) and TOB (bottom row)
separately. 
\label{fig:newtempcomp}
}
\end{center}
\end{figure}


%
All deviations are within what appears to be statistical scatter,
so this comparison does not show statistically significant movements. 
In the TOB, though certain
layers exhibit larger scatter than the others, there is no evidence of any
coherent shift. In the TIB, there are hints of a small systematic shift vs. the
layer number increasing towards outer layers, that could be caused by a 
relative movement between the cylinders or a rotation around the global $z$
axis. No dependence vs. global $z$ is observed, excluding large effects 
of a rotation about the $y$ direction or a twist.    
\end{enumerate}

\subsection{Stability of the Tracker Endcap}
\label{sec:validation-tec-stability}

\begin{figure}[htbp]
  \begin{minipage}[t]{0.48\textwidth}
    \includegraphics[width=\textwidth]{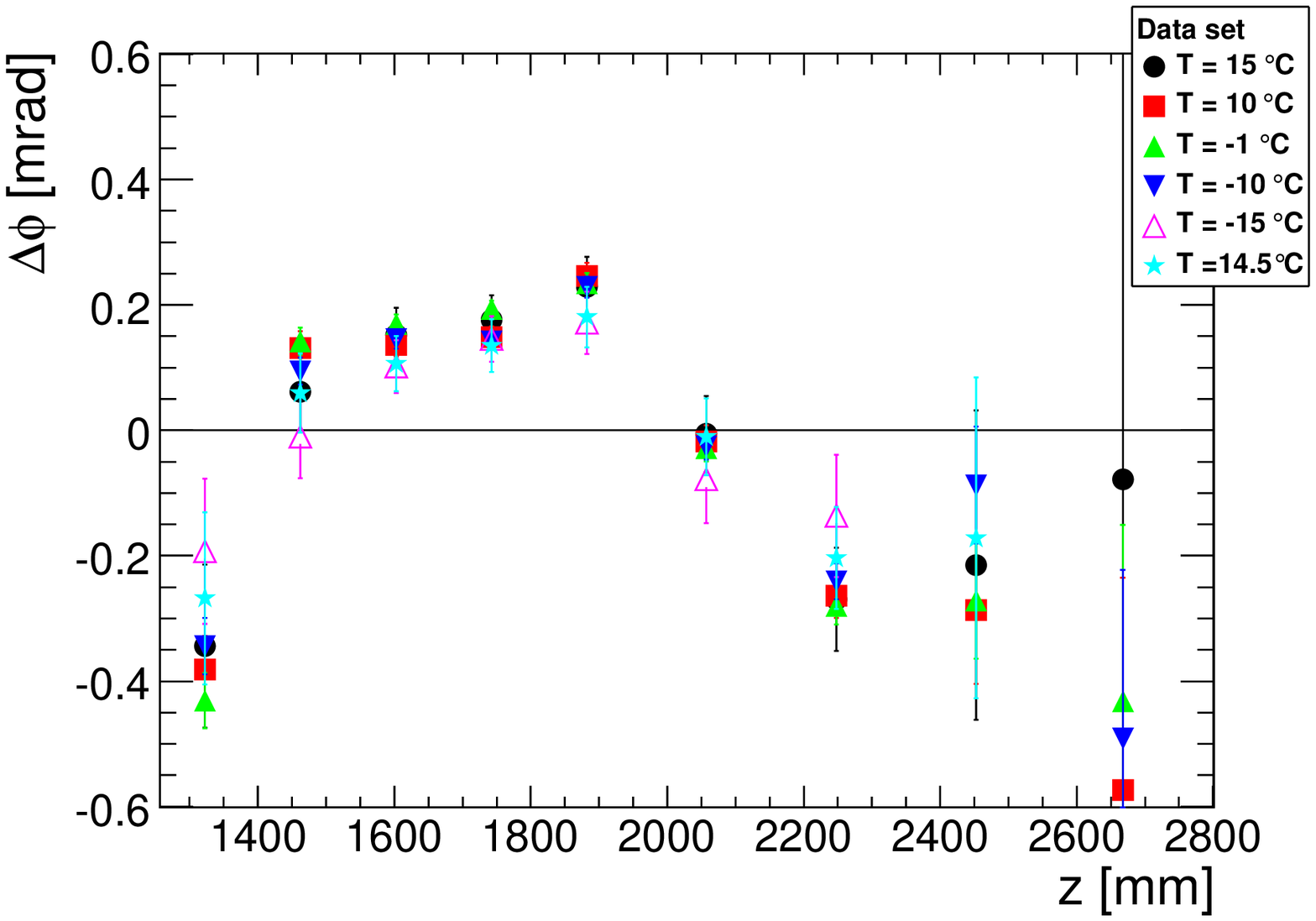}
  \end{minipage} \hfill
  \begin{minipage}[t]{.48\textwidth}
    \includegraphics[width=\textwidth]{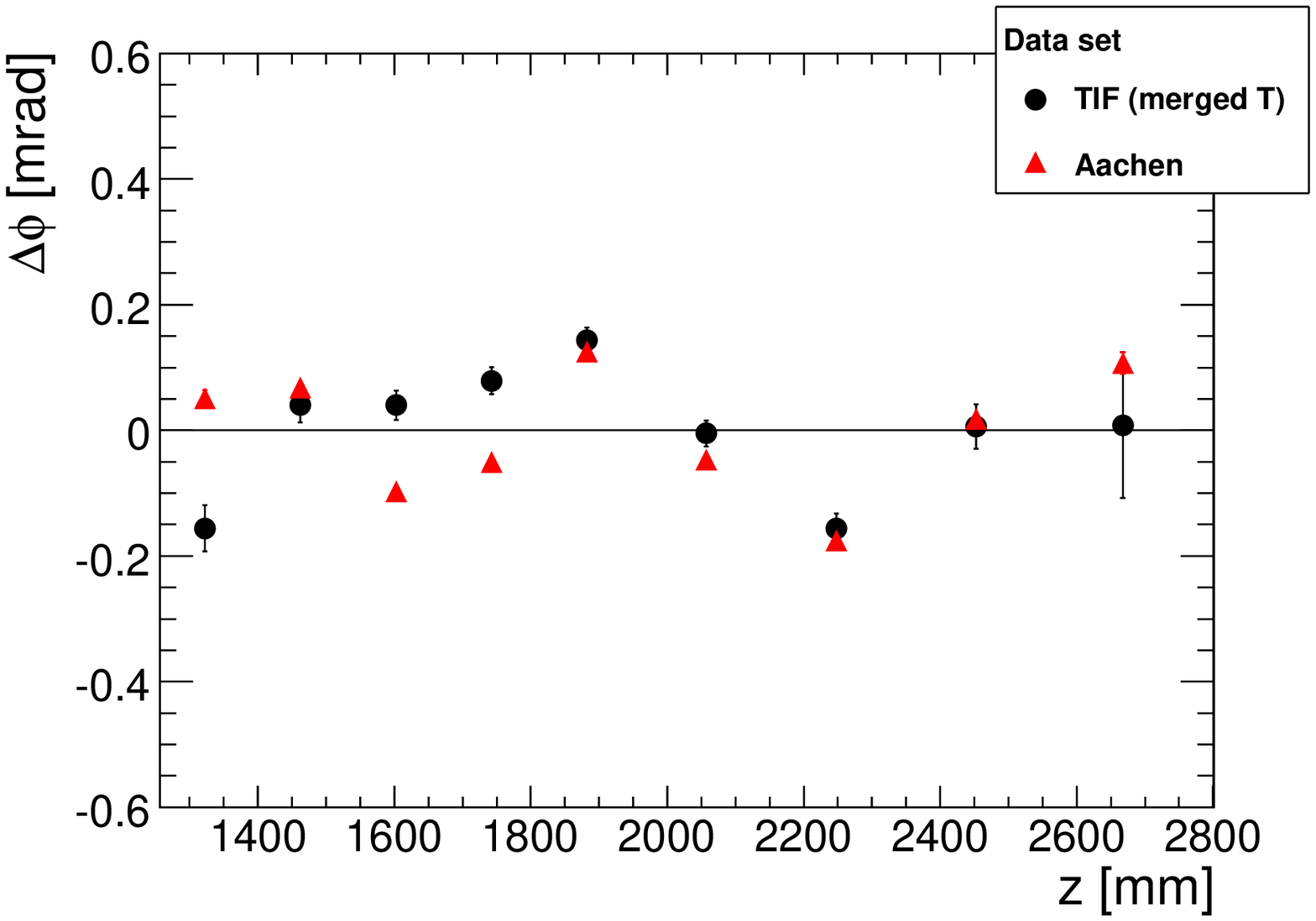}
  \end{minipage}
  \caption{\sl
    Alignment parameter $\Delta \phi$ for TEC disks at six tracker temperature levels (left). Alignment parameter $\Delta \phi$ for TEC disks, determined with TIF data and Aachen data using an obsolete, but common geometry (right).
  }
  \label{fig:tec-comparisons}
\end{figure}

For the TEC stability validation, a comparison is made of the disk alignment
with tracks, using the Kalman filter algorithm, for the temperature levels:
room temperature, 10~$^\circ$C, -1~$^\circ$C, -10~$^\circ$C, -15~$^\circ$C,
and 14.5~$^\circ$C. The alignment parameters calculated with these data sets are
shown on the left of Fig.~\ref{fig:tec-comparisons}. The determined alignment parameters
for the different tracker temperatures agree with each other within their
errors. Disk nine is never hit in the data taken at -15~$^\circ$C or
14.5~$^\circ$C; therefore, there are only eight alignment parameters available at
these temperature levels. At \mbox{-15~$^\circ$C}, the experiment setup changed: Only
the back petals have been activated because there was not enough cooling
power.

In addition, during the TEC integration in Aachen, tracks from cosmic muons
have been recorded. Here, the TEC had been positioned vertically. For each
sector, data had been taken separately after its integration. The processing
of these data had been done using a now obsolete geometry description.
The modules on TEC
rings 2 and 5 are displaced in this geometry by up to 140~$\mu$m.

To create equivalent results, an alignment in $\Delta\phi$ is determined with tracks
from TIF data 
and compared 
with results from 
tracks of sector 2 and 3 of the data from
Aachen. To avoid major differences in the alignment results due to changes in
the geometry, the tracks of the TIF data are reconstructed using the same
geometry as used for Aachen data.
The right of Fig.~\ref{fig:tec-comparisons} shows the alignment parameters gained from
TIF and Aachen data. Except for some changes in disks 1, 3, and 4 of the order
of 0.2~mrad, the TEC seems to have been stable during transportation from
Aachen to Geneva, tilting from a vertical to a horizontal position, and
integration into the tracker. Two petals have been replaced in the active TEC
sectors before taking the TIF data: A back petal of disk 3 and a front petal
of disk 4. Thus, changes in the corrections $\Delta \phi$ of these disks are
expected.

\section{Laser Alignment System Analysis and Discussion}
\label{sec:input-laser-results}

In this section, we discuss results from the Laser Alignment System.
Analysis of the measurements from this system have not been integrated
with the track-based statistical methods. Therefore, we discuss the
data analysis and results independently.

\subsection{Data Taking}

At the TIF, 
data was taken with the laser alignment system.
On the $z^{+}$ side of the tracker, the beams from the alignment tubes of sector 1, 2 and 3 were seen by the barrel modules.
The endcap sectors 2 and 3 were operated with the TEC internal beams and the alignment tubes of those sectors.
Data was taken before cooling the tracker down, during the cooling cycle, and at the end, 
when the tracker was back at room temperature. 

As we mentioned earlier, the Laser Alignment System was designed to measure 
deformations and movements of the tracker support structures.
To do this properly, the whole $\phi$-range of the laser beams needs to be operated.
The fact that only a slice of the tracker was operated during the TIF tests means that no complete picture of the tracker alignment parameters could be obtained.
Nevertheless, the data taking was very useful to verify the proper functioning of the laser beams and the laser data taking.
First of all, the evolution of the measured laser spot positions with temperature was studied. 
Movements could be either due to thermal deformations of the tracker structure, or caused by small movements of the beamsplitter holders.
Then, the data taken in the TEC sectors can be compared to the data obtained during the TEC integration.
Here, observed differences could also have been caused by the handling, transport and insertion of the endcaps.

\subsection{Results from Alignment Tubes}

\begin{figure}[tbp]
\centering
\includegraphics*[angle=0,width=0.85\textwidth]{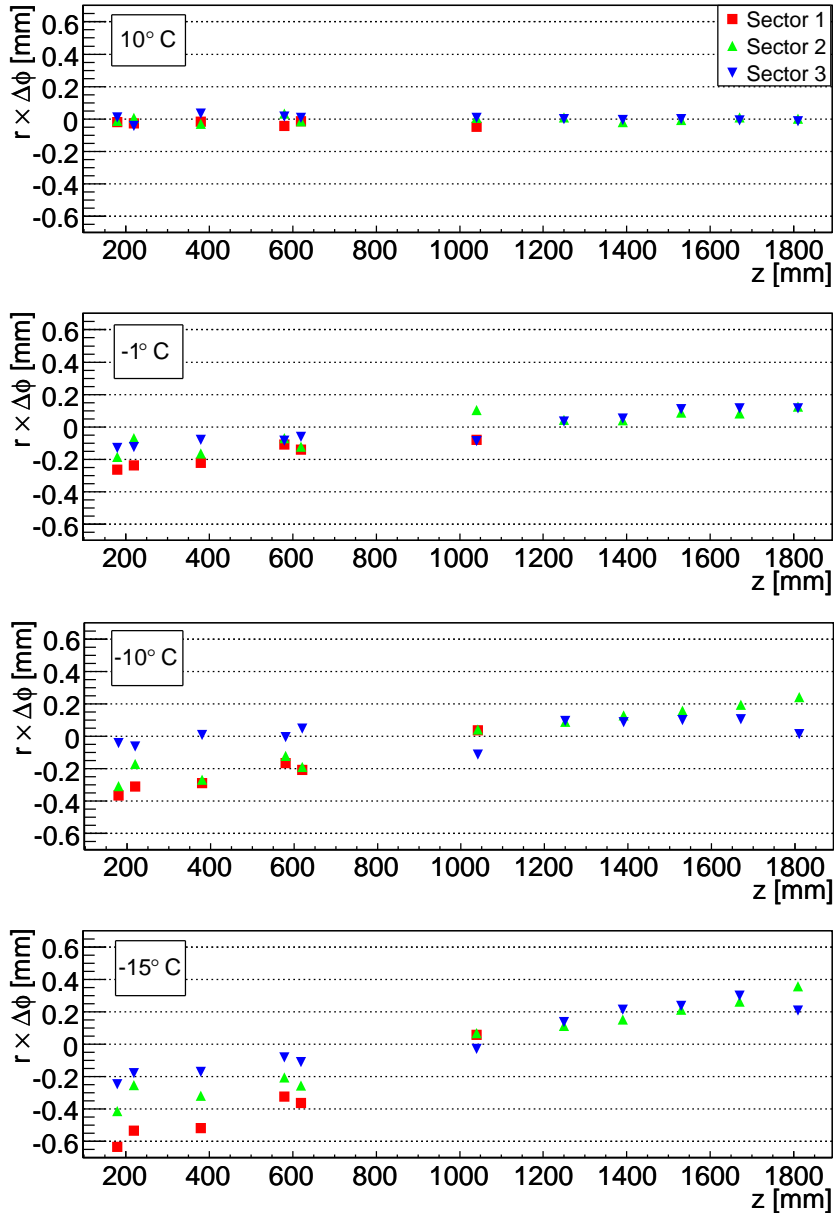}
\caption{Changes in laser spot positions while cooling down the tracker.}
\label{fig:LAS_AT_temp_evolution_max}
\end{figure}

\begin{figure}[tbp]
\centering
\includegraphics*[angle=0,width=0.85\textwidth]{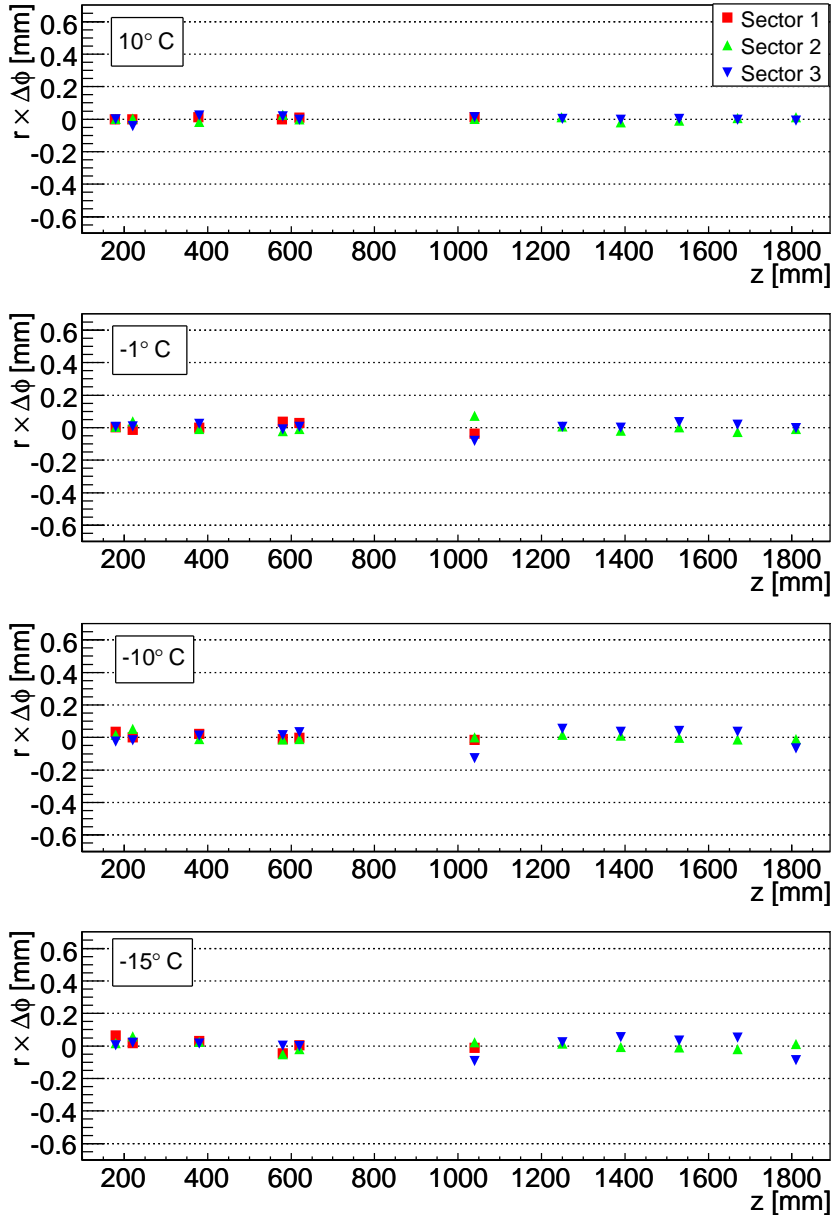}
\caption{Changes in laser spot positions while cooling down the tracker, removing the maximum 
  contribution that can be due to  movements of the alignment tube.}
\label{fig:LAS_AT_temp_evolution_min}
\end{figure}

The alignment tubes were first operated at room temperature. Then, as the tracker was gradually cooled down, they were measured at 10~$^\circ$C, -1~$^\circ$C, -15~$^\circ$C and finally again at room temperature, after the tracker had been warmed up again.
The measured laser spot positions were all compared to the first measurements at room temperature.
The result is shown in Fig.~\ref{fig:LAS_AT_temp_evolution_max}. 
The largest changes of about 600~$\mu$m were observed in the TOB.
The observed movements could come either from movements of the tracker structure, or from movements of the laser beams.
Nevertheless, two bounds can be given.
First, one could assume that all detected movements were due to tracker structure deformations. In this case, we would have observed movements of 600~$\mu$m. 
On the other hand, one could try to absorb as much as of the observed laser spot changes into movements of the laser beams.
In this case, one calculates the tilt of the laser beams and rotations of the alignment tubes that fit best to the observed laser spot movements.
After subtracting this contribution, the remaining laser spot movements would be due to the tracker support deformation.
This is shown in Fig.~\ref{fig:LAS_AT_temp_evolution_min}. Now the maximal movements of the tracker would be less than 100~$\mu$m.

%
%


%
%


\subsection{Comparison of LAS and Track Based Alignment Results}


A comparison is made between the Laser Alignment System residuals and the TEC disk 
alignment results using track based alignment at different temperatures. Corrections are 
applied to the residuals because the beam splitters used by the LAS 
are known to emit two non-perfectly parallel laser beams. Considering the 
laser beamspot radii, the residuals measured at room temperature 
are transformed into disk rotations. The disc corrections, $\Delta \phi$, 
estimated with the Kalman alignment algorithm from cosmic track data 
are used for comparison. There are no significant changes in the TEC 
alignment evaluated with track based alignment at different temperatures, so the track data 
merged from all temperature runs except $T = -15~^\circ$C were used 
to obtain a better precision.

\begin{figure}[ht]
\begin{center}
\centerline{
\epsfig{figure=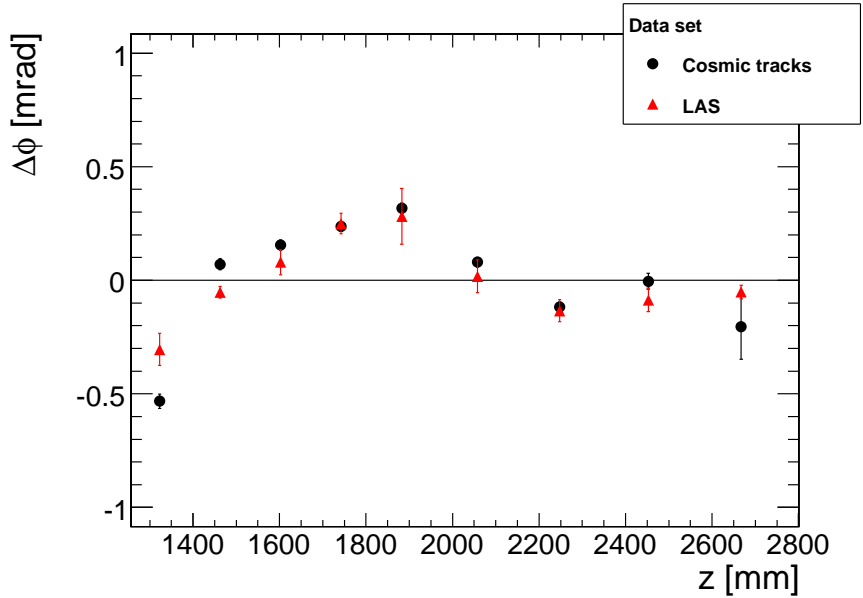,height=8.0cm}
}
\caption{\sl
Corrections $\Delta \phi$ for TEC disks determined with track based alignment and LAS residuals.
\label{fig:lasvskalman}
}
\end{center}
\end{figure}

Because the exact direction of the laser beams is unknown, a linear 
dependence of $\phi$ on $z$ cannot be determined using the LAS residuals. 
Therefore, mean and slope (as a function of $z$) of the corrections to the disc rotations 
are subtracted. The same is done with the results from the Kalman 
alignment algorithm to use a common coordinate system.
The remaining corrections are displayed in Fig.~\ref{fig:lasvskalman}. 
For LAS, the mean and RMS of the four measurements estimated from the four 
active laser beams in the endcap are shown for each disk. 
There are differences among the LAS corrections for the same disk 
of up to 0.7~mrad. These differences are interpreted as misalignment on 
module and petal level. Considering the accuracy of the Kalman alignment
parameters and the spread of the LAS results, the estimated corrections 
show a good agreement.

\section{Summary and Conclusion}
\label{sec:summary}

We have presented results of the CMS tracker alignment analysis 
at the Integration Facility at CERN by means of cosmic tracks, 
optical survey information, and the Laser Alignment System. 
The first alignment of the active silicon modules with three 
different statistical approaches was performed, 
using cosmic track events collected with the
partially active CMS tracker during spring and summer of 2007. 

Optical survey measurements of the tracker
were validated with the track residuals in the active part of the detector.
Clear improvement with respect to the design geometry description was seen.
Overall, further significant improvements
in track $\chi^2$ and track-hit residuals are achieved after track-based
alignment of the tracker at TIF, when compared either to design or
survey geometry.

Detailed studies have been performed on the Tracker Inner and Outer 
Barrel alignment with tracks. The typical achieved precision on module
position measurement in the local $x$ coordinate is estimated to be
about 50 $\mu$m and 80 $\mu$m in the Tracker Outer and Inner Barrels, 
respectively. 
However, since no magnetic field was applied in the tracker,
no momentum estimate of the cosmic tracks was possible.
Therefore, detailed understanding of alignment precision
suffers from uncertainties in multiple scattering of tracks with
unknown momentum, this being the dominant contribution to the
hit resolution. For this reason,
the above alignment precision estimates are based on 
prediction from simulations of hit residuals and may overestimate the detector 
misalignment.

Consistent alignment results have been obtained 
with three different alignment algorithms
Direct comparison of obtained geometries indicate $\sim$150 $\mu$m
consistency in the precisely measured coordinate, consistent with
the indirect interpretation of track residuals.
However, certain $\chi^2$-invariant deformations 
appear in the alignment procedure when using only cosmic tracks.
These $\chi^2$-invariant 
deformations do not affect track residuals and therefore are not visible in
the alignment minimisation, 
thus limiting  understanding of relative position of all modules in space
from the pure geometrical point of view. 

Alignment of the Tracker Endcap was performed at the disk level, both
with tracks and by operating the CMS Laser Alignment 
System and showed good agreement between the two results. 

No significant deformations of the tracker have been 
observed under stress and with variation of temperature, within the 
resolution of the alignment methods. 

The operation of the Laser Alignment System during the TIF slice test 
has shown that the laser beams operate properly. Useful laser signals 
were detected by all modules that were illuminated by the laser beams.
In the worst-case scenario, 
where all observed laser spot shifts are assumed to come from structure 
deformations, the movements would be up to 600~$\mu$m.
Assuming that most of the observed changes 
were coming from laser beam and alignment tube movements,
shifts go down below 100~$\mu$m.
To disentangle the two contributions and get a complete picture 
of the tracker deformations, more beams, distributed around all the 
2$\pi$ $\phi$-range, have to be operated.

Finally, experience gained in alignment analysis of the silicon modules
at the Tracker Integration Facility is valuable in preparation for
the full CMS tracker alignment, which is crucial for high precision necessary 
to achieve the design physics goals of the CMS detector.

\section{Acknowledgments}
\label{sec:acknowledgment}


We thank the administrative staff at CERN and other Tracker
Institutes.  
This work has been supported by:
the Austrian Federal Ministry of Science and Research;
the Belgium Fonds de la Recherche Scientifique and Fonds voor Wetenschappelijk Onderzoek; 
the Academy of Finland and Helsinki Institute of Physics; 
the Institut National de Physique Nucl\'eaire et de Physique des
    Particules~/~CNRS, France; 
the Bundesministerium f\"ur Bildung und Forschung, Germany;
the Istituto Nazionale di Fisica Nucleare, Italy; 
the Swiss Funding Agencies;
the Science and Technology Facilities Council, UK; 
the US Department of Energy, and National Science Foundation.
Individuals have received support from
the Marie-Curie IEF program (European Union) and
the A.~P. Sloan Foundation.

-----------------------------------------------------------------------



\end{document}